\documentclass[11pt,twoside]{report}

\usepackage[authoryear]{natbib}
\usepackage{azb}
\usepackage{amsmath,amssymb}
\usepackage{version}
\usepackage{epsfig}

\bibpunct{(}{)}{;}{a}{}{,}

\newcommand{\aj}{Astron. J.}
\newcommand{\apa}{Astron. Astrophys.}
\newcommand{\apal}{\apa{} Lett.}
\newcommand{\apj}{Astrophys. J.}
\newcommand{\apjl}{\apj{} Lett.}
\newcommand{\apjs}{Astrophys. J. Supp.}
\newcommand{\araa}{Ann. Rev. Astron. Astrophys.}
\newcommand{\mnras}{Mon. Not. R. Astron. Soc.}
\newcommand{\mnrasl}{\mnras{} Lett.}
\newcommand{\nast}{New Astron.}

\newcommand{\pnas}{Proc. Natl. Acad. Sci.}

\newcommand{\ion}[2]{\mbox{#1\,{\sc #2}}}

\newcommand{\hub}{\mbox{km s$^{-1}$ Mpc$^{-1}$}}
\newcommand{\kmps}{\mbox{km s$^{-1}$}}

\newcommand{\kpc}{\mbox{$h^{-1}$kpc}}
\newcommand{\mpc}{\mbox{$h^{-1}$Mpc}}
\newcommand{\mpci}{\mbox{$h$ Mpc$^{-1}$}}
\newcommand{\lya}{Ly$\alpha$}
\newcommand{\oned}{one-dimensional}
\newcommand{\thrd}{three-dimensional}

\title{THE LARGE-SCALE STRUCTURE OF THE UNIVERSE \\[1.8ex] IN ONE DIMENSION}
\author{Hu Zhan}
\director{Li-Zhi Fang, David Burstein, and Daniel Eisenstein}
\begin{document}
\ifx\href\undefined\else\hypersetup{linktocpage=true}\fi
\maketitle

\addtolength{\parindent}{-24pt}
\includegraphics[height=8.5in]{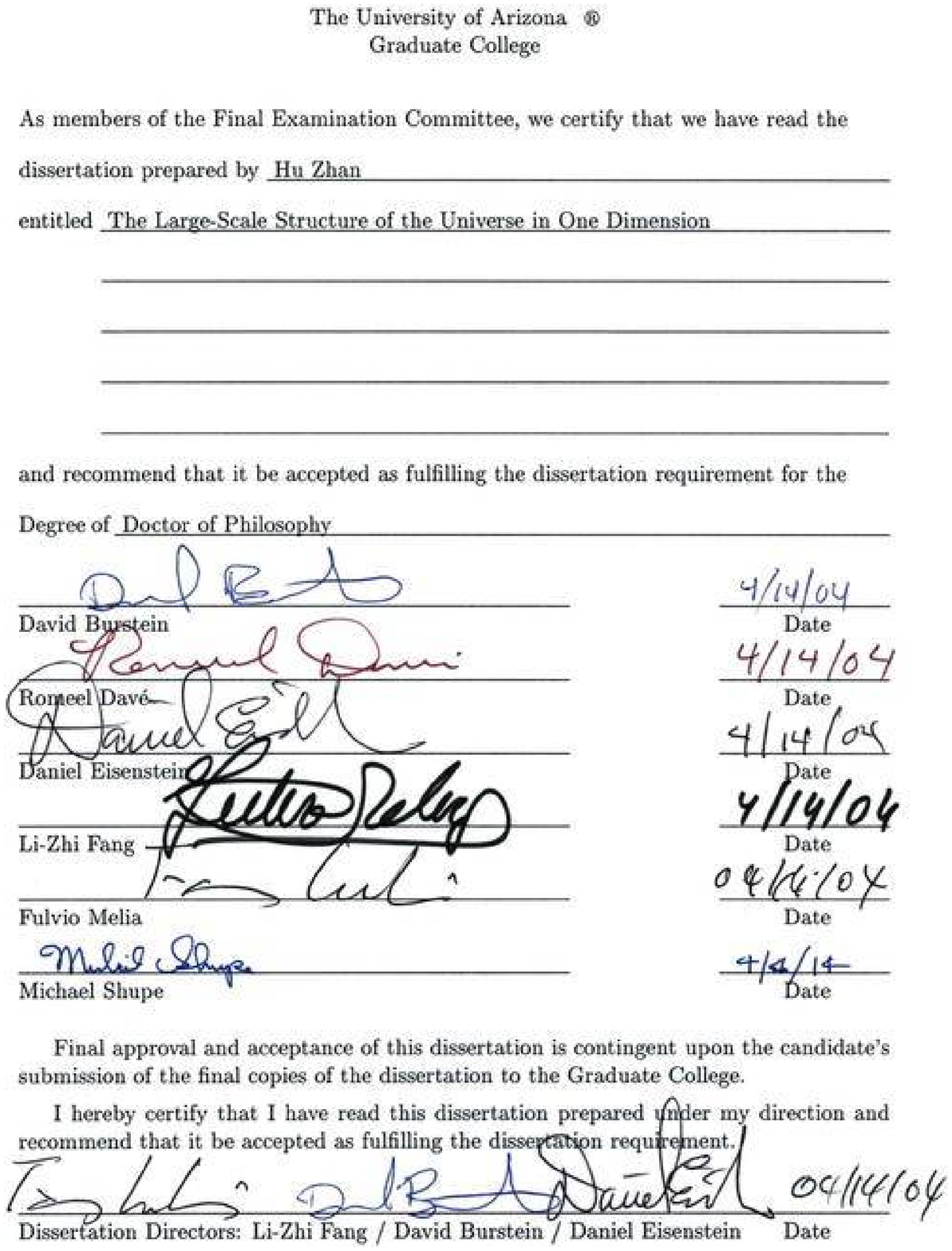}
\addtolength{\parindent}{24pt}
       
\begin{statement}
   This dissertation has been submitted in partial 
fulfillment of requirements for an advanced degree at The 
University of Arizona and is deposited in the University 
Library to be made available to borrowers under rules of the 
Library.

   Brief quotations from this dissertation are allowable 
without special permission, provided that accurate 
acknowledgment of source is made.  Requests for permission 
for extended quotation from or reproduction of this 
manuscript in whole or in part may be granted by the copyright
holder. 

\end{statement}

\begin{acknowledgement}
I thank my advisors David Burstein, Daniel Eisenstein, and Li-Zhi Fang for 
their supervision and guidance over the course of my study.

I also thank my mentors and collaborators, Bruce Barrett, Romeel Dav\'{e}, 
Xiaohui Fan, Priya Jamkhedkar, Fulvio Melia, Petr Navr\'{a}til, 
Andreas Nogga, Michael Shupe, James Vary, Rogier Windhorst, and 
Ann Zabludoff. Without them, I would probably have spent a 
lot more time practicing trial and error.

I appreciate the financial support from Bruce in the Spring and Summer of 
2003, during which ideas in this dissertation germinated. I have benefited 
from grants from the Theoretical Astrophysics Program and registration fee 
waivers from the Physics Department.

I am grateful to my wife, Yue, for her enduring support.
\end{acknowledgement}

\begin{dedication}
This dissertation is dedicated to my parents, Hua'an and Lanzhen, 
and my wife, Yue.
\end{dedication}

\setlength{\parskip}{0ex}
       
\tableofcontents

\listoffigures

\listoftables

\abbreviations{
\medskip
\begin{tabular}{l l}
2dF  & Two Degree Field (Survey) \\
CDM  & Cold Dark Matter \\
CMB  & Cosmic Microwave Background \\
EOS  & Equation of State \\
GRF  & Gaussian Random Field \\
IGM  & Intergalactic Medium \\
LCDM & $\Lambda$ CDM (Model) or Low-Density-and-Flat CDM (Model)\\
LOS  & Line of Sight or Line-of-Sight \\
OCDM & Open CDM (Model) \\
PS   & Power Spectrum \\
QSO  & Quasi-Stellar Object or Quasar \\
SCDM & Standard CDM (Model) \\
SDSS & Sloan Digital Sky Survey \\
SPH  & Smooth-Particle Hydrodynamics \\
TCDM & $\tau$ CDM (Model) or Tilted CDM (Model) \\
TSC  & Triangular-Shaped Cloud (Assignment Function)\\
UV   & Ultraviolet \\
WHIM & Warm-Hot Intergalactic Medium \\
WMAP & Wilkinson Microwave Anisotropy Probe \\
\end{tabular} }
       
\begin{abstract}
\noindent
I investigate statistical properties of \oned{} fields in the universe
such as the \lya{} forest (\lya{} absorptions in the quasar spectrum) 
and inverted line-of-sight densities.
The \lya{} forest has opened a great window for studying the 
large-scale structure of the universe, because it can probe the 
cosmic density field over a wide range of redshift at relatively high 
resolution, which has not been easily accessible with other types of 
observations.

The power spectrum completely characterizes Gaussian random fields. 
However, because of gravitational clustering, the cosmic density field is 
already quite non-Gaussian on scales below 10 \mpc{} at redshift $z=3$. I 
analyze the covariance of the one-dimensional mass power spectrum, which 
involves a fourth-order statistic, the trispectrum. The covariance
indicates that Fourier modes in the cosmic density field are highly 
correlated and that the variance of the measured one-dimensional mass 
power spectrum is much higher than the expectation for Gaussian random
fields. It is found that rare high-density structures contribute
significantly to the covariance. The window function due to the length of
lines of sight introduces additional correlations between different 
Fourier modes. 

In practice, one observes quasar spectra instead of 
one-dimensional density fields. As such, flux power spectrum has been 
the basis of many works. I show that the nonlinear transform between 
density and flux quenches the fluctuations so that the flux power spectrum
is less sensitive to cosmological parameters than the \oned{} 
mass power spectrum. 
The covariance of the flux power spectrum is nearly Gaussian, 
which suggests that higher-order statistics may be less effective for 
the flux.

Finally, I provide a method for inverting Ly$\alpha$ forests and 
obtaining line-of-sight densities, so that statistics can be measured
from \oned{} density fields directly. 
\end{abstract}

\chapter{Introduction}

\noindent
The cosmological principle, i.e.~the assumption of homogeneity 
and isotropy \citep{m35}, has been born out by decades of tests 
\citep[for reviews, see][]{p80,p93}.
As profound as it is, structures do rise from minute 
primordial fluctuations because of gravitational instability 
\citep[e.g.][]{j28,l34}. On scales beyond galaxy clusters, 
the cosmic density field remains in the linear or quasi-linear regime, 
which means the formation and evolution 
of the large-scale structure still bears the initial imprint of 
cosmology. In other words, the large-scale structure is a sensitive 
tool for measuring fundamental properties of the universe. 

Galaxies and clusters of galaxies were the first objects used to 
study the 
large-scale structure of the universe \cite[e.g.][]{h34, s34}. Recent
surveys include the Center for Astrophysics Survey \citep{hgl90}, 
the Automatic Plate-measuring Machine Survey \citep{mes90}, the Las 
Campanas Redshift Survey \citep{slo96}, the Infrared Astronomical 
Satellite Point Source Catalog Redshift Survey \citep{ssm00},
the Two Degree Field (2dF) Galaxy Redshift Survey \citep{cdm01}, and the
Sloan Digital Sky Survey \citep[SDSS, initial release,][]{slb02}. 
One useful statistic is the galaxy clustering power spectrum
\citep[PS,][]{p91, be93, fds93, lks96, pbb01, dnt02}. Because galaxies
reside in high-density peaks of the cosmic density field, galaxy 
statistics may be (slightly) biased 
with respect to the true statistics of the field 
\citep{ds86, gf94, kns97, dl99, bcf00, ps00, sls01, vhp02, wdk04}. 
Galaxy surveys are often limited by depth. For example, the mean redshift
of the above-mentioned surveys is around 0.1 or less, which translates
to a volume on the order of a thousandth of the Hubble volume. Such a 
small survey volume limits our ability to study the large-scale 
structure at high redshift and its evolution.

Quasars, or quasi-stellar objects (QSOs), are far more luminous than 
galaxies, and they have been observed 
near the edge of the visible universe \citep{fnl01, fss03}. The 2dF QSO 
survey \citep{bsc00} and the SDSS will provide large QSO samples for 
statistics such as the QSO clustering correlation function and PS
\citep{s84, sfb87, is88, mf93, sb94, lac98, ohs03}. Like galaxies,
QSOs also live in high-density peaks of the universe, and 
their statistics may be biased as well. In fact, galaxies are 
often found to cluster around QSOs \citep{t86, fbk96, plc01, kh02}. 
Because of large separations, QSOs statistics are more reliable for 
studying the clustering of matter above 10 \mpc, where $h$ is the Hubble 
constant in units of 100 \hub{}.

One often sees numerous \lya{} absorption lines in QSO spectra 
that are resulted from absorptions by the diffusely distributed and 
photoionized intergalactic medium (IGM). Absorption systems that have 
neutral-hydrogen column density $10^{12} \le N_{\rm HI} \le 10^{17}$ 
cm$^{-2}$ are also called the \lya{} forest 
\citep[for a review, see][]{r98}. It has been demonstrated by various 
hydrodynamical simulations that the IGM traces the density of the 
underlying mass field on large scales
\citep{cmo94, zan95, hkw96, mco96, dhw97, bma99}. In other words, 
baryon densities in \lya{} absorption systems are roughly proportional 
to total matter densities for $\rho/\bar{\rho}\lesssim 10$ 
\citep{bd97,gh98,zma98}, where $\rho$ is the density and $\bar{\rho}$ 
is the mean density.
Therefore, the \lya{} forest offers a less biased probe of the 
large-scale structure of the universe. 

The \lya{} forest has been observed up to $z \sim 6$ \citep{bfw01}, 
beyond which neutral hydrogen (\ion{H}{i}) completely absorbs the QSO 
flux at rest wavelength $\lambda_0 = 1216$ \AA{} due to the 
Gunn-Peterson effect \citep{gp65}. For each line of sight (LOS) to a 
QSO, one can sample the density field almost continuously in one dimension 
and obtain more information than the clustering of QSOs. With enough LOS's 
covering $0 \leq z \lesssim 6$, the \lya{} forest will enable us to 
establish a more complete picture of the universe and its evolution. 
In fact, statistics of the \lya{} forest have been applied to many aspects of 
large-scale structure studies such as recovering the initial linear mass 
PS \citep{cwk98, cwp99, h99, ff00, mmr00, gh02, zsh03}, measuring the 
flux PS and bispectrum \citep{hbs01, mms03, kvh04, vmh04}, estimating
cosmological parameters \citep{mm99, zht01, cwb02, smm03, m03, vmt03},
inverting the \lya{} forest \citep{nh99, pvr01, z03}, finding the 
applicable range of the hierarchical clustering model \citep{fpf01, zjf01}, 
and estimating the velocity field \citep{zf02}. These studies show that 
the \lya{} forest has provided an important complement to studies 
based on galaxy and QSO samples.

The unique nature of the \lya{} forest brings its own 
subtleties. First, the \lya{} forest probes well into the nonlinear
regime, e.g.~from several \mpc{} down to tens of \kpc{}. The 
non-Gaussianity on such small scales plays an important role in 
statistics, and cosmological hydrodynamical simulations are needed for 
the study. Second, the direct observable is the \lya{} flux rather than 
the density. Statistics of the density, which are more fundamental to 
cosmology, have to be recovered from flux statistics. Third, the 
observed \lya{} forest may be affected by many elements such as the 
continuum fitting \citep{hbs01}, the UV ionization background 
\citep{rms97, sbdk00, mm01, mw04},
saturated \lya{} absorptions \citep{z03, vhc04}, metal-line
contaminations, and redshift distortion. At low redshift, additional 
uncertainties arise from the thinning of the \lya{} forest 
\citep{rpm98, tle98} and the shock-heated warm-hot IGM 
\citep[WHIM,][]{dhk99, dt01, dco01}. 

This dissertation focuses on theoretical and numerical aspects of 
\oned{} statistics of the large-scale structure. It is organized as 
follows. Chapter \ref{ch:p1d} discusses the relation between the 
\oned{} PS and the \thrd{} PS with an emphasis on spatial average.
The covariance of the \oned{} mass PS is derived for  Gaussian 
random fields (GRFs) and the cosmic density field in \mbox{Chapter 
\ref{ch:cov}}. One-dimensional mass PS's and their covariances are 
calculated in Chapter \ref{ch:mps} for simulated density fields. Flux 
PS's and their covariances are measured from simulated as well as 
observed \lya{} forests in  \mbox{Chapter \ref{ch:fps}}, and they are 
compared with corresponding mass PS's and covariances. 
Effects of the UV ionization background and the WHIM are also 
discussed there.  \mbox{Chapter \ref{ch:inv}} introduces the mean 
density--width relation \citep{z03} and its application in 
extracting density information from saturated \lya{} absorptions. 
\mbox{Chapter \ref{ch:con}} concludes the dissertation.

\chapter{One-Dimensional Power Spectrum} \label{ch:p1d}
\noindent
The one-dimensional mass PS and its relation to the
three-dimensional mass PS have been frequently utilized to recover the
linear mass PS from the Ly$\alpha$ forest \citep{cwk98, cwp99, cwb02,
gh02}. This opens a great window for studying the large-scale
structure of the universe over a wide range of redshift. 

For an ensemble of isotropic fields, the one-dimensional mass PS $P_{\rm
1D}(k)$ is a simple integral of the three-dimensional mass PS $P(k)$
\citep{lhp89},
\begin{equation} \label{eq:iso321}
P_{\rm 1D}(k)=\frac{1}{2\pi}\int_k^{\infty} P(k^{\prime})k^{\prime}\,
{\rm d}k^{\prime},
\end{equation}
where $k$ is the LOS wavenumber. However, there is only one
observable universe. The ensemble average has to be replaced by a
spatial average. For instance, one may sample multiple LOS
densities from the three-dimensional cosmic density field and use the
average PS of the one-dimensional densities in place of the
ensemble-average quantity. Such spatial average alters the relation
between the \oned{} mass PS and the \thrd{} mass PS. I discuss the 
modification to equation (\ref{eq:iso321}) and its significance in this 
chapter. Although I use the density field as an example, the results 
obtained here are generic.

\section{Three-Dimensional Power Spectrum} \label{sec:p3D}

It is beneficial to review the \thrd{} PS before we look at the \oned{}
PS. I assume the following convention of Fourier transforms for a cubic
density field $\rho({\bf x})$ in the volume $V = B^3$:
\begin{eqnarray} \label{eq:Fourier}
\delta({\bf x}) &=& \frac{1}{V} \sum_{{\bf n}=-\infty}^{\infty}
 \hat{\delta}({\bf n}) \, e^{2\pi i {\bf n}\cdot{\bf x}/B}, \\ \nonumber
\hat{\delta}({\bf n}) &=& \int_V \delta({\bf x}) \,
e^{-2\pi i {\bf n}\cdot{\bf x}/B}\,{\rm d}{\bf x},
\end{eqnarray}
where $\delta({\bf x}) = [\rho({\bf x})-\bar{\rho}]/\bar{\rho}$ is the
overdensity, $\hat{\delta}({\bf n})$ is the Fourier transform of 
$\delta({\bf x})$,
the summation $\sum_{{\bf n}=-\infty}^{\infty}$ is an abbreviation for
$\sum_{n_1,n_2,n_3=-\infty}^{\infty}$, and the wavevector 
${\bf k}$ equals $2\pi{\bf n}/B$ with integral $n_1$, $n_2$, and $n_3$. 
I choose this form of Fourier transforms because real
surveys always have a finite volume and, after all, the visible universe 
is finite. Equation (\ref{eq:Fourier}) will approach the limit of 
continuous Fourier transforms as the volume increases.

With the understanding that Fourier modes exist only at
discrete wavenumbers, i.e.~$n_1$, $n_2$, and $n_3$ are integers, 
one may use ${\bf k}$ and ${\bf n}$ interchangeably
for convenience. To be complete, the orthonormality relations are
\begin{eqnarray} \label{eq:orth_n}
\frac{1}{V} \int_V e^{-2\pi i ({\bf n}-{\bf n}')\cdot {\bf x}/B} 
{\rm d} {\bf x} &=& \delta^{\rm K}_{{\bf n},{\bf n}'}, \\
\label{eq:orth_x} \frac{1}{V} \sum_{{\bf n}=-\infty}^{\infty} 
e^{2\pi i {\bf n} \cdot ({\bf x} - {\bf x}')/B} 
&=& \delta^{\rm D}({\bf x}-{\bf x}'), 
\end{eqnarray}
where $\delta^{\rm K}_{{\bf n},{\bf n}'}$ is the three-dimensional
Kronecker delta function, and $\delta^{\rm D}({\bf x}-{\bf x}')$ the Dirac
delta function.

The three-dimensional mass PS of the universe is defined through
\begin{equation} \label{eq:powsp_n}
\langle \hat{\delta}({\bf k})\hat{\delta}^{*}({\bf k}') \rangle
= P({\bf k}) V \delta^{\rm K}_{{\bf n},{\bf n}'},
\end{equation}
where $\langle \ldots \rangle$ stands for an ensemble average. One may
define an observed PS, 
\begin{equation}
P'({\bf k}) = |\hat{\delta}({\bf k})|^2 / V, 
\end{equation}
so that $P({\bf k}) = \langle P'({\bf k}) \rangle$. For an 
isotropic field, $P({\bf k})$ is a function of the length of 
${\bf k}$ only, i.e.~$P({\bf k}) \equiv P(k)$. 
 A shot-noise term should 
be included if the PS is measured from discrete objects, and it is 
inversely proportional to the mean number density of the objects. I 
neglect the shot-noise in this dissertation. An alternative definition 
of $P(k)$ utilizes the correlation function $\xi(r)$, e.g. 
\begin{eqnarray} \label{eq:xi=dd}
\xi(r) &=& \langle \delta({\bf x}+{\bf r})\delta({\bf x})\rangle, \\
P(k) &=& \int_V \xi(r)\, e^{-i {\bf k}\cdot{\bf r}}\, {\rm d}{\bf r}.
\label{eq:p=xi}
\end{eqnarray}

The four-point function of $\delta({\bf x})$ is 
\begin{equation} \label{eq:x4pt}
\langle \delta({\bf x}_{\rm a})\delta({\bf x}_{\rm b})
        \delta({\bf x}_{\rm c})\delta({\bf x}_{\rm d}) \rangle =
\xi_{\rm ab}\xi_{\rm cd} + \xi_{\rm ac}\xi_{\rm bd}
+\xi_{\rm ad}\xi_{\rm bc} + \eta,
\end{equation}
where $\xi_{\rm ab}$ stands for 
$\xi(|{\bf x}_{\rm a} - {\bf x}_{\rm b}|)$, and $\eta$ is the reduced 
four-point correlation function that has six degrees of freedom arising 
from relative coordinates of the four points under the constraint of
homogeneity \citep{p80}. From equations (\ref{eq:p=xi}) and 
(\ref{eq:x4pt}), we have the four-point function of 
$\hat{\delta}({\bf k})$
\begin{eqnarray} \label{eq:4point}
\langle \hat{\delta}({\bf k}_{\rm a})\hat{\delta}^{*}({\bf k}_{\rm b}) 
\hat{\delta}({\bf k}_{\rm c})\hat{\delta}^{*}({\bf k}_{\rm d})\rangle &=& 
V^2 \big [P(k_{\rm a})P(k_{\rm c})
\delta^{\rm K}_{{\bf n}_{\rm a}, {\bf n}_{\rm b}}
\delta^{\rm K}_{{\bf n}_{\rm c}, {\bf n}_{\rm d}} + 
P(k_{\rm a})P(k_{\rm b})\delta^{\rm K}_{{\bf n}_{\rm a}, {\bf n}_{\rm d}}
\delta^{\rm K}_{{\bf n}_{\rm b}, {\bf n}_{\rm c}} \big] \nonumber \\
& & + VT({\bf k}_{\rm a}, -{\bf k}_{\rm b}, {\bf k}_{\rm c}, 
-{\bf k}_{\rm d}),
\end{eqnarray}
where 
\[
T({\bf k}_{\rm a}, {\bf k}_{\rm b}, {\bf k}_{\rm c}, {\bf k}_{\rm d})= 
\frac{1}{V} \int_V \eta\, \exp [{-i{\bf k}_{\rm a}\cdot{\bf x}_{\rm a}
-i{\bf k}_{\rm b}\cdot{\bf x}_{\rm b}
-i{\bf k}_{\rm c}\cdot{\bf x}_{\rm c} 
-i{\bf k}_{\rm d}\cdot{\bf x}_{\rm d}}]
\,{\rm d}{\bf x}_{\rm a}\, {\rm d}{\bf x}_{\rm b}
\, {\rm d}{\bf x}_{\rm c} \, {\rm d}{\bf x}_{\rm d}
\]
is the trispectrum, and the wavevectors are restricted 
to be in the same hemisphere so that the term 
$P(k_{\rm a})P(k_{\rm b})\delta^{\rm K}_{{\bf n}_{\rm a}, 
-{\bf n}_{\rm c}} \delta^{\rm K}_{{\bf n}_{\rm b}, -{\bf n}_{\rm d}}$ 
does not appear. Since the reduced four-point correlation function 
$\eta$ is a six-variable function, there is a redundancy in the variables 
of the trispectrum. It is evident from the four-point function that the 
covariance of the three-dimensional mass PS is \citep[see also][]{mw99, ch01}
\begin{eqnarray} \nonumber
\sigma^2({\bf k},{\bf k}') &=& \langle 
[P'({\bf k})-P({\bf k})] [P'({\bf k}')-P({\bf k}')]\rangle
 = \langle \hat{\delta}({\bf k})\hat{\delta}^{*}({\bf k}) 
\hat{\delta}({\bf k}')\hat{\delta}^{*}({\bf k}')\rangle V^{-2}
 - P(k)P(k') \\
&=& P^2(k) \delta^{\rm K}_{{\bf n},{\bf n}'} + 
V^{-1}T({\bf k}, -{\bf k}, {\bf k}', -{\bf k}').
\end{eqnarray}
For GRFs, the trispectrum vanishes, and 
\begin{equation}
\sigma^2({\bf k}, {\bf k'}) = 
P^2(k) \delta^{\rm K}_{{\bf n},{\bf n}'}.
\end{equation}
Since the survey volume is always smaller than the observable universe, 
the covariance will be modified according to \citet{fkp94}, and 
the non-Gaussianity will further alter the covariance through the 
trispectrum.

\section{One-Dimensional Power Spectrum} \label{sec:321}

For a LOS density that is along the $x_3$-axis and sampled at
$(x_1,x_2)$, the one-dimensional Fourier transform gives
\begin{equation} \label{eq:Fourier1D}
\tilde{\delta}({\bf x}_\perp,n_3) = \int_0^B \delta({\bf x})\, 
e^{-2\pi i n_3x_3/B}\, {\rm d} x_3 =
\frac{1}{B^2} \sum_{{\bf n}_\perp=-\infty}^{\infty} 
\hat{\delta}({\bf n}_\perp,n_3)\, 
e^{2\pi i {\bf n}_\perp \cdot {\bf x}_\perp / B},
\end{equation}
where the subscript $\perp$ signifies the first two components of a
vector, i.e.~${\bf x}_\perp = (x_1,x_2)$. Similar to the three-dimensional
mass PS, the one-dimensional mass PS is expected to follow
\begin{equation} \label{eq:ens1d}
\langle \tilde{\delta}({\bf x}_\perp, n_3)
\tilde{\delta}^{*}({\bf x}_\perp, n'_3) \rangle
= P_{\rm 1D}(n_3) B \delta^{\rm K}_{n_3, n'_3}.
\end{equation}
Substituting equation (\ref{eq:Fourier1D}) in equation (\ref{eq:ens1d})
and making use of equation (\ref{eq:powsp_n}), one finds the relation
between the one-dimensional mass PS and the three-dimensional mass PS,
\begin{equation} \label{eq:321sum}
P_{\rm 1D}(n_3) = \frac{1}{B^2} \sum_{{\bf n}_\perp=-\infty}^{\infty}
P({\bf n}_\perp, n_3),
\end{equation}
which is a discrete analog of equation (\ref{eq:iso321}).

Practically, one measures PS's of LOS densities sampled at 
some locations, e.g.
\begin{equation}
P'_{\rm 1D}({\bf x}_\perp, k_3) =
|\tilde{\delta}({\bf x}_\perp, k_3)|^2 / B.
\end{equation}
A simple estimator of the one-dimensional mass PS may be constructed by 
a spatial average over many LOS's, i.e. 
\begin{equation} \label{eq:p1d-x-def}
P'_{\rm 1D}(k_3) = \langle P'_{\rm 1D}({\bf x}_\perp, k_3) 
\rangle_{{\bf x}_\perp}, 
\end{equation}
where $\langle \ldots \rangle_{{\bf x}_\perp}$ stands for a spatial
average over all LOS's in the sample. To assess the
performance of the estimator, two questions need to be addressed: (1) How
does $P'_{\rm 1D}(k_3)$ relate itself to $P'({\bf k})$; and (2) What is
the covariance of $P'_{\rm 1D}(k_3)$ with respect to $P_{\rm 1D}(k_3)$?
The rest of this chapter answers the former, and Chapter \ref{ch:cov} 
the latter.

For simplicity, I assume that LOS's are sampled regularly in transverse
directions at an interval of $b = B / m$, where $m$ is an integer. Each
LOS has a length of $B$, and ${\bf x}_\perp = b{\bf l}_\perp$ with $l_1,
l_2 = 0, \ldots, m-1$. The estimated one-dimensional mass PS is then
\begin{eqnarray} \label{eq:raw-P1D}
P'_{\rm 1D}(n_3) &=& \frac{1}{m^2} \sum_{{\bf l}_\perp=0}^{m-1}
|\tilde{\delta}(b{\bf l}_\perp, n_3)|^2/B \nonumber \\
&=& \frac{1}{m^2 B^5} \sum_{{\bf l}_\perp=0}^{m-1} 
\Big |\sum_{{\bf n}_\perp = -\infty}^{\infty} 
\hat{\delta}({\bf n}_\perp,n_3)\, 
e^{2\pi i {\bf n}_\perp \cdot {\bf l}_\perp/m} \Big |^2
\nonumber \\
&=& \frac{1}{B^5} \sum_{{\bf n}_\perp,\, {\bf j}_\perp=-\infty}^{\infty}
\hat{\delta}({\bf n}_\perp,n_3)\,
\hat{\delta}^*({\bf n}_\perp\! + m{\bf j}_\perp,n_3),
\end{eqnarray}
where the equality, 
\begin{equation} \label{eq:suml}
\frac{1}{m} \sum_{l=0}^{m-1} e^{2 \pi i (n-n') l / m} = 
\delta^{\rm K}_{n,n'+mj} \quad j=0,\pm 1, \ldots, \pm \infty,
\end{equation}
has been used to obtain the last line of equation (\ref{eq:raw-P1D}). It
is easy to show with equation (\ref{eq:raw-P1D}) that 
\begin{equation} \label{eq:p1d-ens-x}
P_{\rm 1D}(k_3) = \langle P_{\rm 1D}'(k_3) \rangle = 
\langle P_{\rm 1D}'({\bf x}_\perp, k_3) \rangle.
\end{equation}
Note that there is no spatial average on the far right side of equation 
(\ref{eq:p1d-ens-x}). It seems that only in the limit 
$m\to \infty$ do $P'_{\rm 1D}(k_3)$
and $P'({\bf k})$ follow the same relation as equations (\ref{eq:iso321})
and (\ref{eq:321sum}), i.e.~$P'_{\rm 1D}(k_3) = \sum_{{\bf
n}_\perp=-\infty}^{\infty} P'({\bf k}_\perp, k_3) / B^2$, where I have 
assumed $\hat{\delta}(\infty) = 0$ so that only ${\bf j}_\perp = (0, 0)$ 
terms contribute. In other words, one would need a high sampling rate 
in the transverse direction to accurately recover $P'({\bf k})$ 
from $P'_{\rm 1D}(k_3)$ regardless the scale of interest. 
This apparently contradicts the new sampling paradigm\footnote{A major 
difference between the classic sampling theorem \citep{s49} and the new
sampling paradigm is that the former assumes bandlimited signals while 
the latter has relaxed such assumption and developed a variety of 
prefiltering, sampling, and reconstruction techniques.} \citep{u00}, 
according to which one only has to prefilter signals and sample them at 
a minimum rate of twice the frequency (wavenumber) of interest. 
However, the \oned{} mass PS is a projection of the \thrd{} mass PS, so 
that even a prefiltering of transverse modes on small scales will lead to 
a distortion of the measured \oned{} mass PS on large scales. 

\section{Aliasing}
Without losing generality, one may choose $m$ to be even, so that equation
(\ref{eq:raw-P1D}) can be re-arranged as
\begin{eqnarray} \label{eq:est-P1D}
P'_{\rm 1D}(n_3) &=& \frac{1}{B^2} \sum_{{\bf n}_\perp=-m/2}
^{m/2} P_{\rm a}({\bf n}_\perp, n_3) \\
&=& \frac{1}{B^2} \sum_{{\bf n}_\perp=-\infty}^{\infty} 
P'({\bf n}_\perp, n_3) + A(n_3), \nonumber
\end{eqnarray}
where
\begin{equation} \label{eq:aliasPS}
P_{\rm a}({\bf n}_\perp, n_3) = \frac{1}{V} 
\Big|\sum_{{\bf j}_\perp=-\infty}^{\infty}
\hat{\delta}({\bf n}_\perp\! +m{\bf j}_\perp,n_3)\Big|^2,
\end{equation}
and
\begin{equation} \label{eq:A}
A(n_3) = \frac{1}{B^5} \sum_{\mbox{\scriptsize $\begin{array}{c}
{\bf j}_\perp,{\bf j}'_\perp\! = -\infty \\ 
{\bf j}'_\perp \!\ne 0 \end{array}$}}^{\infty}
\sum_{{\bf n}_\perp\!=-m/2}^{m/2} 
\hat{\delta}({\bf n}_\perp\! + m{\bf j}_\perp,n_3) \,
\hat{\delta}^*({\bf n}_\perp \! + m{\bf j}_\perp
\! + m{\bf j}'_\perp,n_3).
\end{equation}
Equation (\ref{eq:est-P1D}) provides two equivalent views of the sampling
effect. On one hand, the one-dimensional mass PS is a sum of the aliased
three-dimensional mass PS, $P_{\rm a}({\bf k})$, that is sampled by a 
grid of $m\times m$ LOS's. 
On the other hand, it is a complete sum of the underlying
three-dimensional mass PS with an extra term $A(k_3)$ that is determined 
by the sampling rate and properties of the density field.

The discrete Fourier transform cannot distinguish a principal mode at 
$|k| \le k_{\rm Nyq}$ from its aliases at $k\pm 2k_{\rm Nyq},
k\pm 4k_{\rm Nyq},\ldots$, where $k_{\rm Nyq}$ is the sampling Nyquist 
wavenumber. For example, $k_{\rm Nyq} = \pi / b$ if one samples the 
field at an equal spacing of $b$. The alias modes will be added to the 
principal mode if they are not properly filtered out before sampling
\citep[for details, see][]{he81}. Since the cosmic density field is not
bandlimited, aliasing can distort statistics of the field. The 
distortion on the \thrd{} mass PS cannot be quantified \emph{a priori}, 
because it depends on relative phases between the principal mode 
and its aliases. The alias effect is less pronounced, if amplitudes of 
the alias modes are much smaller than that of the principal mode. Since 
the three-dimensional mass PS decreases towards small scales, a high 
sampling rate (or $k_{\rm Nyq}$) can suppress aliasing for modes with 
$k \ll k_{\rm Nyq}$. Alternatively, one 
can use anti-aliasing filters to reduce the alias effect
\citep[for instance, with wavelets][]{fangfeng00}. 

Aliasing occurs in the estimated \oned{} mass PS because the continuous 
density field contains significant Fourier modes at wavenumbers that 
are greater than the sampling Nyquist wavenumber in $x_1$ and $x_2$ 
directions. This is evident in equation (\ref{eq:aliasPS}). Unlike the
\thrd{} case, prefiltering small-scale transverse modes may not improve 
the estimated \oned{} mass PS, because each mode of the theoretical 
\oned{} mass PS contains contributions of transverse modes in the
\thrd{} cosmic density field on all scales. 

If the Fourier modes of the density field are uncorrelated, the term 
$A(k_3)$ may neglected for a finite number of LOS's and, therefore, 
validate equation (\ref{eq:321sum}). Strictly speaking, $A(k_3)$ 
vanishes only as an ensemble average over many GRFs, but since there 
are so many independent modes in a shell of radius around 
$k$, especially at large wavenumbers, the summation in $A(k_3)$ 
will tend to vanish even for a single GRF. The cosmic density field is
more Gaussian at higher redshift, so equation (\ref{eq:321sum}) may be a
good approximation then. At low redshift, however, one might only be able 
to recover a heavily aliased three-dimensional mass PS from a sparse 
sample of LOS's.

\chapter{Covariance of the One-Dimensional Power \\ 
Spectrum: Gaussian Random Fields}
\label{ch:cov} \noindent
As one starts to attempt precision cosmology using the Ly$\alpha$ forest 
\citep{cwb02, mms03}, it becomes necessary to quantify systematic 
uncertainties of the PS analysis. The covariance of the PS is of interest 
because it tells us the sample variance error of the measured PS and how 
much the modes on different scales are correlated. For GRFs, the variance
of the PS (without binning and averaging) equals the PS itself. The 
non-Gaussianity of the cosmic density field will strongly affect the 
covariance of the \oned{} mass PS on both large and small scales because the 
\oned{} mass PS is an integral of the \thrd{} mass PS to the smallest scale possible.

The covariance of the spatial-average mass PS can differ from that of the
ensemble-average mass PS for at least two reasons. First, LOS's are no longer
independent of each other. Correlations between the LOS's could increase
the variance of the one-dimensional mass PS. Second, the length of each 
LOS is always less than the size 
of the universe, so that false correlations between two different modes
are introduced in the covariance (for the three-dimensional case, see
Feldman et al. 1994).
%\citep[for the three-dimensional case, see][]{fkp94}. 

\section{Covariance of the One-dimensional Power Spectrum} \label{sec:cov}

For the one-dimensional mass PS, the covariance is defined as
\begin{equation} \label{eq:cov-def}
\sigma^2_{\rm 1D}(k_3, k_3') = \langle [P'_{\rm 1D}(k_3) - P_{\rm 1D}(k_3)]
[P'_{\rm 1D}(k_3') - P_{\rm 1D}(k_3')] \rangle,
\end{equation}
where $P'_{\rm 1D}(k_3) = \langle P'_{\rm 1D}({\bf x}_\perp, k_3) 
\rangle_{{\bf x}_\perp}$ and $P_{\rm 1D}(k_3) = \langle P_{\rm 1D}'(k_3) 
\rangle = \langle P_{\rm 1D}'({\bf x}_\perp, k_3) \rangle$ from Chapter
\ref{ch:p1d}.
The covariance of the mean PS of $N$ LOS's can be expanded into a sum of 
pair-wise covariances between two LOS's separated by 
${\bf s}^{jl}_\perp = {\bf x}^j_\perp - {\bf x}^l_\perp$, e.g.
\begin{equation} \label{eq:cov2}
\sigma^2_{\rm 1D}(k_3, k_3') = \frac{1}{N^2} \sum_{j, l = 1}^{N} 
\sigma^2_{\rm 1D}(k_3, k_3'; {\bf s}^{jl}_\perp),
\end{equation}
where 
\begin{eqnarray} \nonumber
\sigma^2_{\rm 1D}(k_3, k_3'; {\bf s}^{jl}_\perp)
 &=& \langle [P'_{\rm 1D}({\bf x}^j_\perp, k_3) - P_{\rm 1D}(k_3)]
[P'_{\rm 1D}({\bf x}^l_\perp, k_3') - P_{\rm 1D}(k_3')] \rangle \\
&=& \langle P'_{\rm 1D}({\bf x}^j_\perp, k_3) 
P'_{\rm 1D}({\bf x}^l_\perp, k_3')\rangle 
- P_{\rm 1D}(k_3)P_{\rm 1D}(k_3'), \nonumber
\end{eqnarray}
and ${\bf x}^j_\perp$ is the location of the $j$th LOS in $x_1$--$x_2$
plane.

For GRFs, the four-point function [equation (\ref{eq:4point})] helps
reduce the pair-wise covariance to
\begin{eqnarray} \label{eq:cov2-xi} \nonumber
\sigma^2_{\rm 1D}(n_3, n_3';{\bf s}) &=& \frac{1}{B^{10}} 
\!\sum_{{\bf n}_\perp, {\bf n}'_\perp, {\bf n}''_\perp, {\bf n}'''_\perp 
= -\infty}^{\infty} \!\!\!\langle \hat{\delta}({\bf n}_\perp, n_3)
\hat{\delta}^*({\bf n}'_\perp, n_3)\hat{\delta}({\bf n}''_\perp, n'_3)
\hat{\delta}^*({\bf n}'''_\perp, n'_3) \rangle \\ \nonumber
&& \qquad\qquad\quad \times e^{2\pi i [({\bf n}_\perp - {\bf n}'_\perp)
\cdot {\bf x}_\perp^j + ({\bf n}''_\perp - {\bf n}'''_\perp)\cdot 
{\bf x}_\perp^l]/B} - P_{\rm 1D}(n_3)P_{\rm 1D}(n_3') \\
&=& \Big |\frac{1}{B^2} \sum_{{\bf n}_\perp = -\infty}^{\infty} 
P({\bf n}_\perp, n_3)\, e^{2\pi i {\bf n}_\perp \cdot\, {\bf s} / B} 
\Big |^2 \delta^{\rm K}_{n_3, n_3'} = |\xi({\bf s}, n_3)|^2
\delta^{\rm K}_{n_3, n_3'},
\end{eqnarray} 
where the subscript and the superscript are dropped for ${\bf s}$. 
Fourier transforms
of $\xi({\bf s}, k_3)$ will give the three-dimensional mass PS and correlation
function of the density field, and $P_{\rm 1D}(k_3) = \xi(0, k_3)$
\citep[see also][]{vmm02}. Because of isotropy, the pair-wise covariance 
$\sigma^2_{\rm 1D}(k_3, k_3';{\bf s})$ and $\xi({\bf s}, k_3)$ depend 
only on the magnitude of the separation, i.e. 
$\sigma^2_{\rm 1D}(k_3, k_3';{\bf s}) \equiv \sigma^2_{\rm
1D}(k_3, k_3';s)$ and $\xi({\bf s}, k_3) \equiv \xi(s, k_3)$.

If one LOS is sampled from each GRF, the variance of the measured
one-dimensional mass PS, $P'_{\rm 1D}(k_3)$, is $\sigma^2_{\rm 1D}(k_3, k_3) =
\sigma^2_{\rm 1D}(k_3, k_3; 0) = P^2_{\rm 1D}(k_3)$, analogous to the
three-dimensional case. If $N=m^2$ LOS's are sampled in each GRF on a
regular grid as in Chapter \ref{ch:p1d}, e.g. ${\bf s}$ = $({\bf l}_\perp
- {\bf l}'_\perp) b$, the covariance becomes
\begin{eqnarray} \label{eq:cov-sum}
\sigma^2_{\rm 1D}(n_3, n_3') &=& \frac{1}{N^2}
\sum_{{\bf l}_\perp, {\bf l}'_\perp=0}^{m-1} 
\sigma^2_{\rm 1D}[n_3, n_3'; ({\bf l}_\perp - {\bf l}'_\perp)b] \nonumber \\
&=& \frac{1}{N^2 B^4} \sum_{{\bf n}_\perp, {\bf n}'_\perp = -\infty}^{\infty}
\!\!\!\!\! P({\bf n}_\perp, n_3) P({\bf n}'_\perp, n_3) \,
\Big | \sum_{{\bf l}_\perp=0}^{m-1} 
e^{2\pi i ({\bf n}_\perp - {\bf n}'_\perp) \cdot {\bf l}_\perp / m} \Big |^2
\delta^{\rm K}_{n_3, n_3'} \nonumber \\ &=& \frac{1}{B^4} 
\sum_{{\bf n}_\perp,\, {\bf j}_\perp = -\infty}^{\infty}
\!\!\!\!\! P({\bf n}_\perp, n_3) P({\bf n}_\perp+m{\bf j}_\perp, n_3) 
\,\delta^{\rm K}_{n_3, n_3'},
\end{eqnarray}
where I have used equation (\ref{eq:suml}) to reach the last line. This
result can be easily obtained using equations (\ref{eq:4point}) and 
(\ref{eq:raw-P1D}) as well.
It is seen that the summation in the last line of equation
(\ref{eq:cov-sum}) runs over only one mode out of every $N$ modes in the
Fourier space. If the three-dimensional mass PS were constant, the variance of
the mean PS of the $N$ LOS's would be $N$ times smaller than the variance
of the PS of a single LOS. This is coincident
with the theory of the variance of the mean. Since the cosmic density
field is not a GRF in general, equation (\ref{eq:cov-sum}) is not expected
to give an accurate estimate.

If the $N$ LOS's are sampled randomly in each GRF, the summation over
LOS's in the second line of equation (\ref{eq:cov-sum}) should be
re-cast to read
\begin{equation} \label{eq:cov-ran}
\sigma^2_{\rm 1D}(n_3, n_3') = \frac{1}{N^2 B^4} 
\sum_{{\bf n}_\perp, {\bf n}'_\perp = -\infty}^{\infty}
\!\!\!\!\! P({\bf n}_\perp, n_3) P({\bf n}'_\perp, n_3)
\sum_{j,\, l=1}^{N} e^{2\pi i ({\bf n}_\perp - {\bf n}'_\perp) \cdot 
({\bf x}^j_\perp - {\bf x}^l_\perp)/B} \,\delta^{\rm K}_{n_3, n_3'}.
\end{equation}
Since ${\bf x}^j_\perp$ is randomly distributed, with a large number of
LOS's the second sum in equation (\ref{eq:cov-ran}) tends to vanish
except for $j=l$. Thus, one obtains
\begin{equation} \label{eq:cov-ran2}
\sigma^2_{\rm 1D}(n_3, n_3') \simeq \frac{1}{N B^4} 
\sum_{{\bf n}_\perp, {\bf n}'_\perp = -\infty}^{\infty}
\!\!\!\!\! P({\bf n}_\perp, n_3) P({\bf n}'_\perp, n_3)\,
\delta^{\rm K}_{n_3, n_3'} = \frac{1}{N} P^2_{\rm 1D}(n_3) \,
\delta^{\rm K}_{n_3, n_3'}.
\end{equation}
Again, it shows that the variance of $P'_{\rm 1D}(k_3)$ is inversely
proportional to the number of LOS's, but this is valid only for a large
number of LOS's randomly sampled in a GRF.

Through the Gaussian case one can find the lowest bound of the 
uncertainty for a measured one-dimensional mass PS and the minimum number of 
LOS's needed for a target precision. For example, to measure the \oned{} 
mass PS of GRFs accurate to $5\%$ on \emph{every} mode, one needs at least 400 
LOS's. However, the cosmic density field has a far more complex 
covariance due to its non-vanishing trispectrum, which can 
significantly increase the variance of the measured \oned{} mass PS. The 
trispectrum introduces an extra term to the covariance of the \oned{} mass PS 
in addition to the Gaussian piece [equation (\ref{eq:cov-sum})], i.e.
\begin{equation} \label{eq:cov-tri}
\sigma^2_{\rm 1D}(n_3, n_3') = \frac{1}{B^4} 
\sum_{{\bf n}_\perp,\, {\bf j}_\perp = -\infty}^{\infty}
\!\!\!\!\! P({\bf n}_\perp, n_3) P({\bf n}_\perp+m{\bf j}_\perp, n_3) 
\,\delta^{\rm K}_{n_3, n_3'} + T_{\rm 1D}(n_3, n_3'),
\end{equation}
where 
\begin{equation} \label{eq:T1D}
T_{\rm 1D}(n_3, n_3') = \frac{1}{B^4} 
\sum_{{\bf n}_\perp, \,{\bf j}_\perp,\, {\bf n}'_\perp, 
\,{\bf j}'_\perp = -\infty}^{\infty} 
T({\bf n}, -{\bf n} - m{\bf j}, {\bf n}', -{\bf n}' - m{\bf j}'),
\end{equation}
${\bf n}_\perp$, ${\bf j}_\perp$, ${\bf n}'_\perp$, and 
${\bf j}'_\perp$ are the transverse components of 
${\bf n}$, ${\bf j}$, ${\bf n}'$, and 
${\bf j}'$, respectively, and $j_3 = j'_3 = 0$. 
In deriving equation (\ref{eq:cov-tri}), I have made use of
equations (\ref{eq:4point}) and (\ref{eq:raw-P1D}).

\section{Line-of-Sight Length} \label{sec:short}

Observationally, the length of a LOS is always much less than the size of
the observable universe. In this case, the one-dimensional mass PS and its
covariance must be re-formulated.

For a LOS from $({\bf x}_\perp, 0)$ to $({\bf x}_\perp, L)$, its Fourier
transform is
\begin{eqnarray} 
\tilde{\delta}_{\rm L}({\bf x}_\perp, \tilde{n}_3) 
&= & \int_0^L \delta ({\bf x}_\perp, x_3) 
e^{-2\pi i \tilde{n}_3 x_3 / L} {\rm d} x_3 \nonumber \\ 
&=& \frac{L}{B^3} \sum_{{\bf n} = -\infty}^{\infty} 
\hat{\delta}({\bf n}) e^{2\pi i {\bf n}_\perp \cdot
{\bf x}_\perp / B} 
\frac{e^{2\pi i (n_3 L / B - \tilde{n}_3)} - 1} 
{2\pi i (n_3 L / B - \tilde{n}_3)},
\end{eqnarray}
so that
\begin{equation}
\langle \tilde{\delta}_{\rm L}({\bf x}_\perp, \tilde{n}_3)
\tilde{\delta}^*_{\rm L}({\bf x}_\perp, \tilde{n}'_3) \rangle = 
\frac{L^2}{B^3} \sum_{{\bf n} = -\infty}^{\infty} 
(-1)^{\tilde{n}'_3 - \tilde{n}_3} P({\bf n}) w(n_3, \tilde{n}_3)
w(n_3, \tilde{n}'_3),
\end{equation}
where 
\begin{equation} \label{eq:windlen}
w(n_3, \tilde{n}_3) = \frac{\sin[(n_3L/B-\tilde{n}_3)\pi]}
{\pi (n_3L/B-\tilde{n}_3)}.
\end{equation}
The factor $w(n_3, \tilde{n}_3)$ acts like a window function that mixes
Fourier modes of the density field along $n_3$ direction into the
one-dimensional Fourier mode at $\tilde{n}_3$. Because $w(n_3,
\tilde{n}_3)$ and $w(n_3, \tilde{n}'_3)$ are not orthogonal to each other,
the term $\langle \tilde{\delta}_{\rm L}({\bf x}_\perp, \tilde{n}_3)
\tilde{\delta}^*_{\rm L}({\bf x}_\perp, \tilde{n}'_3) \rangle$ is no
longer diagonal with respect to $\tilde{n}_3$ and $\tilde{n}'_3$. However,
it remains diagonal-dominant.

One may define an observed one-dimensional mass PS at ${\bf x}_\perp$ as
\begin{equation}
\widetilde{P}_{\rm 1D} ({\bf x}_\perp, \tilde{n}_3) = \langle
|\tilde{\delta}_{\rm L} ({\bf x}_\perp, \tilde{n}_3)|^2 \rangle / L. 
\end{equation}
The ensemble-averaged one-dimensional mass PS is
\begin{equation}
\widetilde{P}_{\rm 1D}(\tilde{n}_3) = \langle 
\widetilde{P}_{\rm 1D}({\bf x}_\perp, \tilde{n}_3) \rangle =
\frac{L}{B^3} \sum_{{\bf n} = -\infty}^{\infty} 
P({\bf n}) w^2(n_3, \tilde{n}_3).
\end{equation}
The covariance of $\widetilde{P}_{\rm 1D} (\tilde{n}_3)$ for GRFs is
\begin{eqnarray} \label{eq:cov-short}
\tilde{\sigma}^2_{\rm 1D}(\tilde{n}_3, \tilde{n}'_3) &= &
\langle [\widetilde{P}_{\rm 1D}({\bf x}_\perp, \tilde{n}_3) -
\widetilde{P}_{\rm 1D}(\tilde{n}_3)]
[\widetilde{P}_{\rm 1D}({\bf x}_\perp, \tilde{n}'_3) -
\widetilde{P}_{\rm 1D}(\tilde{n}'_3)] \rangle \nonumber \\ &=&
\frac{L^2}{B^6} \!\sum_{{\bf n},{\bf n}' = -\infty}^{\infty} \!\!
P({\bf n})P({\bf n}') w(n_3, \tilde{n}_3) w(n_3, \tilde{n}'_3)
 w(n'_3, \tilde{n}_3) w(n'_3, \tilde{n}'_3).
\end{eqnarray}
As expected, the covariance is not diagonal because of the window function
$w(n_3, \tilde{n}_3)$.

\chapter{Covariance of the One-Dimensional Power \\
Spectrum: the Cosmic Density Field} \label{ch:mps}
\noindent
Even at moderately high redshift, the cosmic density field is already 
quite non-Gaussian on scales below 10 \mpc{} %\citep{ff00, zjf01, zf02}.
(Feng \& Fang 2000; Zhan et al. 2001; Zhan \& Fang 2002).
When projected in one dimension, i.e. equation (\ref{eq:iso321}), the 
small-scale non-Gaussianity will obviously affect the measured \oned{}
mass PS on much larger scales. A non-vanishing trispectrum 
arises because of the non-Gaussianity, and it introduces an extra term, 
$T_{\rm 1D}(k_3, k'_3)$, to the covariance of the \oned{} mass PS
[see equation (\ref{eq:cov-tri})]. Although 
it is possible to derive the trispectrum based on the halo model 
\citep[e.g.][]{ch01}, the \oned{} projection, unfortunately, obscures 
the contribution of the trispectrum to the \oned{} mass PS. 
Therefore, numerical simulations are necessary for the study.

\section{Simulations}
Three $N$-body simulations of 256$^3$ cold dark matter (CDM) particles are
used to quantify the covariance of the one-dimensional mass PS. The model
parameters are largely consistent with \emph{WMAP} results \citep{svp03},
e.g. ($\Omega$, $\Omega_{\rm b}$, $\Omega_\Lambda$, $h$, $\sigma_8$, $n$) 
= (0.27, 0.04, 0.73, 0.71, 0.85, 1), where $\Omega$ is the cosmic matter
density parameter, $\Omega_{\rm b}$ the baryon density parameter, 
$\Omega_\Lambda$ the energy density parameter associated with the 
cosmological constant, $\sigma_8$ the rms density
fluctuation within a radius of 8 \mbox{$h^{-1}$Mpc}, and $n$ is the power
spectral index. The box sizes of the simulations are 128
\mbox{$h^{-1}$Mpc} (labeled as B128), 256 \mbox{$h^{-1}$Mpc} (B256), and
512 \mbox{$h^{-1}$Mpc} (B512). The baryon density $\Omega_{\rm b}$ is
used only for the purpose of calculating the transfer function using
\textsc{linger} \citep{mb95}, which is then read by \textsc{grafic2}
\citep{b01} to generate the initial condition. The CDM particles are
evolved from $z = 44.5$ to present using \textsc{gadget} \citep{syw01}.

The simulation produces snapshots at $z=3$ and 0. For each snapshot, the
particles are assigned to a density grid of 512$^3$ nodes using the
triangular-shaped-cloud (TSC) scheme \citep{he81}. LOS's are then sampled
along the $x_3$-axis. The particle Nyquist wavenumber ($k_{\rm p} = 2\pi,
\pi, \pi/2 \ h$ Mpc$^{-1}$) sets a nominal cut-off of the wavenumber
beyond which the Fourier modes of the density field cannot be represented
by discrete particles. Of course, since the simulation code retains
higher-wavenumber perturbations to the particles, the actual cut-off can
be somewhat higher. Nevertheless, the Nyquist wavenumber of the density
grid ($2k_{\rm p}$) should be sufficient to accurately recover most of the
Fourier modes contained in the particle distribution.

Because the density grid has a finite resolution, both the coordinates and
the wavenumbers are discrete and finite. The equations in Chapters
\ref{ch:p1d}--\ref{ch:cov} need to be modified accordingly, so that
they do not sum over non-existing modes. For example, equation
(\ref{eq:321sum}) becomes
\begin{equation} \label{eq:321fsum}
P_{\rm 1D}(n_3) = \frac{1}{B^2} 
\sum_{{\bf n}_\perp=-M/2}^{M/2}P({\bf n}_\perp,n_3),
\end{equation}
where $M = 512$ is the number of nodes of the density grid in transverse
directions.

For brevity and the purpose of comparing covariances, I
introduce the following normalized covariances:
\begin{equation} \label{eq:Cs}
C(k_3, k_3'; s) = \sigma^2_{\rm 1D}(k_3, k_3'; s) 
[P_{\rm 1D}(k_3)P_{\rm 1D}(k_3')]^{-1},
\end{equation}
which is the pair-wise covariance between two LOS's separated 
transversely by a distance $s$ and scaled by the one-dimensional mass PS;
\begin{equation} \label{eq:C}
C(k_3, k_3') = \sigma^2_{\rm 1D}(k_3, k_3') 
[P_{\rm 1D}(k_3)P_{\rm 1D}(k_3')]^{-1},
\end{equation}
which is the covariance of the estimated one-dimensional mass PS, e.g. 
equation (\ref{eq:raw-P1D}), scaled by the one-dimensional mass PS; and
\begin{equation} \label{eq:Chat}
\hat{C}(k_3, k_3') = C(k_3, k_3') [C(k_3, k_3)C(k_3', k_3')]^{-1/2}. 
\end{equation} 
For GRFs, all these covariances are diagonal in matrix representation, 
if the length of LOS's is the same as the size of the simulation box. In
addition, $C(k_3, k_3; 0) = 1$ and $C(k_3, k_3) = N^{-1}$, 
where $N$ is the number of LOS's that are sampled for estimating the 
one-dimensional mass PS. 
The advantage of $\hat{C}(k_3, k_3')$ is that $\hat{C}(k_3, k_3) =
1$ for all fields, so that they can be compared with each other in a
single (grey) scale.

\section{Pair-Wise Covariance}

\begin{figure}
\centering
\includegraphics[width=5.8in]{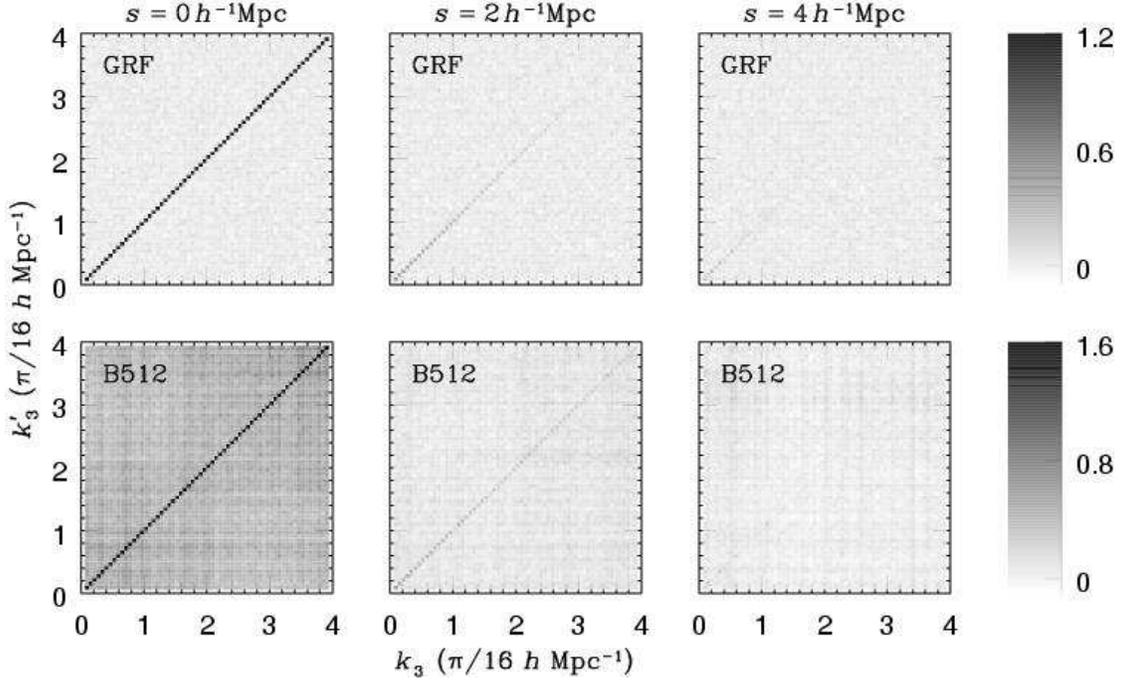}
\caption[Normalized pair-wise covariances $C(k_3, k_3'; s)$ 
in grey scale]{
Normalized pair-wise covariances $C(k_3, k_3'; s)$ 
[see equation (\ref{eq:Cs})] in grey scale. 
The covariances are calculated over 2000 pairs of LOS's sampled at a 
fixed separation $s$ from the simulation B512 at $z=3$ (lower panels)
and from 2000 GRFs that have the same three-dimensional mass PS as the 
simulation (upper panels).
The grey scale extends to $-0.1$, which corresponds to white.
\label{fig:cov2}}
\end{figure}

\begin{figure} 
\centering
\includegraphics[width=5.5in]{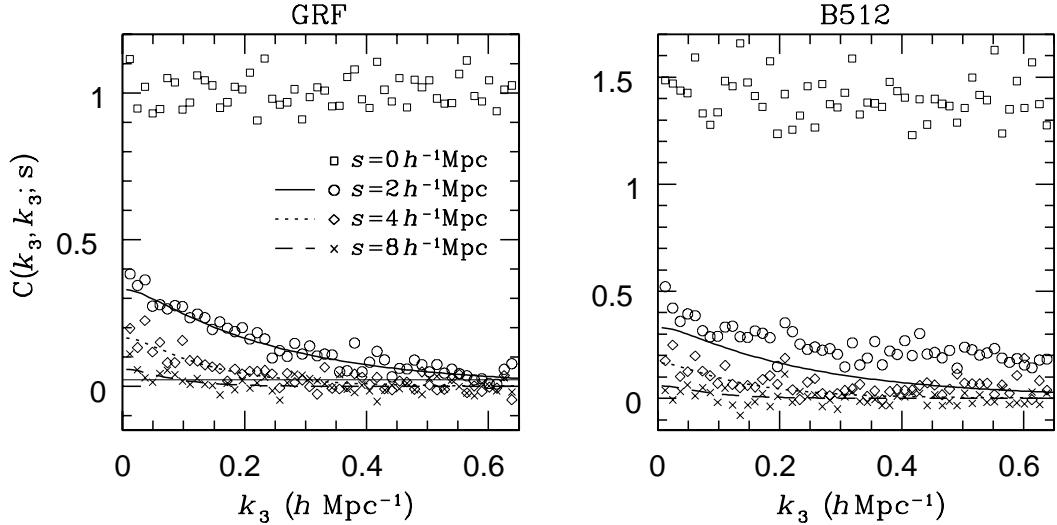}
\caption[Diagonal elements of normalized pair-wise covariance matrices 
$C(k_3, k_3'; s)$]{
Diagonal elements of the covariance matrix $C(k_3, k_3'; s)$. The horizontal
solid line in the left panel marks the rms value of the off-diagonal
elements in the $s = 8\ h^{-1}$Mpc case. Other lines are from direct
summations of the three-dimensional mass PS using equation (\ref{eq:cov2-xi}).
The symbols are measured from GRFs (left) and the simulation B512 at $z=3$
(right). For GRFs, $C(k_3, k_3; 0)$ is expected to be unity.
\label{fig:xi}}
\end{figure}

The normalized pair-wise covariance $C(k_3, k_3'; s)$ is shown in
Fig.~\ref{fig:cov2} for GRFs and the simulation B512 at $z = 3$. The
covariances for the GRFs are averaged over an ensemble of 2000 random
realizations, while those for the simulation are averaged over 2000 pairs
of LOS's from a single field. The behavior of the covariances is consistent
with the expectation. Namely, $C(k_3, k_3'; 0) \simeq \delta^{\rm K}_{n_3,
n_3'}$ and $C(k_3, k_3; s)$ decreases as the separation $s$ increases. The
simulation does deviate from GRFs because of the non-Gaussianity, which 
increases the variance $C(k_3, k_3; s)$. Fig.~\ref{fig:xi}
compares $C(k_3, k_3; s)$ with that from equation (\ref{eq:cov2-xi}). 
I note in passing that the expected values of $C(k_3, k_3;8\ \mpc)$ are 
so close to 0 that they are even below the rms value of the off-diagonal 
elements of $C(k_3, k'_3;8\ \mpc)$ for 2000 GRFs.
Hence, it is practically difficult to recover three-dimensional
statistics from $\sigma_{\rm 1D}(k_3, k_3; s)$ or $\xi(s, k_3)$ if the
LOS's are too far apart. In theory, the modes of two LOS's are always 
correlated as $k \to 0$, regardless of their separation. However, the 
correlation for $s \gtrsim 8\ h^{-1}$Mpc is so weak that it will not be 
easily detected against statistical uncertainties. Thus, 
Figs.~\ref{fig:cov2} and \ref{fig:xi} suggest that LOS's sampled in a 
single cosmic density field are practically independent of each other as 
long as $s \gtrsim 8\ h^{-1}$Mpc.

\section{Covariance}

\begin{figure} 
\centering
\includegraphics[width=5.5in]{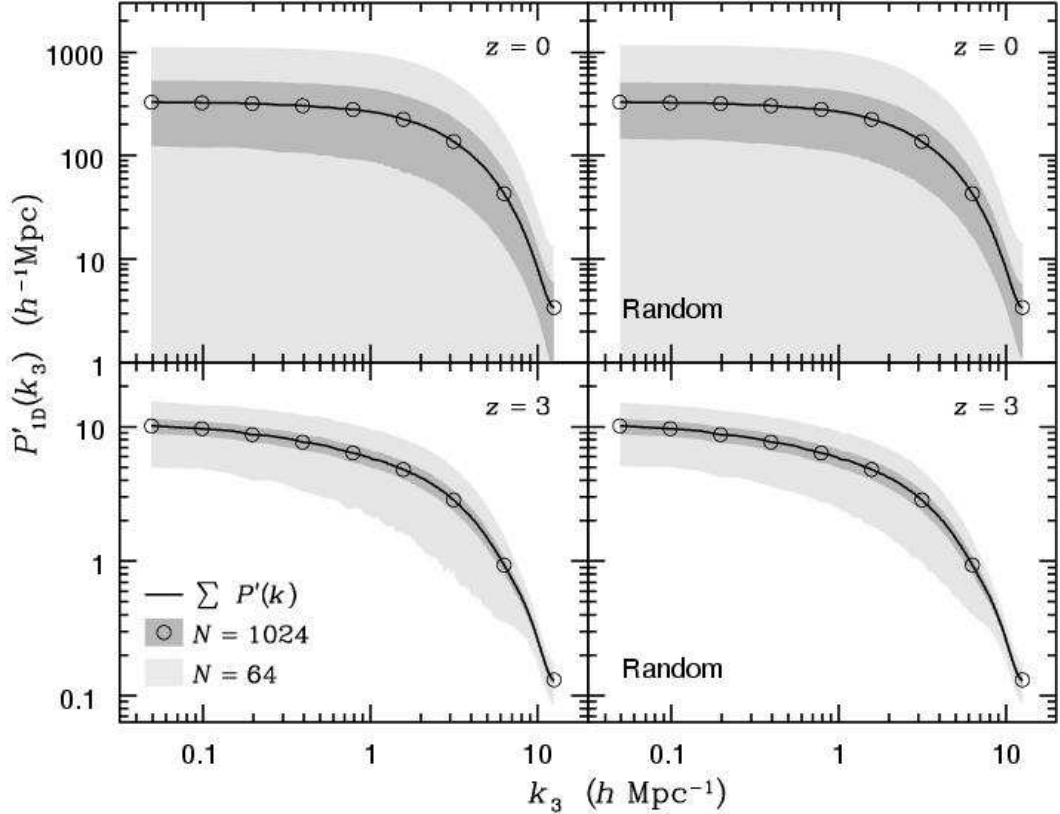}
\caption[Estimated one-dimensional mass power spectra and their variances]{
Estimated one-dimensional mass PS's $P'_{\rm 1D}(k_3)$. The PS's are 
measured by
averaging over 64 (light grey) and 1024 (dark grey) LOS's that are drawn
from the B128 simulation. Shaded areas mark $1 \sigma$ dispersion of
the PS among 2000 (light grey) and 256 (dark grey, as $256 \times 1024$
exhausts all the $512^2$ LOS's) distinct drawings. Circles are the
mean PS's for dark grey areas, i.e. the mean of $512^2$ LOS's. Solid
lines are results of a direct summation of the three-dimensional mass PS using
equation (\ref{eq:321fsum}). The LOS's are sampled on a grid with fixed 
spacing in the left panels but drawn randomly in the right panels.
\label{fig:var1d}}
\end{figure}

\begin{figure} 
\centering
\includegraphics[width=5.8in]{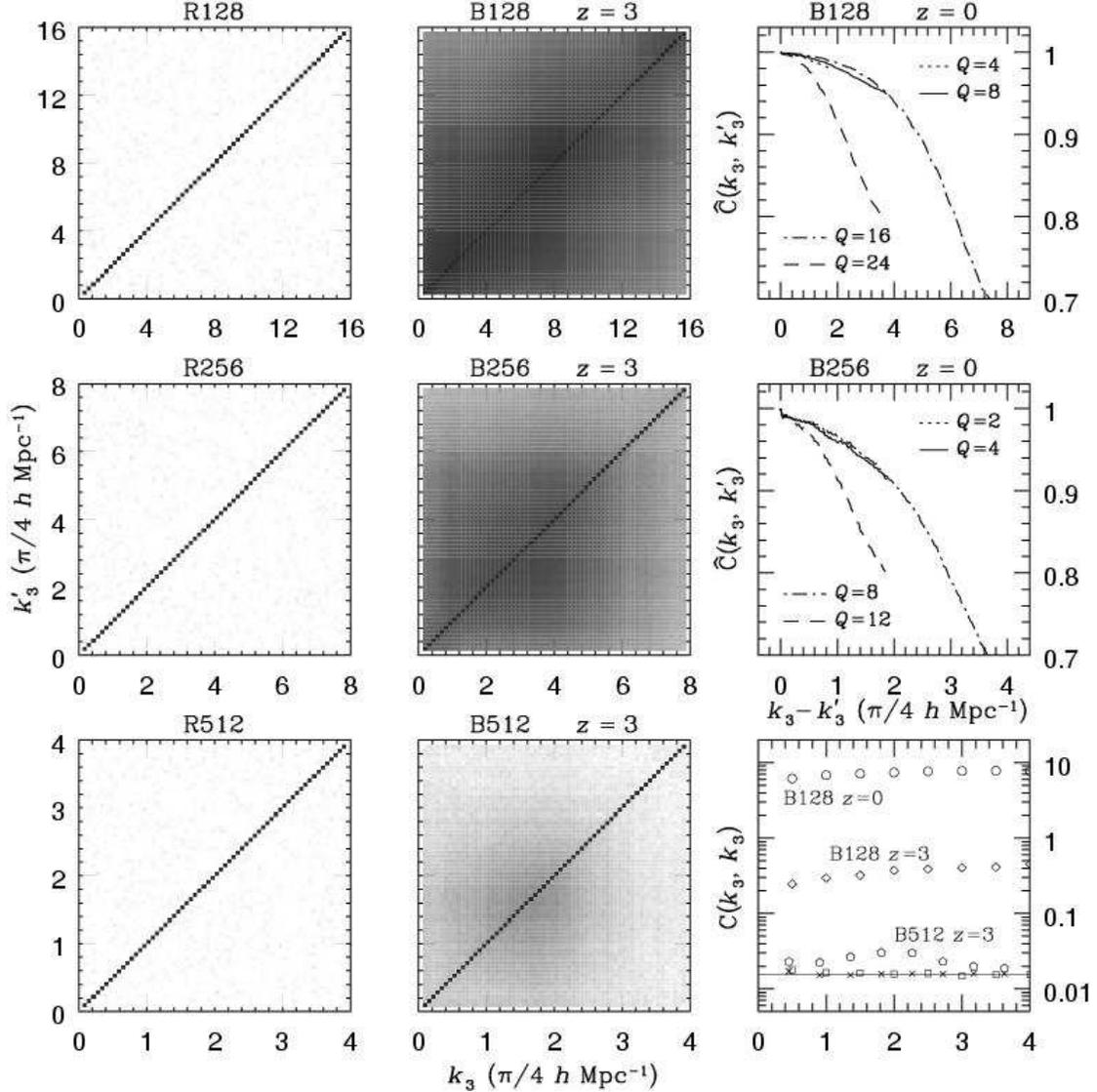}
\caption[Normalized covariances of \oned{} mass power spectra averaged over 
2000 groups, each of which consists of 64 lines of sight]{
Normalized covariances $\hat{C}(k_3,k'_3)$ 
[see equation (\ref{eq:Chat})] averaged over 2000 groups,
each of which consists of 64 LOS's ($N=64$). For each panel in the left,
the LOS's are randomly drawn from a single GRF that has a box size of 128
\mbox{$h^{-1}$Mpc} (R128), or 256 \mbox{$h^{-1}$Mpc} (R256), or 512
\mbox{$h^{-1}$Mpc} (R512). The GRFs have the same three-dimensional mass PS as
their corresponding simulations at $z=0$, but note that
$\hat{C}(k_3,k'_3)$ is independent of redshift for GRFs. Similarly, the
middle column is for simulations at $z=3$. The covariances are shown
in a linear grey scale with black being $1.2$ and white less than or equal
to 0. At $z = 0$, the normalized covariances become much less diagonally
dominant than the B128 $z=3$ panel in the same grey scale, so only 4 cross
sections along $Q=(k_3+k_3')/(\pi/4\ h\ \mbox{Mpc}^{-1})$ are plotted for
B128 and B256. Diagonal elements $C(k_3, k_3)$ 
[see equation (\ref{eq:C})] of R128 (squares), B128
at $z=0$ (circles), B128 at $z=3$ (diamonds), R512 (crosses), and B512 at
$z=3$ (pentagons) are shown in the lower right panel along with a thin
solid line marking the value $1/64$, i.e. the expectation for GRFs.
\label{fig:cov64}}
\end{figure}

\begin{figure} 
\centering
\includegraphics[width=5.8in]{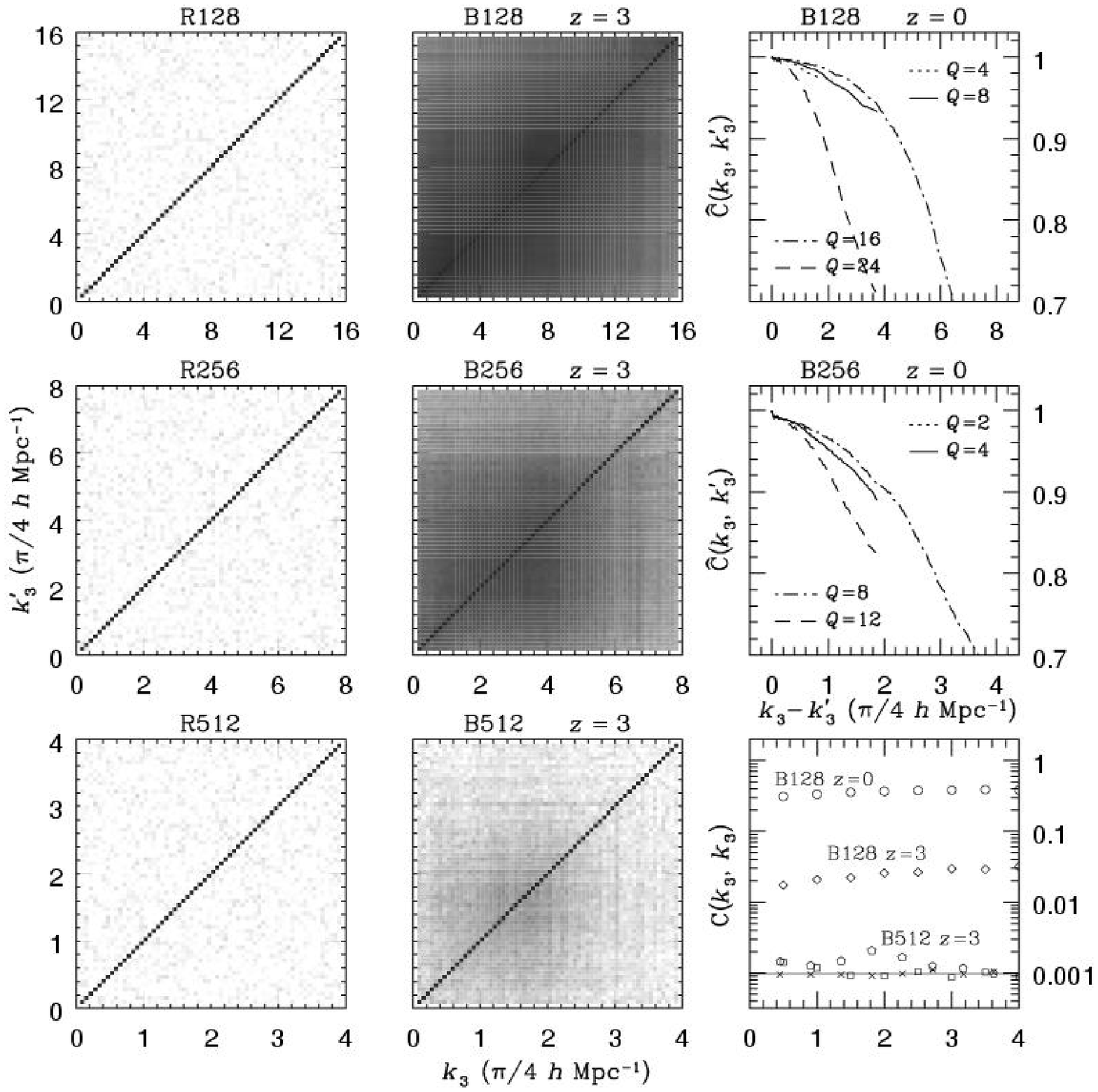}
\caption[Normalized covariances of \oned{} mass power spectra averaged 
over 256 groups, each of which consists of 1024 lines of sight]{
The same as Fig.~\ref{fig:cov64}, except that the normalized covariances
$\hat{C}(k_3,k'_3)$ are averaged over 256 groups, each of which consists of
1024 LOS's ($N=1024$). The thin solid line in the lower right panel marks
the value of $1/1024$.
\label{fig:cov1024}}
\end{figure}

Fig.~\ref{fig:var1d} shows the one-dimensional mass PS measured by averaging
over 64 and 1024 LOS's from the B128 simulation. Although the mean PS of
all the $512^2$ LOS's agrees with the result of a direct summation of the
three-dimensional mass PS using equation (\ref{eq:321fsum}), the deviation of
the PS for any particular group of 64 or 1024 LOS's is substantial,
especially at $z = 0$. The variance is smaller at $z = 3$ than at 
$z = 0$ because the cosmic density field is more Gaussian earlier on. 
It is roughly 
inversely proportional to the number of LOS's. This can seen better by 
comparing the lower right panels of Figs.~\ref{fig:cov64} and 
\ref{fig:cov1024}. However, even at $z = 3$ the variance of the \oned{} 
mass PS is still much higher than $N^{-1}P_{\rm 1D}^2(k_3)$, i.e. the 
variance for GRFs, which indicates a heavy contribution from the 
trispectrum. The formulae in
Chapters \ref{ch:p1d} and \ref{ch:cov} often assume that LOS's are
sampled on a grid with fixed spacing, which may not be applicable to 
realistic data
such as inverted densities from the Ly$\alpha$ forest \citep{nh99,z03}.
Therefore, I sample the LOS's in two ways in Fig.~\ref{fig:var1d}: grid 
sampling and random sampling. Since no significant difference is observed, 
random sampling can be safely applied in the rest of this dissertation.

\begin{figure} 
\centering
\includegraphics[width=5.8in]{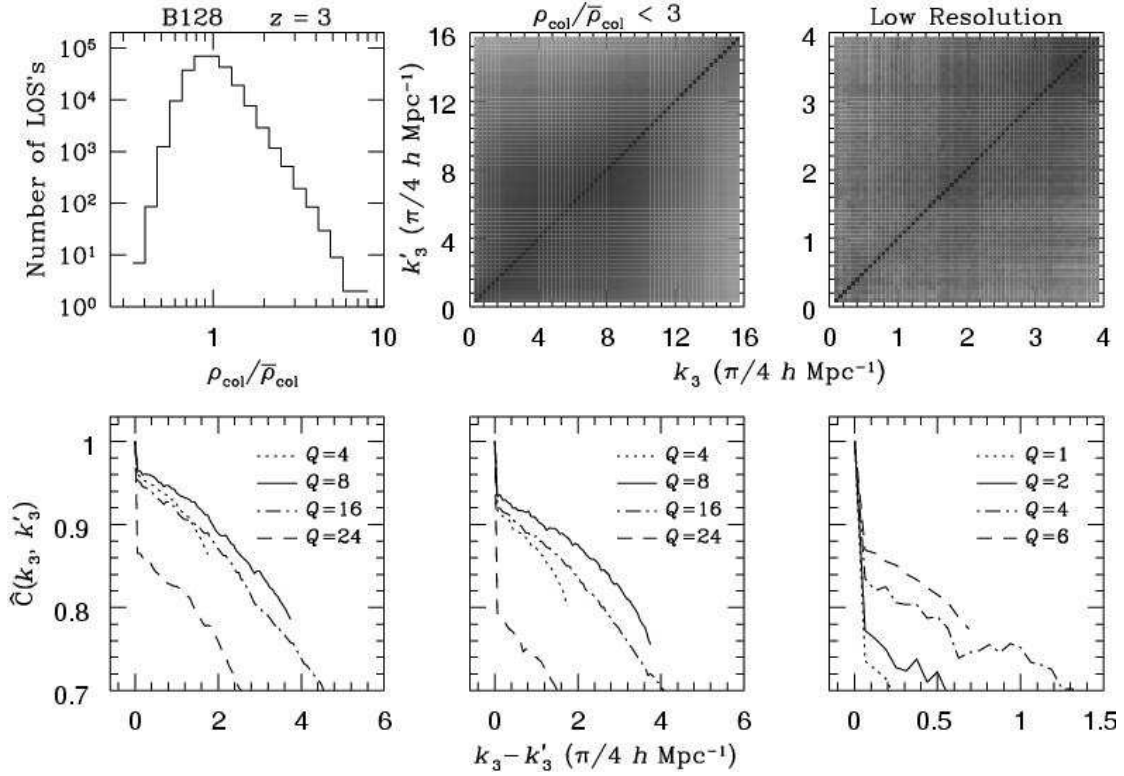}
\caption[Effect of non-Gaussianity and resolution on the covariance
of the \oned{} mass power spectrum]{
The effect of non-Gaussianity and resolution on the covariance. The left
column shows the distribution function of column density $\rho_{\rm
col}/\bar{\rho}_{\rm col}$ (upper panel) and cross sections of the
normalized covariance $\hat{C}(k_3, k_3')$ along $Q=(k_3+k_3')/(\pi/4\ h\
\mbox{Mpc}^{-1})$ (lower panel) for B128 at $z = 3$. The middle column
shows $\hat{C}(k_3, k_3')$ (upper panel) and its cross sections (lower
panel) for the same simulation output but with an exclusion of 318 LOS's 
that have $\rho_{\rm col}/\bar{\rho}_{\rm col} \ge 3$. The normalized
covariances in the first two columns are calculated in the same way as in
Fig.~\ref{fig:cov64}, i.e. with 2000 groups and $N=64$. Thus, about half
of the 318 LOS's would be selected without the criterion $\rho_{\rm
col}/\bar{\rho}_{\rm col} < 3$. The right column is similar to the middle
column, but the density is assigned on a grid of $128^3$ nodes. All the
$128^2$ LOS's are selected and divided into 256 groups with $N=64$. The
grey scale is the same as Fig.~\ref{fig:cov64}.
\label{fig:cov64l}}
\end{figure}

The normalized covariances $\hat{C}(k_3, k_3')$ and $C(k_3,k_3')$ are
quantified in Figs.~\ref{fig:cov64} ($N=64$) and \ref{fig:cov1024} 
($N=1024$). The left column in each figure is the covariances of 
the spatially averaged one-dimensional mass PS from a single GRF that 
has the same box size and three-dimensional mass PS as
its corresponding simulation. Clearly, the covariances based on 
spatial average are nearly diagonal with unity diagonal elements. This is 
in agreement with the expectations based on ensemble average for GRFs and 
is consistent with the
ergodicity argument. The middle column is similar to the left column
except that the density fields are from simulations at $z = 3$. The modes
in the simulated density fields are strongly correlated, so that the
covariances are no longer diagonal. In other words, the trispectrum is
non-vanishing for the cosmic density field, as it is the only term that 
contributes to off-diagonal elements in the covariance.
The cosmic density field becomes so
non-Gaussian at $z=0$ that grey scale figures of the covariances will not
be readable. Hence, I only plot four
cross sections perpendicular the diagonal with 
$Q=(k_3+k_3')/(\pi/4\ h\ \mbox{Mpc}^{-1})$ for B128 
and B256 in the right column. The dominance of the diagonals suggested by 
these cross sections is actually weaker than that in the middle column, 
which can be seen by contrasting the cross sections for B128 at $z=0$ in 
Fig.~\ref{fig:cov64} with that for B128 at $z=3$ in Fig.~\ref{fig:cov64l}.

The non-Gaussianity is reflected not only in
the correlations between different modes but also in the
variance of the one-dimensional mass PS, i.e. the diagonal elements of the 
normalized covariance $C(k_3, k'_3)$. The lower
right panels of Figs.~\ref{fig:cov64} and \ref{fig:cov1024} compare
$C(k_3, k_3)$ for five different density fields. The variance from 
the simulation B128 is orders of magnitude higher than 
$N^{-1}P^2_{\rm 1D}(k_3)$, and it grows
as the non-Gaussianity becomes stronger toward $z = 0$. As a result,
the sample variance error estimated for GRFs is much lower than what
one can actually measure from the cosmic density field. According to 
equation (\ref{eq:cov-tri}), both aliasing and the trispectrum contribute
to the variance. Since the GRFs have the same three-dimensional mass PS 
and are sampled in the same way as 
the simulations, their near-Gaussian variances of spatially averaged
one-dimensional mass PS's suggest that the contribution of the aliasing 
effect is negligible. Comparisons of
$C(k_3, k_3)$ for the same density field but with different sizes of 
sample ($N=64$ and 1024) confirm the observation in 
Fig.~\ref{fig:var1d} that the 
variance of the \oned{} mass PS scales roughly as $N^{-1}$. This means even
though the non-Gaussianity drives up the sample variance error of the
measured \oned{} mass PS, one can still reduce the error by sampling a 
large number  of LOS's.

\begin{figure} 
\centering
\includegraphics[width=5in]{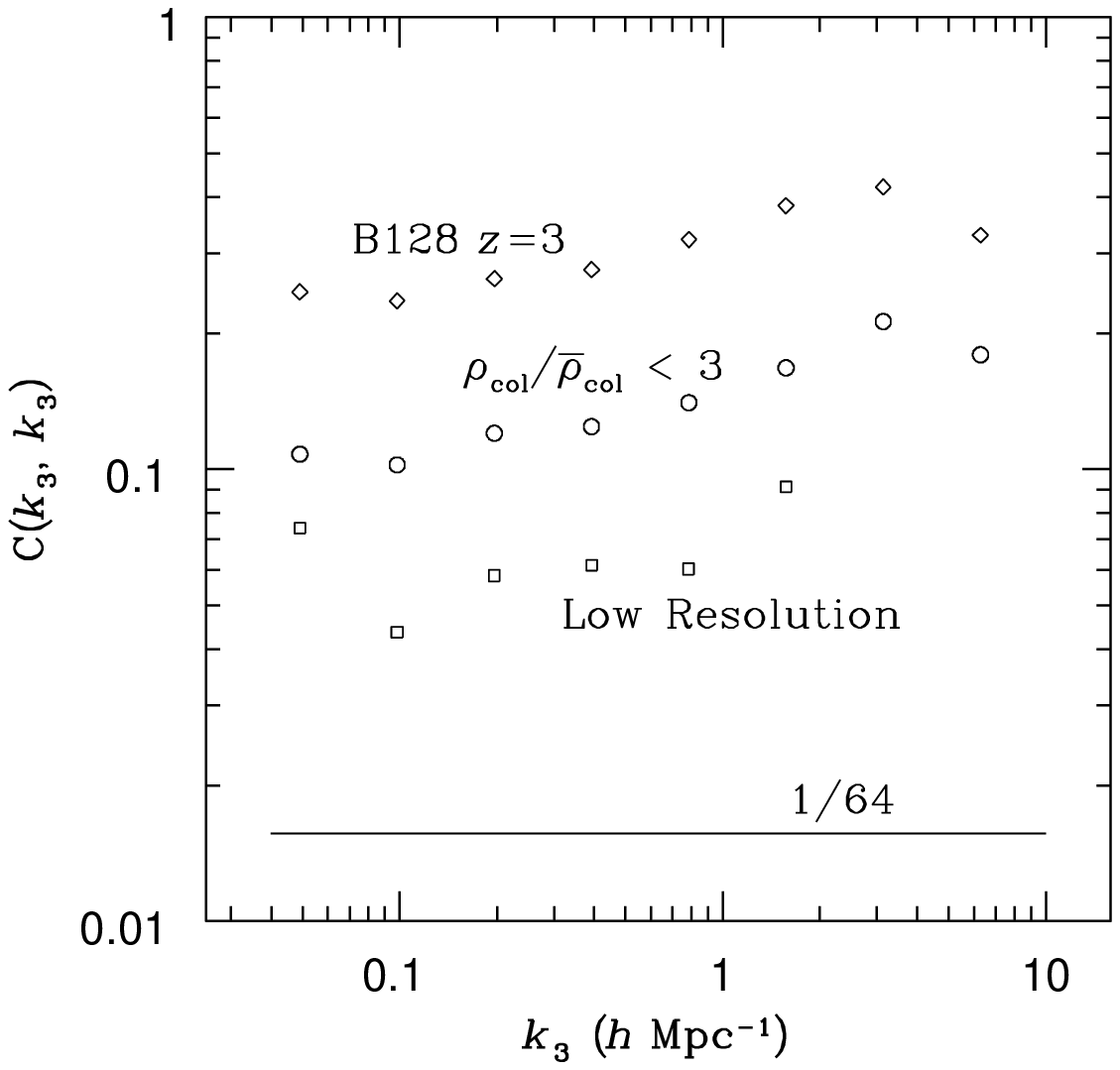}
\caption[Effect of non-Gaussianity and resolution on the variance of 
the \oned{} mass power spectrum]{
The effect of non-Gaussianity and resolution on the variance, i.e. 
diagonal elements of the covariance $\sigma^2_{\rm 1D}(k_3, k'_3)$. 
Diamonds are the normalized variance $C(k_3, k'_3)$ for B128 at $z=3$.
Circles correspond to the middle column of Fig.~\ref{fig:cov64l}, which 
imposes the selection criterion 
$\rho_{\rm col} / \bar{\rho}_{\rm col} < 3$. Squares are from the 
low-resolution calculation in the right panel of Fig.~\ref{fig:cov64l}.
The horizontal line marks the Gaussian value of $1/64$.
\label{fig:var64l}}
\end{figure}

The distribution function of line-of-sight column densities (upper left
panel of Fig.~\ref{fig:cov64l}) suggests that the long tail of
high-column-density LOS's may affect the
covariance. I re-calculate the covariance $\hat{C}(k_3, k_3')$ with a
selection criterion that the column density of each LOS $\rho_{\rm col} /
\bar{\rho}_{\rm col} < 3$. The result is shown in the upper middle panel
of Fig.~\ref{fig:cov64l}. The diagonal elements are more
dominant than those in the B128 $z = 3$ panel in Fig.~\ref{fig:cov64}. The
first two lower panels are the cross sections of $\hat{C}(k_3, k_3')$ with
(middle) or without (left) the selection criterion. They provide a more
quantitative comparison, which suggests that rare high-column-density
LOS's do increase the correlations between different modes of
fluctuations. Meanwhile, Fig.~\ref{fig:var64l} clearly demonstrates that
these LOS's also increase the variance of the \oned{} mass PS by at least a 
factor of 2 on all scales.

For a fixed number of particles, the simulation box sets a cut-off scale,
below which the fluctuations cannot be represented. In other words, the
number of particles and the size of the simulation determine the
highest-wavenumber modes that are included in calculations of the
one-dimensional mass PS and its covariance. Since the non-Gaussianity is
stronger at smaller scales, a larger simulation box cuts off more 
small-scale fluctuations and may cause the correlation
to appear weaker in Figs.~\ref{fig:cov64} and \ref{fig:cov1024}.
To test this, I assign the density field of the simulation B128 at $z=3$
on a grid of $128^3$ nodes. The spatial resolution is the same as the
simulation B512 on a grid of $512^3$ nodes. The covariance $\hat{C}(k_3,
k_3')$ is calculated in the same way as those in Fig.~\ref{fig:cov64} but
with fewer groups of LOS's. Each group still has 64 LOS's. The results are
shown in the right column of Fig.~\ref{fig:cov64l} and in 
Fig.~\ref{fig:var64l}. Evidently, there is a 
significant reduction of the correlations between different modes as well
as the variance of the \oned{} mass PS. Indeed, the variance from the 
low-resolution calculation is close to that from the B512 simulation.
Thus, the apparent resemblance between B512 and R512 in 
Figs.~\ref{fig:cov64} and \ref{fig:cov1024} is mostly due to 
the low resolution of the large-box simulation.

\begin{figure} 
\centering
\includegraphics[width=5.8in]{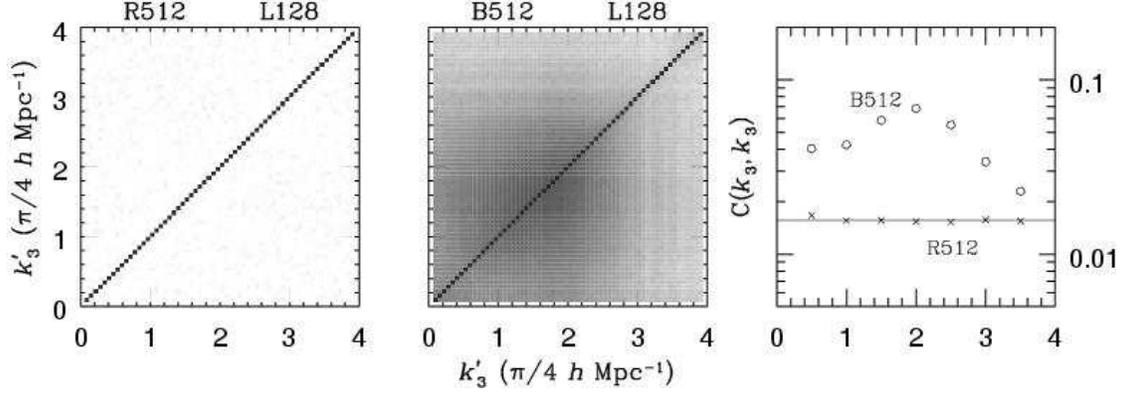}
\caption[Effect of the length of lines of sight]{
The same as the bottom row of Fig.~\ref{fig:cov64}, but with length of
LOS's $L = 128\ h^{-1}$Mpc.
\label{fig:covlen}}
\end{figure}

It is expected from equation (\ref{eq:cov-short}) that the covariance
matrix will not be diagonal if the length of LOS's is less than the size
of the simulation box. Fig.~\ref{fig:covlen} shows the covariances that
are calculated in the same way as those in Fig.~\ref{fig:cov64} for the
GRF R512 and the simulation B512, except that each LOS is only 128
\mbox{$h^{-1}$Mpc} long. The effect of the length is not visible for the
GRF, but it does increase the correlation between different modes and 
doubles the variance of the one-dimensional mass PS for the simulation B512.
For real observations, the line-of-sight length is always much less than
the size of the observable universe, so that the window function in the
line-of-sight direction will cause stronger mixing of modes and more 
pronounced increase of correlation and variance.

\chapter[Covariance of the One-Dimensional Power \\
Spectrum: the \lya{}
Forest]{Covariance of the One-Dimensional Power Spectrum: 
the Ly$\boldsymbol{\alpha}$ Forest} \label{ch:fps}
\noindent
Previous chapters have focused on the mass PS and its covariance. 
However, one does not directly observe \oned{} density fields. 
What can be observed instead is the \lya{} flux. Therefore, many 
works have been based on flux statistics, which are then used along 
with simulations to constrain cosmology. Here I examine the 
flux PS and its covariance using simulated \lya{} forests.

\section[Simulated \lya{} Forests]{Simulated Ly$\boldsymbol{\alpha}$ 
Forests}
The \lya{} forest probes deeply into the nonlinear regime of the cosmic
density field. This has made numerical simulations an indispensable
tool for understanding the nature of the \lya{} forest and inferring
cosmological parameters from flux statistics. Two types of 
cosmological simulations have been commonly used 
to simulate the \lya{} forest. One is pure CDM simulations 
($N$-body simulations) that assume baryons to trace dark matter 
(e.g.~Petitjean, M\"{u}cket \& Kates 1995; Riediger et al.~1998).
%\citep[e.g.][]{pmk95, rpm98}. 
%(e.g.~Petitjean et al.~1995; Riediger et al.~1998). 
The other is hydrodynamical simulations 
(e.g.~Cen et al.~1994; Zhang et al.~1995; Hernquist et al.~1996).
%\citep[e.g.][]{cmo94, zan95, hkw96}. 
Other types of simulations 
exist as well. For example, the simple log-normal model \citep{bd97} is 
already able to reproduce some \lya{} flux statistics. 

I use both $N$-body simulations and hydrodynamical simulations to 
investigate the flux PS and its covariance. Readers are 
referred to the above references for details of simulation techniques, 
and I describe the procedure of extracting \lya{} forests from 
simulations in this section.

\subsection{Hydrodynamical Simulations with Photoionization}
\label{sec:hyd-lya}
Two hydrodynamical simulations (HYDRO1 \& HYDRO2) were provided by 
Romeel Dav\'{e}. They are both variants of the 
low-density-and-flat CDM (LCDM) model with a slight tilt of the power 
spectral index $n$ (see Table \ref{tab:para1}). 
HYDRO1 evolves $128^3$ CDM particles and $128^3$ gas particles from 
$z = 49$ to 0 using Parallel TreeSPH \citep{ddh97}. HYDRO2 differs 
from HYDRO1 only in cosmological parameters, and it has snapshots 
available down to $z = 2$. 
The box size is 22.222 \mpc{} in each dimension with a 5 \kpc{} 
resolution. The two simulations also include star formation with feedback 
and photoionization \citep{kwh96}. The UV ionization background is from
\citet{hm96}. 
\begin{table}
\centering
\caption[Parameters of the simulations]{Parameters of the simulations.}
\label{tab:para1}
\begin{tabular}{lccccccc}
\hline
Model & Type & $\Omega$ & $\Omega_{\rm b}^{\ a}$ & 
$\Omega_\Lambda$ & $h$ & $n$ & $\sigma_\mathrm{8}$ \\ 
\hline
HYDRO1 & Hydro. & 0.4 & 0.05 & 0.6 & 0.65 & 0.95 & 0.8 \\
HYDRO2 & Hydro. & 0.3 & 0.04 & 0.7 & 0.7 & 0.95 & 0.8 \\
HIGH$n$ & $N$-Body & 0.3 & 0.04 & 0.7 & 0.7 & 1.1  & 0.8 \\
HIGH$\sigma$ & $N$-Body & 0.3 & 0.04 & 0.7 & 0.7 & 1.0  & 1.0 \\
LCDM   & $N$-Body & 0.3 & 0.04 & 0.7 & 0.7 & 1.0  & 0.8 \\
OCDM  & $N$-Body & 0.3 & 0.04 & 0   & 0.7 & 1.0  & 0.8 \\
\hline
\end{tabular}

%\medskip
\parbox{0.65 \textwidth}{
\begin{itemize}
\setlength{\leftskip}{-0.2in}
\setlength{\itemsep}{0ex}
\item[$^a$] With exceptions of HYDRO1 \& HYDRO2, the baryon density 
parameter is used only for generating the initial mass power spectrum.
\end{itemize}}
\end{table}

\subsubsection{Density Grid}
Snapshots of the simulations contain the position ${\bf r}_i$ and 
velocity ${\bf v}_i$ of each particle, where $i$ is the label of the
$i$th particle. Smooth-particle hydrodynamics (SPH) defines the baryon
density $\rho_{\rm b}({\bf x})$ at any location to be a sum of 
contributions from all nearby gas particles, i.e.
\begin{equation}
\rho_{\rm b}({\bf x}) = \sum_{i = 1}^{N_{\rm p}} 
m_i\, w(|{\bf x}-{\bf r}_i|, \epsilon^j_i),
\end{equation}
where $N_{\rm p}$ is the total number of particles, $w$ is the density 
kernel or the assignment function, $m_i$ is the mass of particle $i$, 
and $\epsilon^j_i$ is the smoothing length determined by the distance 
between particle $i$ and its $j$th neighbor ($j = 32$ in this chapter). 
In practice, densities are assigned on a discrete grid for further 
analysis.

\begin{figure} 
\centering
\includegraphics[width=5in]{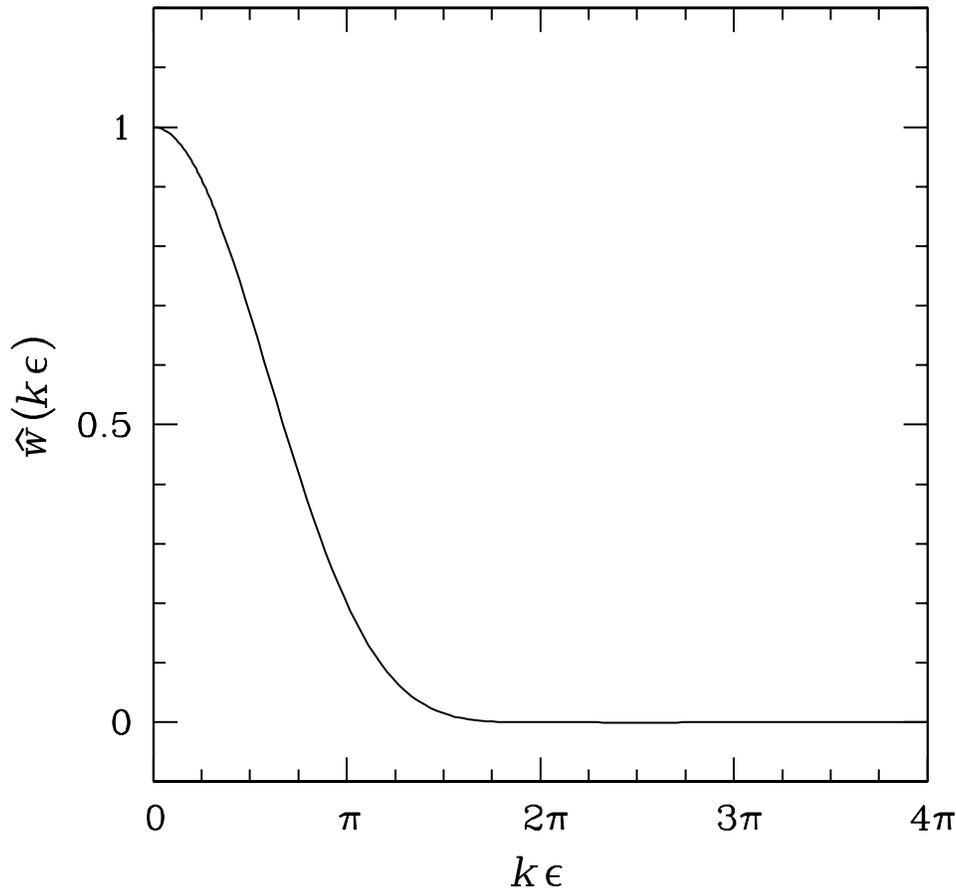}
\caption[Fourier transform of the assignment function]{
Fourier transform of the assignment function $w(r, \epsilon)$.
\label{fig:faf}}
\end{figure}

I adopt a spherically symmetric spline kernel from 
\citet[][with a typo corrected]{ml85}, which is also used in TreeSPH 
for force calculations. It has the form
\begin{equation}
w(r, \epsilon) = \frac{1}{\pi \epsilon^3}\left\{ \begin{array}{ll}
1 - \frac{3}{2} \left(\frac{r}{\epsilon}\right)^2 
+ \frac{3}{4} \left(\frac{r}{\epsilon}\right)^3 & 
\quad 0 \leq r < \epsilon \\ 
\frac{1}{4}\left[2- \frac{r}{\epsilon}\right]^3 & 
\quad \epsilon \leq r < 2 \epsilon \\ 
0 & \quad r \geq 2 \epsilon, \\ \end{array} \right.
\end{equation}
which vanishes beyond the radius $2\epsilon$ and has a smooth gradient 
everywhere. The Fourier transform of the kernel is
\begin{equation}
\hat{w}(k,\epsilon) \equiv \hat{w}(k\epsilon) = \frac{48}{(k\epsilon)^6}
\left[(4-k\epsilon\sin k\epsilon) \sin^2\frac{k\epsilon}{2}
-\sin^2 k\epsilon \right].
\end{equation}
Fig.~\ref{fig:faf} shows that the density kernel is an effective 
low-pass filter that suppresses fluctuations on scales smaller than
$2\epsilon$ ($k > \pi/\epsilon$). This filtering is necessary because
fluctuations on scales less than the physical size of an SPH particle are
unphysical. Furthermore, to reduce the alias effect (see Chapter 
\ref{ch:p1d}), the smoothing length $\epsilon$ is required to be greater 
than or equal to the spacing of the density grid. Other kernels such as 
the TSC and wavelet scaling functions have also been used for similar 
purpose.

It is important to realize that the filtering scale should be adjusted to 
particle con-centration---the denser the environment, the smaller the 
filtering scale. A particle in an empty region does not represent a 
condensed clump of matter sitting in vacuum but rather a dilute 
distribution that fills the space between the particle and its distant 
neighbors. Thus, particles in empty regions should not contribute to 
small-scale fluctuations. On the other hand, particles in high density
regions contain more small-scale information, and their filtering scale
should be smaller, i.e.~a higher cutoff wavenumber. 
Density kernels based on neighbor distances satisfy such 
requirement, and they are widely used in SPH simulations.
When the scale of interest is larger than or comparable to the mean 
inter-particle distance ($k \lesssim k_{\rm p}$, see Chapter 
\ref{ch:mps}), kernels with an indiscriminating filtering 
scale for all particles are just as good. Otherwise, a density-dependent
kernel must be used. 

The density field $\rho_{\rm b}({\bf x})$ can be readily constructed 
once the smoothing length is determined for each particle. In principle, 
LOS's may be sampled in any random direction, but for computational 
simplicity I assign the density on a grid of $256^3$ nodes,
and then extract \oned{} densities randomly from this grid. Meanwhile,
particle temperatures are also assigned to each node with weights 
proportional to each particle's contribution of density at that node.

\subsubsection{Ly$\boldsymbol{\alpha}$ Flux} \label{sec:lyaf}
I assume a universal hydrogen fraction of 0.76 to convert baryon 
densities to hydrogen densities. For each density node, the equilibrium 
\ion{H}{i} fraction is calculated from the balance between cooling and 
heating, which include adiabatic cooling (cosmic expansion), 
photoionization, recombination, collisional excitation and ionization, 
thermal Bremsstrahlung, and Compton scattering with cosmic microwave 
background (CMB) photons (for details, see Katz et al.~1996).
%\citep[for details, see][]{kwh96}. 
R.~Dav\'{e} provided codes for this calculation. The assumption of ionization 
equilibrium certainly breaks down in very dynamic regions such as shocks. 
However, since the equilibrium \ion{H}{i} fraction calculated for shocks is 
already considerably lower than that elsewhere, there will not be much an 
effect on 
simulated \lya{} forests even if additional shock physics can 
further reduce the \ion{H}{i} fraction by orders of magnitude. Besides,
shock fronts, unlike shocked gases, occupy only a small fraction of the 
total simulation volume, so they could not have a great impact on 
the \lya{} forest.

With the \ion{H}{i} fraction and hydrogen density along the LOS, one 
can determine the \lya{} optical depth $\tau$ and transmitted \lya{} 
flux $F$ of each pixel, i.e.~each node of the density grid. The mean 
flux $\bar{F}$ of all LOS's is well determined by observations. I adjust 
the intensity of the UV ionization background $\Gamma_{\rm UV}$ so that 
the mean flux of all pixels in the simulations follows
\begin{equation} \label{eq:mflx}
\bar{F}(z) \simeq \left\{ \begin{array}{ll}
\exp \left[-0.032\,(1+z)^{3.37\pm 0.2}\right] & \quad 1.5 \leq z \leq 4 \\
0.97-0.0252\,z\pm (0.003+0.0054\,z) & \quad 0 \leq z < 1.5, \\
\end{array} \right.
\end{equation}
which is adapted from \citet{kcc02} and \citet{dhk99}. 
This mean flux formula is also consistent with other observations 
\citep{lsw96, rms97, mmr00}. There is a slight
inconsistency that HYDRO1 and HYDRO2 have already included the UV 
ionization background, yet I need to adjust the intensity of the UV 
radiation on outputs of the simulations to fit mean fluxes. 
This inconsistency does not significantly affect the results that follow 
because, first, the simulation outputs are able to reproduce the observed 
mean flux with their internal UV ionization background \citep{dhk99}.
External adjustments are only needed to vary the mean flux within the 
given observational and numerical uncertainties.
Second, the UV ionization background has an important role in the 
evolution of the IGM temperature (Katz et al.~1996), %\citep{kwh96}, 
but the large-scale distribution of baryons is driven by gravity. 
Thus, even if the intensities of the externally adjusted UV 
background were used internally in the simulations, LOS 
\lya{} absorptions should not change greatly.

The mean temperature of the IGM is on the order of $10^4$ K, so thermal 
broadening of absorptions must be taken into account. At a temperature
$T_i$, the flux decrement $D_i = 1-F_i$ of pixel $i$ is spread to pixel 
$j$ as
\begin{equation}
D_{ji} =  \frac{D_i}{\sqrt{\pi T_{4,i}}} \frac{H d}{b}
\int_{j-1/2}^{j+1/2}\, e^{-(i-t)^2 (H d)^2 / b^2 T_{4,i}}\,{\rm d} t, 
\end{equation} 
where $b\simeq 13$ \kmps{}, $H$ is the Hubble constant at corresponding 
redshift, $d$ is the physical separation mapped by two adjacent pixels, 
and $T_{4,i} = T_i / (10^4\ \mbox{K})$. The ionization-equilibrium 
temperature can be uniquely determined from density, and it is 
obtained for each pixel while \ion{H}{i} fraction is calculated 
recursively. Because of shock heating, especially at low redshift,
ionization-equilibrium temperatures are often lower than density-weighted 
SPH temperatures of the density grid. In this case, the latter is used for 
thermal broadening. The broadened flux $\widetilde{F}$ is then
\begin{equation}
\widetilde{F}_i= 1-\sum_{j = 1}^{M}D_{ij},
\end{equation}
where $M$ is the number of pixels along the LOS, but practically the 
summation is over a much smaller number of pixels where $D_{ij}$ is
significant. 

\begin{figure}
\centering
\includegraphics[width=5.5in]{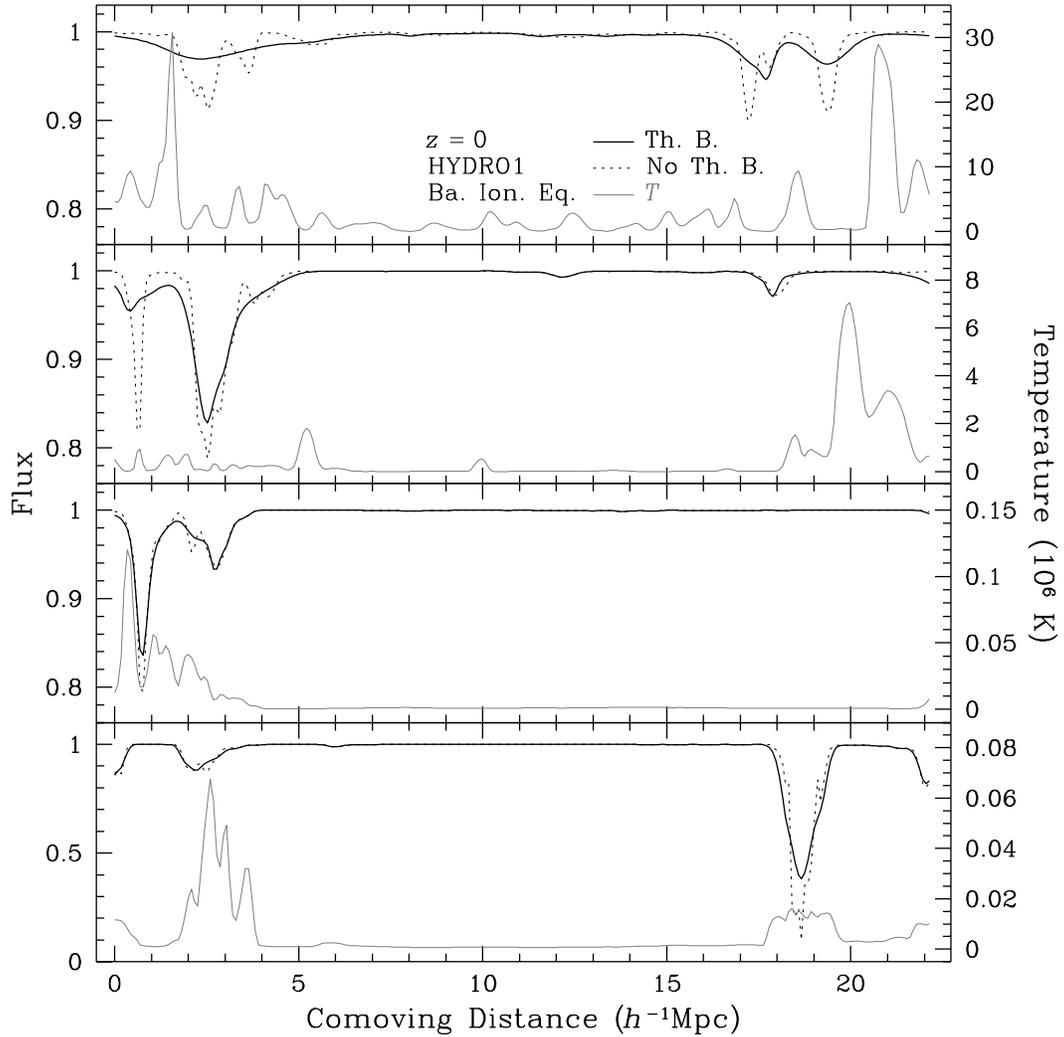}
\caption[Thermal broadening of \lya{} absorptions at $z=0$]{
Thermal broadening of \lya{} absorptions at $z=0$. Solid lines
are \lya{} absorptions with thermal broadening, and dotted lines 
without. Grey lines are temperatures along LOS's. The \lya{} forests 
are generated from baryon distributions in HYDRO1 with the assumption
of ionization equilibrium.
\label{fig:tb0}}
\end{figure}

\begin{figure}
\centering
\includegraphics[width=5.5in]{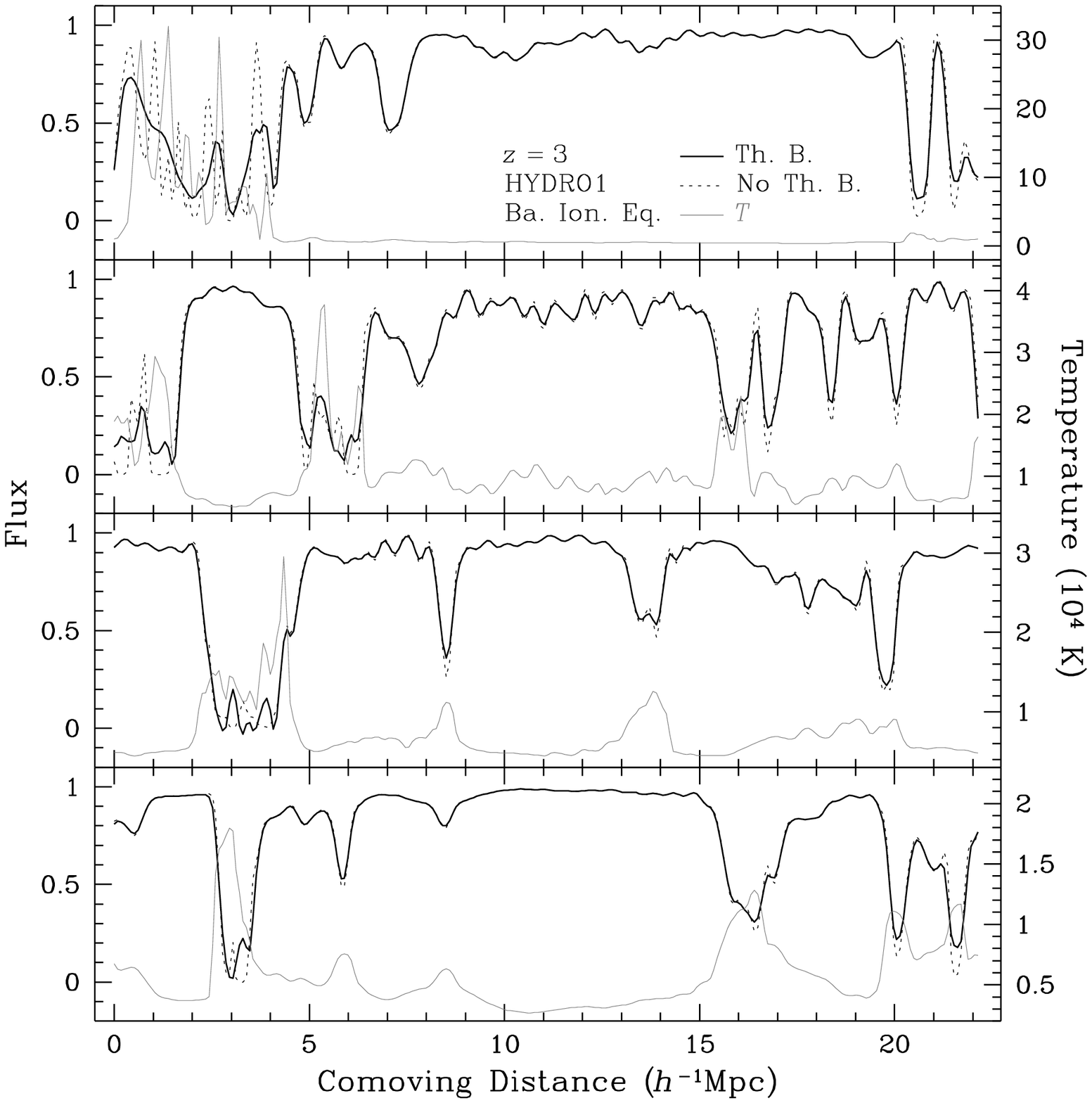}
\caption[Thermal broadening of \lya{} absorptions at $z=3$]{
Thermal broadening of \lya{} absorptions at $z=3$. Legends are
the same as in Fig.~\ref{fig:tb0}.
\label{fig:tb3}}
\end{figure}

Thermal broadening 
smoothes out small-scale fluctuations in the \lya{} forest without 
altering the mean flux. Therefore, it preferentially reduces the flux 
power on small scales. At $z=3$, the majority of \lya{} absorptions 
arise from regions of density $\rho/\bar{\rho} \lesssim 10$, where
temperatures are not much above $10^4$ K and broadening widths 
$b\sqrt{T_4}$ are a few tens of \kmps. At $z = 0$, however, \lya{} 
absorptions are produced in much hotter regions where broadening 
widths can be as high as hundreds of \kmps. 
In other words, thermal broadening is a much stronger effect at 
lower redshift, which can be seen by comparing simulated \lya{} forests
at $z= 0$ (Fig.~\ref{fig:tb0}) with those at $z = 3$ 
(Fig.~\ref{fig:tb3}). 

There is a general correlation between the temperature and \lya{} 
absorption due to the temperature--baryon density--\ion{H}{i} fraction 
relation (see Section \ref{sec:eos}). However, high-temperature regions 
($T\gtrsim 10^6$ K) at $z = 0$ are not always related to \lya{} 
absorptions because they are often too hot and too dilute to maintain an 
appreciable \ion{H}{i} column density. These high-temperature regions 
are categorized as the shock-heated $10^5$--$10^7$ K WHIM 
\citep{dhk99, dt01, dco01}. 

The cosmic density field becomes more and more clustered so that for any 
random LOS the chance of passing through relatively high density regions 
that can produce detectable \lya{} absorptions is much smaller at $z = 0$.
This is reflected in Figs.~\ref{fig:tb0} and \ref{fig:tb3} that 
low-redshift \lya{} forests have much fewer absorptions per unit comoving
distance than high-redshift ones. It is also interesting to note that 
there is a relatively deep \lya{} 
absorption in the last panel of Fig.~\ref{fig:tb0} despite the high
mean flux of 0.97 at $z = 0$. The LOS in this panel is the same as that
in the second and fourth panels of Fig.~\ref{fig:rhodmbal0}, where 
LOS baryon and dark matter densities are shown in real space 
and redshift space, respectively. The deep absorption actually arises
from a nearly virialized cluster of real-space density 
$\rho/\bar{\rho} \sim 160$. 

\subsubsection{Line of Sight in Redshift Space}
So far I have not mentioned the fact that the \lya{} forest is observed 
in redshift space where the distribution of matter is distorted by peculiar 
velocities. A simple way to approximate redshift distortion is to 
displace each particle a distance $v_3/H$ along the LOS before 
constructing the density grid, i.e.
\begin{equation}
{\bf r}^S_i = {\bf r}_i + \frac{{\bf v}_i \cdot \hat{x}_3}{H} \hat{x}_3,
\end{equation}
where the superscript $S$ stands for redshift space, and I have made 
use of the plan-parallel approximation and assumed $\hat{x}_3$ to be 
in the LOS direction. Subsequently, the redshift-space density 
$\rho^S_{\rm b}({\bf x})$ is
\begin{equation}
\rho^S_{\rm b}({\bf x}) = \sum_{i = 1}^{N_{\rm p}} m_i \,
w(|{\bf x}-{\bf r}^S_i|, \epsilon^{32}_i),
\end{equation}
where the smoothing length of each particle is also obtained in 
redshift space. Then, \lya{} forests can be extracted from 
$\rho^S_{\rm b}({\bf x})$ in the same way as discussed above. A more 
detailed discussion on the effect of redshift distortion is given in 
Section \ref{sec:badm}.

\subsection{Pseudo-Hydro Technique}
Although full hydrodynamical simulations are well suited for studies
of the \lya{} forest, they are currently too time-consuming to explore 
a large cosmological parameter space as one often desires. Whereas, 
$N$-body simulations run much faster, and they can be used to cover a 
wide range of cosmological models in practical time. 

\citet{cwk98} proposed a pseudo-hydro technique for generating \lya{} 
forests from $N$-body simulations. It is based on two 
important theoretical expectations that are supported by hydrodynamical 
simulations: (1) In general, baryons trace dark matter 
above the Jeans scale \citep[e.g.][]{gh98}; and (2) In ionization 
equilibrium, the equation of state (EOS) of the IGM gives rise to an 
approximate temperature--density relation
\begin{equation} \label{eq:eos}
T=T_0(\rho_{\rm b}/\bar{\rho}_{\rm b})^\alpha, 
\end{equation}
where $T_0 \sim 10^4$ K, $0.3\leq \alpha \leq 0.6$, and 
$\rho_{\rm b}/\bar{\rho}_{\rm b}\lesssim 10$ \citep{hg97}. Since the 
\lya{} optical depth is proportional to $\rho_{\rm b}^2\, T^{-0.7}$ in 
regions around the mean density, one finds
\begin{equation} \label{eq:lyaf}
F \simeq e^{-A (\rho_{\rm b}/\bar{\rho}_{\rm b})^\gamma} 
\simeq e^{-A (\rho_{\rm d}/\bar{\rho}_{\rm d})^\gamma},
\end{equation}
where $A \propto \Omega_{\rm b}^2\,\Gamma_{\rm UV}^{-1}T_0^{-0.7}$, 
$\gamma = 2 - 0.7 \alpha$, and $\rho_{\rm d}$ is the dark matter density.
Usually the constant $A$ is left as a fitting parameter adjusted to 
reproduce the observed mean flux.

\begin{figure}
\centering
\includegraphics[width=5.in]{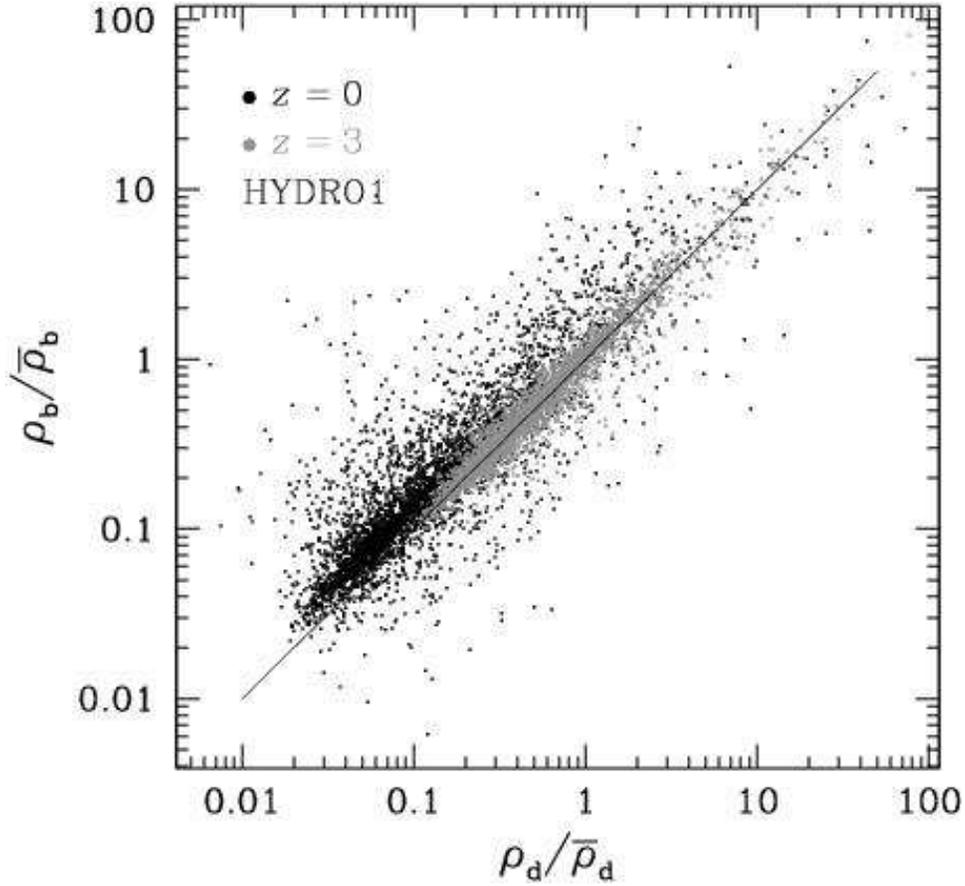}
\caption[Correlation between baryon and dark matter densities in the 
simulation]
{The correlation between baryon and dark matter densities. Grey dots are 
for $z=3$, and black dots $z = 0$. Each group consists of 4000 pairs of 
density values randomly selected from the density grid of the simulation 
HYDRO1. The diagonal line follows 
$\rho_{\rm b}/\bar{\rho}_{\rm b} = \rho_{\rm d}/\bar{\rho}_{\rm d}$.
\label{fig:rhodmba}}
\end{figure}

\subsubsection{Correlation between Baryons and Dark Matter} 
\label{sec:badm}
To justify the pseudo-hydro technique, one must show that baryons and
dark matter trace each other on large scales. For a simple test, 
Fig.~\ref{fig:rhodmba} compares baryon densities with dark matter 
densities from the same set of 4000 randomly selected density nodes of
HYDRO1. There is clearly a strong correlation between baryons
and dark matter \citep[see also][]{gh98}. The correlation has 
larger scatter at lower redshift because gravity is no longer the
dominant driving force behind strong hydrodynamical events such as 
shocks, which occur more frequently at lower redshift. 
Baryons are slightly denser than dark 
matter below the mean density at $z = 0$ because of finite pressure of
SPH particles \citep{gh98}. In other words, SPH particles have much 
larger smoothing radii than CDM particles, so that SPH particles in a
low-density region receive less acceleration toward a nearby 
high-density region than CDM particles.

The Jeans length sets a characteristic scale, below which baryon 
pressure will resist the growth of gravitational perturbations and 
cause baryons not to trace dark matter. By a simple comparison between 
the dynamical time and sound travel time, one finds the comoving Jeans 
length 
\begin{eqnarray} \label{eq:jeans}
L_\mathrm{J} &=&  (1+z) \sqrt{\frac{\pi\,c_\mathrm{s}^2}{G\rho}}=
\sqrt{\frac{40\pi^2 k_\mathrm{B} T}{9\mu m_\mathrm{p} 
H_\mathrm{0}^2 \Omega\,\rho\, (1+z)}} 
\nonumber \\ 
&=& 780\ T_\mathrm{4}^{1/2} 
\left[\Omega\, (\rho/\bar{\rho}) (1+z)\right]^{-1/2} \kpc,
\end{eqnarray}
where $c_\mathrm{s}$ is the speed of sound, $G$ is the gravitational 
constant, $k_\mathrm{B}$ is 
the Boltzmann constant, $\mu\simeq 0.59$ is the mean molecular weight, 
$m_\mathrm{p}$ is the proton mass, and $H_0 = 100\ h$ 
\mbox{km s$^{-1}$ Mpc$^{-1}$}. With the EOS equation 
(\ref{eq:eos}), the Jeans length can be rewritten as 
\begin{equation}
L_\mathrm{J}=780\ \left(\frac{T_0}{10^4\ {\rm K}}\right)^{1/2} 
\left[\frac{1}{\Omega\, (1+z)}\right]^{1/2} 
\left(\frac{\bar{\rho}}{\rho}\right)^{(1-\alpha)/2}\kpc.
\end{equation}
For $T_0 = 10^4$ K and $\Omega = 0.4$ (HYDRO1), it is expected that
baryons in regions around the mean density to generally follow dark 
matter above 1.2 \mpc{} (0.6 \mpc{}) at $z = 0$ ($z = 3$). Of course, 
in vast low density regions the Jeans length may be much larger.

The Jeans length analysis idealizes the state of baryons and neglects
external forces. Thus, it is not applicable in very dynamic regions 
such as shock fronts. For example, in the spherical collapse case, even 
though baryons may initially follow dark matter, shocks could eventually 
develop and allow baryons to separate from dark matter. 

\begin{figure}
\centering
\includegraphics[width=5.5in]{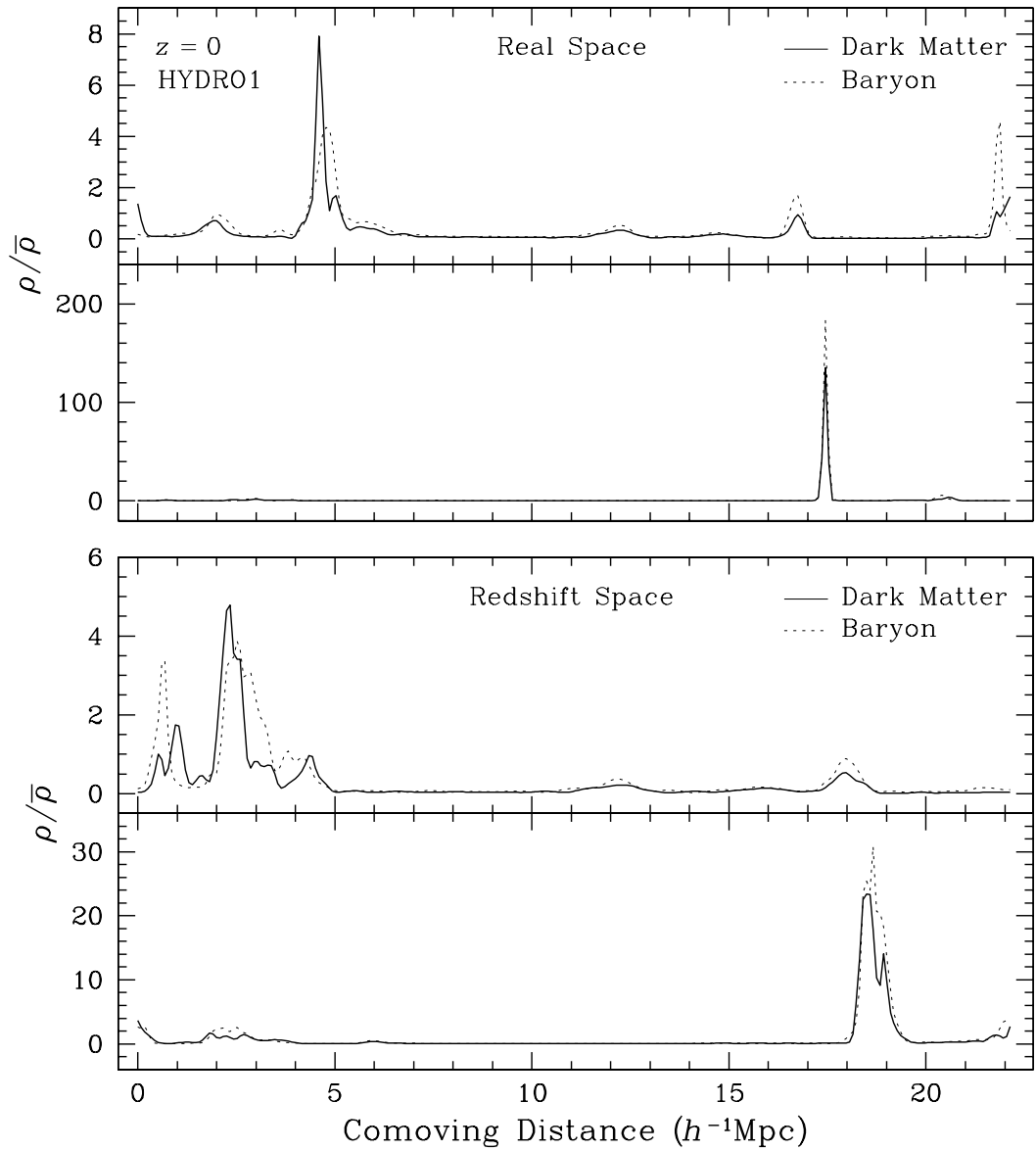}
\caption[Line-of-sight baryon densities and dark matter densities at 
$z=0$]{LOS baryon densities (dashed liens) and dark matter 
densities (solid lines) at $z=0$. The upper two panels show LOS's in 
real space, while the lower two the same LOS's in redshift space. 
All densities are extracted 
from HYDRO1. The two LOS's here correspond to those in the second and 
last panels of Fig.~\ref{fig:tb0}.
\label{fig:rhodmbal0}}
\end{figure}

\begin{figure}
\centering
\includegraphics[width=5.5in]{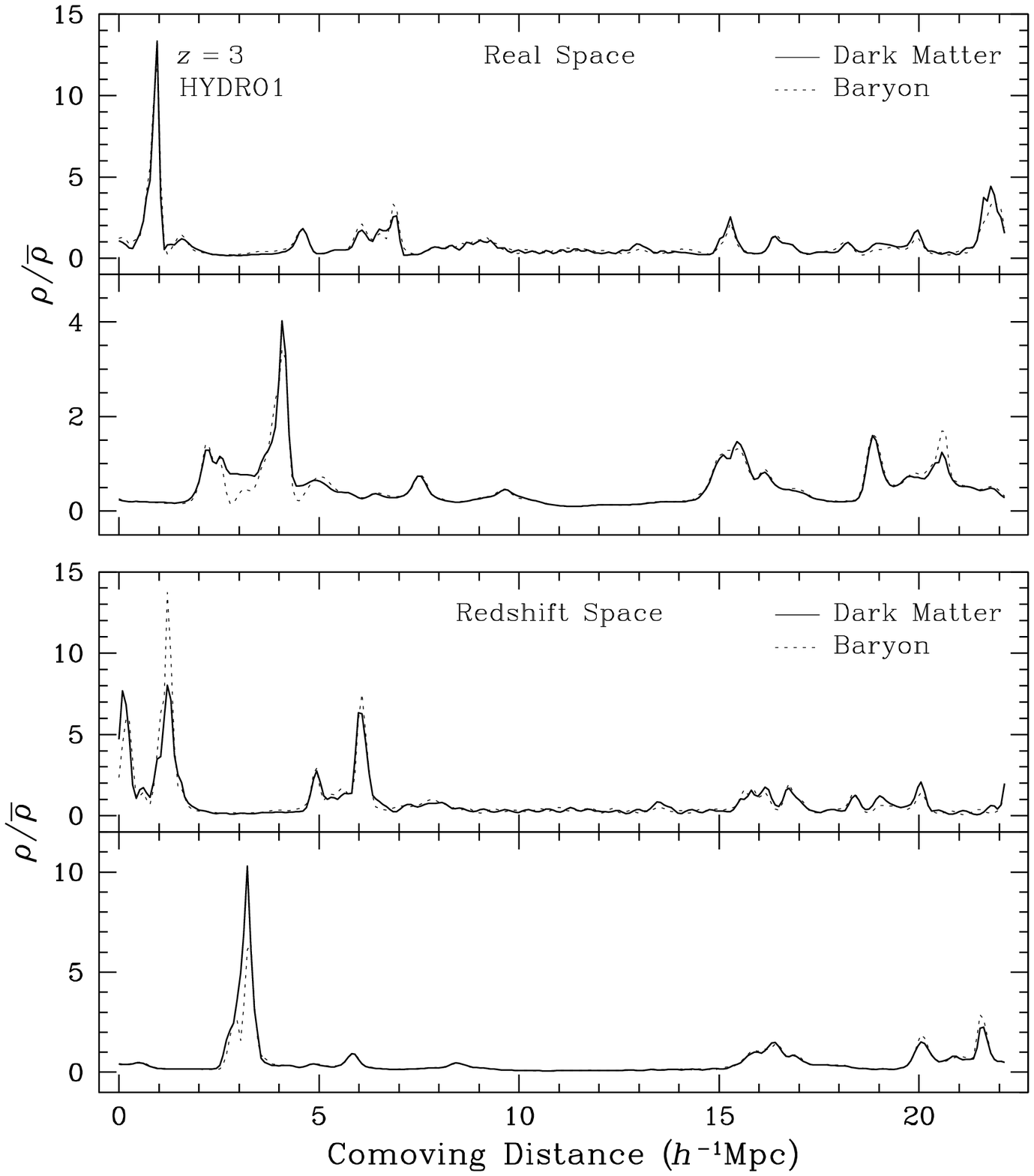}
\caption[Line-of-sight baryon densities and dark matter densities at
$z=3$]{The same as Fig.~\ref{fig:rhodmbal0} but at $z = 3$.
The two LOS's here correspond to those in the second and last panels 
of Fig.~\ref{fig:tb3}.
\label{fig:rhodmbal3}}
\end{figure}

Figs.~\ref{fig:rhodmbal0} and \ref{fig:rhodmbal3} examine baryon and 
dark matter densities along LOS's in both real space and redshift space.
The two LOS's in Fig.~\ref{fig:rhodmbal0} (Fig.~\ref{fig:rhodmbal3}) 
correspond to the LOS's in the second and last panels of Fig.~\ref{fig:tb0}
(Fig.~\ref{fig:tb3}). Baryon densities and dark matter densities have 
almost one-to-one correspondence in real space at $z = 3$. However, they
do not share the same velocity structure because SPH particles receive not 
only gravitational accelerations but also hydrodynamical accelerations. 
This leads to the departure of baryons from dark matter in redshift space. 
At $z = 0$, the difference between baryon and dark matter distributions is 
already seen in real space, and it does not seem to be amplified 
by redshift distortion. 

\begin{figure}
\centering
\includegraphics[width=5.5in]{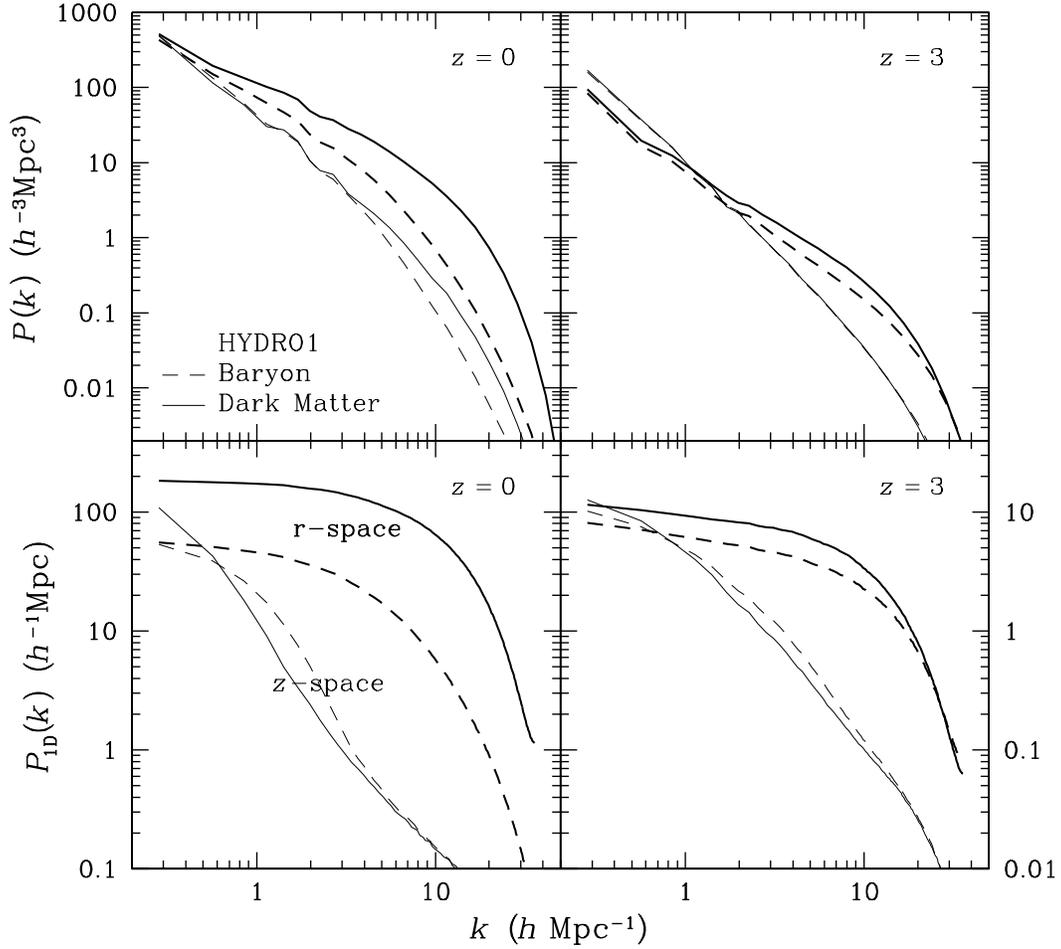}
\caption[Three-dimensional and \oned{} mass power spectra of baryons 
and dark matter in real space and redshift space]
{Three-dimensional and \oned{} mass PS's of baryons (dashed lines) and 
dark matter (solid lines) from HYDRO1. Redshift-space PS's are shown in 
thick lines, and real-space PS's thin lines.
\label{fig:psdmba}}
\end{figure}

The LOS densities of baryons and dark matter are qualitatively 
consistent with the expectation from Jeans length analysis, and statistics 
are needed to quantify how well baryons and dark matter trace each other. 
Fig.~\ref{fig:psdmba} compares baryon and dark matter PS's in both real 
space and redshift space. At $z = 3$, the \thrd{} baryon PS has already
departed from the \thrd{} dark matter PS on scales below 6 \mpc{} 
($k \gtrsim 1$ \mpci{}), which is ten times larger than the Jeans 
length in mean-density regions. This discrepancy can be attributed to 
the fact that the vast volume of the universe is well below the mean 
density and have much larger Jeans lengths. Therefore, the \thrd{} baryon 
PS is somewhat lower than that of dark matter for $k \gtrsim 1$ \mpci, 
even though the Jeans length in mean-density regions corresponds to 
$k_{\rm J} \simeq 10$ \mpci.
At $z = 0$, shocks have heated parts of the IGM to much higher 
temperatures and led to more reduction of the baryon PS with respect
to the dark matter PS.

The linear redshift distortion is first derived by \citet{k87},
\begin{equation}
P^S({\bf k}) = \left(1+\beta\,\mu^2_\theta\right)^2P({\bf k}),
\end{equation}
where $P^S({\bf k})$ is the \thrd{} mass PS in redshift space, 
$\beta \simeq \Omega^{0.6}(z)$ \citep{lrl91}, and $\mu_\theta$ is the cosine
of the angle between the LOS and the wavevector ${\bf k}$. The monopole 
of $P^S({\bf k})$ is 
\begin{equation}
P^S_0(k) = \Big(1 + \frac{2}{3}\beta + \frac{1}{5}\beta^2\Big) P(k),
\end{equation}
which is boosted with respect to $P(k)$ by a factor of 1.9 at high redshift. 
The nonlinear redshift distortion reduces the power of Fourier modes 
along the LOS, and the reduction is more severe on smaller scales. 
These effects are illustrated in Fig.~\ref{fig:psdmba}. Both the 
constant boost on large scales and the reduction of power on small 
scales can be seen from \thrd{} mass PS's at $z = 3$. Whereas, 
the nonlinear scale of redshift distortion has 
evolved beyond the size of the simulation box at $z = 0$, so that the 
monopole of the redshift-space PS is always below the real-space PS. 
The nonlinear redshift distortion is also reflected in 
Fig.~\ref{fig:rhodmbal0}, where a real-space structure of
$\rho/\bar{\rho} \sim 160$ is smoothed to $\rho/\bar{\rho} \sim 20$ in
redshift space. 
Since the nonlinear redshift distortion is equivalent to a small-scale
filter, differences between PS's are expected to be smaller in redshift
space than in real space. However, the agreement of $P^S_0(k)$ between 
baryons and dark matter at $z = 3$ is probably coincidental. 

The real-space \oned{} PS of baryons can differ significantly from that 
of dark matter even on large scales, because the \oned{} mass PS is an 
integral of the \thrd{} mass PS. The difference should be a constant on 
scales that baryons have exactly the same \thrd{} PS as dark matter. 
In redshift space, the \oned{} mass PS is
\begin{equation} \label{eq:p1dz}
P^S_{\rm 1D}(k_3) = \frac{1}{4\pi^2}\int_0^\infty P^S({\bf k}_\perp, k_3) 
\,{\rm d} {\bf k}_\perp.
\end{equation}
The anisotropy of the redshift-space \thrd{} mass PS has made redshift 
distortion less intuitive for the \oned{} mass PS.

\subsubsection{Equation of State} \label{sec:eos}

\begin{figure}
\centering
\includegraphics[width=5in]{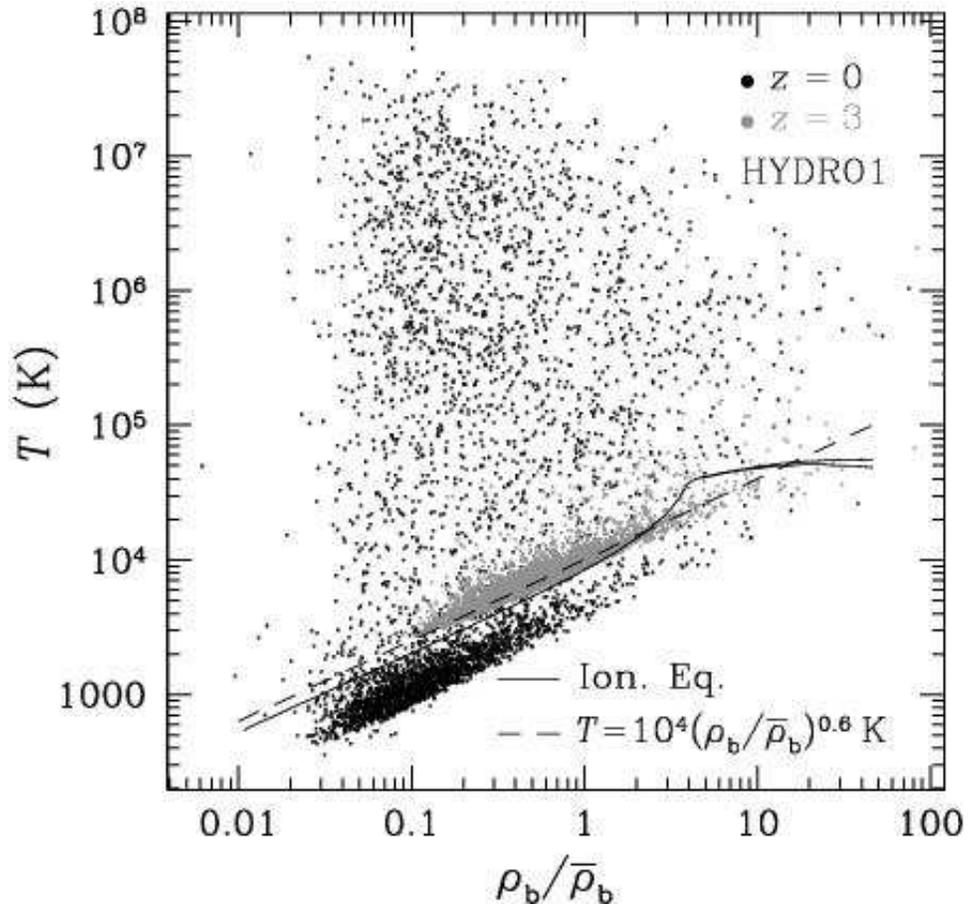}
\caption[Temperature--density relation of baryons in the simulation]
{The temperature--density relation of baryons. Grey dots are for $z=3$, and 
black dots $z = 0$. Each group consists of 4000 pairs of temperature 
and density values randomly selected from the density grid of 
HYDRO1. The dashed line represents an EOS 
$T=10^4(\rho_{\rm b}/\bar{\rho}_{\rm b})^{0.6}$ K. The solid lines are
calculated from ionization equilibrium with a factor of 2.5 difference in 
the UV background intensity.
\label{fig:eos}}
\end{figure}

Fig.~\ref{fig:eos} demonstrates the 
correlation between temperature and baryon density for the simulation 
HYDRO1. 
The temperature--density relation is fairly tight at $z = 3$, and it 
follows the EOS $T = T_0(\rho_{\rm b}/\bar{\rho}_{\rm b})^{0.6}$ closely. 
Two theoretical temperature--density curves for ionization equilibrium at 
$z = 3$ are also provided in Fig.~\ref{fig:eos}. The UV background
intensities of the two curves are adjusted to fit mean fluxes of 
$0.77$ and $0.64$. Although the UV intensities differ by 
a factor of 2.5, there is no visible difference between the two curves 
at $\rho_{\rm b}/\bar{\rho}_{\rm b} < 5$. Compared to the
simulation and the curves of ionization equilibrium, the simple EOS is 
indeed a good approximation.

At $z=0$, the IGM develops multiple phases; the bulk of the IGM has 
become cooler, while some gases are shock-heated to temperatures up 
to a few times $10^7$ K---the WHIM. 
These WHIM gases are generally too
hot and too dilute to produce appreciable \lya{} absorptions. They
explain why many high-temperature regions at $z = 0$ (see the first two 
panels of Fig.~\ref{fig:tb0}) do not correspond to any absorptions at all, 
even though high-temperature regions would be naively thought of as 
high-density regions from the simple EOS. 

Note that the temperatures and densities in Fig.~\ref{fig:eos} are 
randomly selected from the density grid. Since smoothing lengths 
of SPH particles are required to be at least the size of the grid 
spacing, one will not get a high-density tail that is often seen in 
a particle temperature--particle density plot.
The sharp edges of temperature--density distributions are due to the 
internal UV background of the simulation, which keeps SPH particles
from cooling below the ionization-equilibrium temperature. 

\clearpage
\subsubsection{Variants of the Pseudo-Hydro Technique}
\begin{table}
\centering
\caption[Methods for generating the \lya{} forest]
{Methods for generating the \lya{} forest.}
\label{tab:psh}
\begin{tabular}{ccccc}
\hline
Method & Particle & $\rho_{\rm b}$ & $T_{\rm th}^{\ \ a}$ & $\tau$ \\ 
\hline
BI & SPH &  & Max$[T_{\rm sph}, T_{\rm ie}(\rho_{\rm b})]$  & 
	Ion. Eq. \\
BE & SPH &  & $T_0 (\rho_{\rm b}/\bar{\rho}_{\rm b})^\alpha$ & 
	$\propto (\rho_{\rm b}/\bar{\rho}_{\rm b})^\gamma$ \\
DI & CDM & $\propto \rho_{\rm d}$ & $T_{\rm ie}(\rho_{\rm b})$  & Ion. Eq. \\
DE & CDM & $\propto \rho_{\rm d}$ & 
	$T_0 (\rho_{\rm b}/\bar{\rho}_{\rm b})^\alpha$ & 
	$\propto (\rho_{\rm b}/\bar{\rho}_{\rm b})^\gamma$ \\
\hline
\end{tabular}

%\medskip
\parbox{0.7 \textwidth}{
\begin{itemize}
\setlength{\leftskip}{-0.2in}
\setlength{\itemsep}{0.0ex}
\item[$^a$] $T_{\rm th}$ is the temperature used for thermal broadening,
$T_{\rm sph}$ is the grid temperature based on temperatures of SPH particles 
in hydrodynamical simulations, and $T_{\rm ie}(\rho_{\rm b})$ is the 
ionization-equilibrium temperature of a gas with density $\rho_{\rm b}$.
\end{itemize}}
\end{table}

Separately, Petitjean et al.~(1995) %\citet{pmk95} 
developed a slightly different pseudo-hydro 
technique. They assume baryons to trace dark matter but do not use the 
simple EOS. Temperatures and optical depths of baryons are calculated from 
ionization equilibrium in their method (labeled as DI). 
As seen in Fig.~\ref{fig:eos}, the ionization-equilibrium temperature is 
usually lower than the density-weighted SPH temperature. As such, 
small-scale fluctuations may not be sufficiently 
filtered by thermal broadening in this method.

One may devise another variant of the pseudo-hydro technique by applying
the simple EOS and equation (\ref{eq:lyaf}) to baryons in hydrodynamical 
simulations. I refer to this method as BE and that of \citet{cwk98} DE.
Similarly, the name BI is given to the full-hydrodynamical approach 
described in Section \ref{sec:hyd-lya}. One can assess the importance of 
shocks by comparing methods BE with BI, while the difference between methods 
BE and DE must arise from the difference between baryon distributions and 
dark matter distributions. The four methods are summarized in Table 
\ref{tab:psh}.

\begin{figure}
\centering
\includegraphics[width=5.5in]{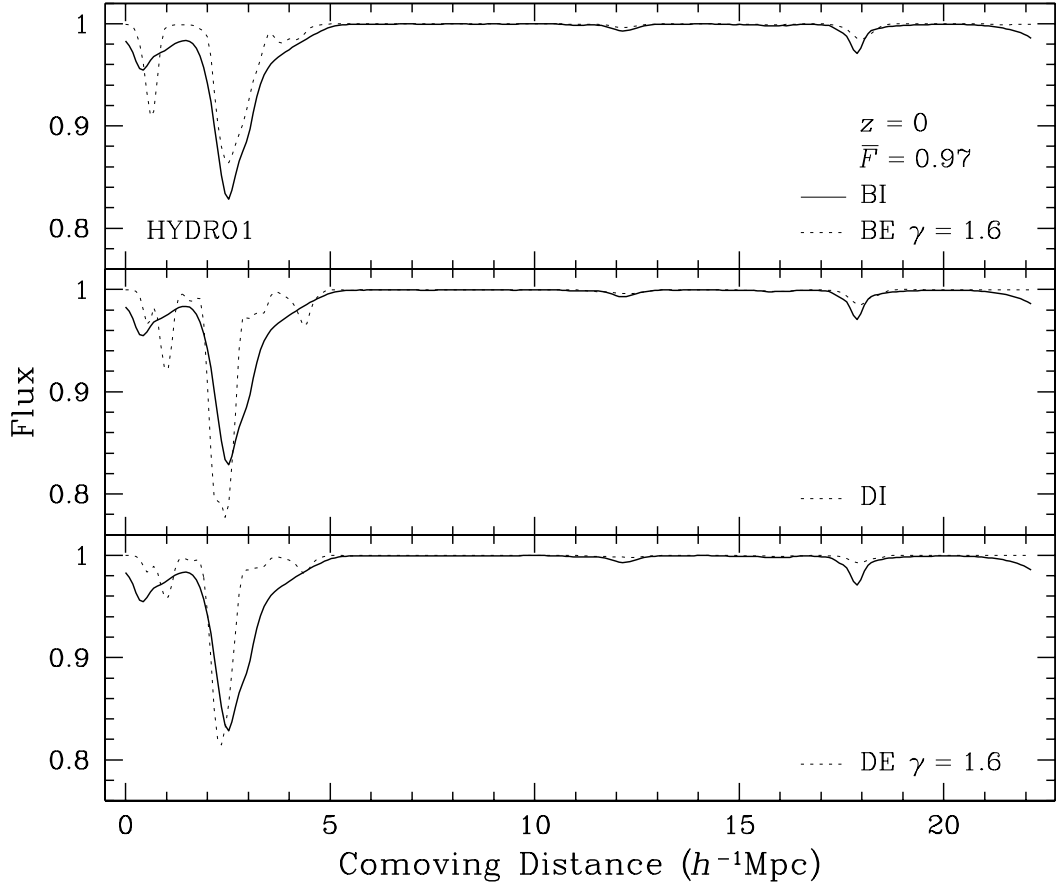}
\caption[Transmitted \lya{} flux based on baryon and dark matter 
distributions at $z = 0$]
{Transmitted \lya{} flux based on baryon and dark matter distributions 
at $z = 0$. The three panels compare the same \lya{} forest generated 
from baryons in ionization equilibrium (BI) to those from baryons with 
an EOS (BE), dark-matter-converted baryons in ionization equilibrium 
(DI), and dark-matter-converted baryons with an EOS (DE). All the four 
methods are required to reproduce the same mean flux of $0.97$.
\label{fig:fdmba0}}
\end{figure}

\begin{figure}
\centering
\includegraphics[width=5.5in]{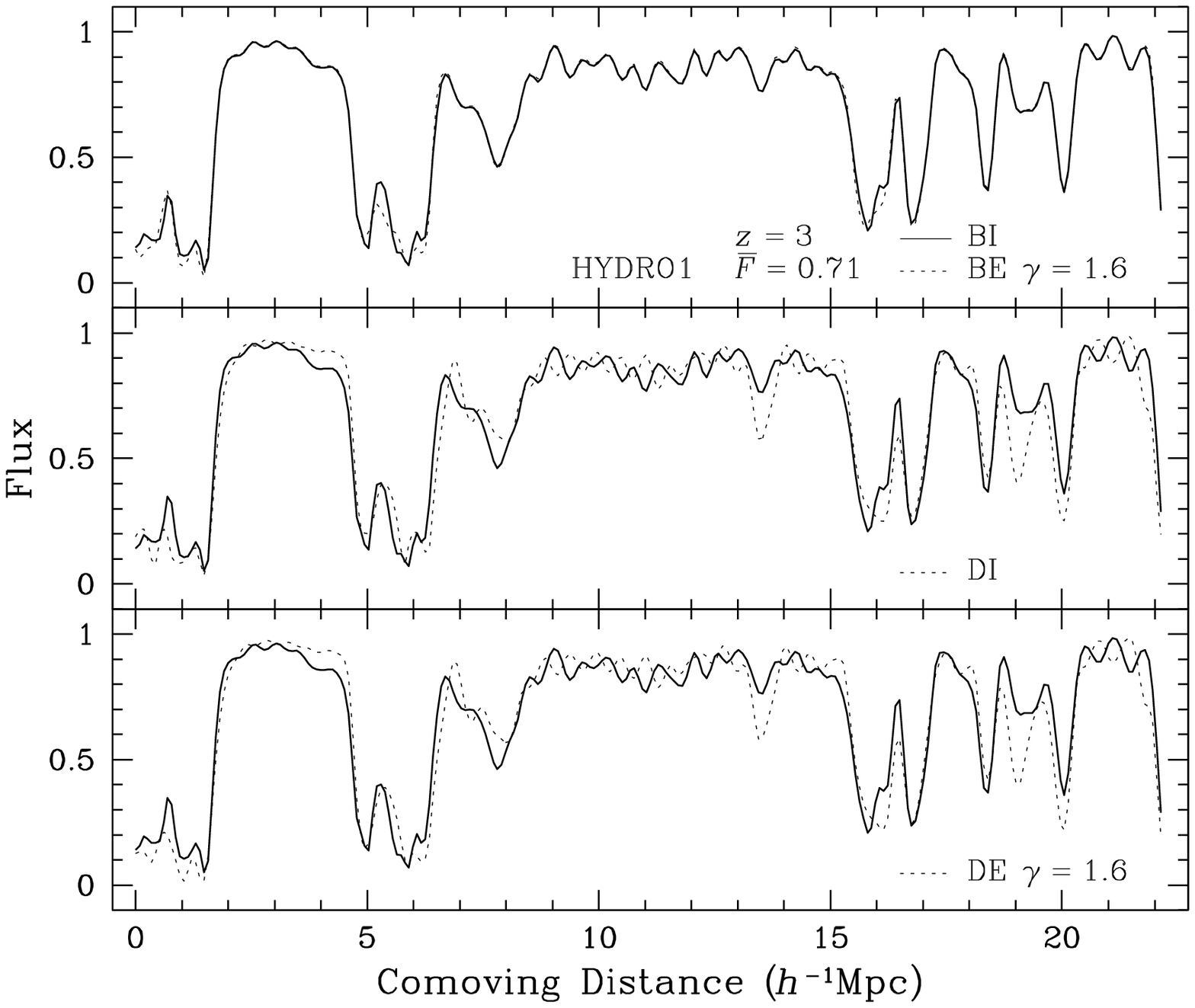}
\caption[Transmitted \lya{} flux based on baryon and dark matter 
distributions at $z = 3$]{The same as Fig.~\ref{fig:fdmba0}, except 
that \lya{} forests are generated from baryon and dark matter 
distributions at $z = 3$ and the mean flux is $0.71$.
\label{fig:fdmba3}}
\end{figure}

\subsection{Comparison}
To give a visual impression of pseudo-hydro techniques, 
I show in Figs.~\ref{fig:fdmba0} and \ref{fig:fdmba3} \lya{} forests 
obtained along the same LOS using the four methods,
BI, BE, DI, and DE. The mean flux over all $256^2$ \lya{} forests
by the four methods are required to match the observed mean flux of 
$0.97$ at $z = 0$ and $0.71$ at $z = 3$, but the mean flux of a single 
LOS is not necessarily the same across the methods. Since a simple EOS 
does not take into account the substantial amount of WHIM at $z = 0$
(see Fig.~\ref{fig:eos}), pseudo-hydro techniques are expected to be less
accurate at lower redshift. This is seen in Fig.~\ref{fig:fdmba0}. On the
other hand, at $z = 3$, methods involving ionization equilibrium and the EOS 
generate almost identical \lya{} forests, and the difference between \lya{} 
forests generated from baryons and those from dark matter is also small.

\begin{figure}
\centering
\includegraphics[width=5.8in]{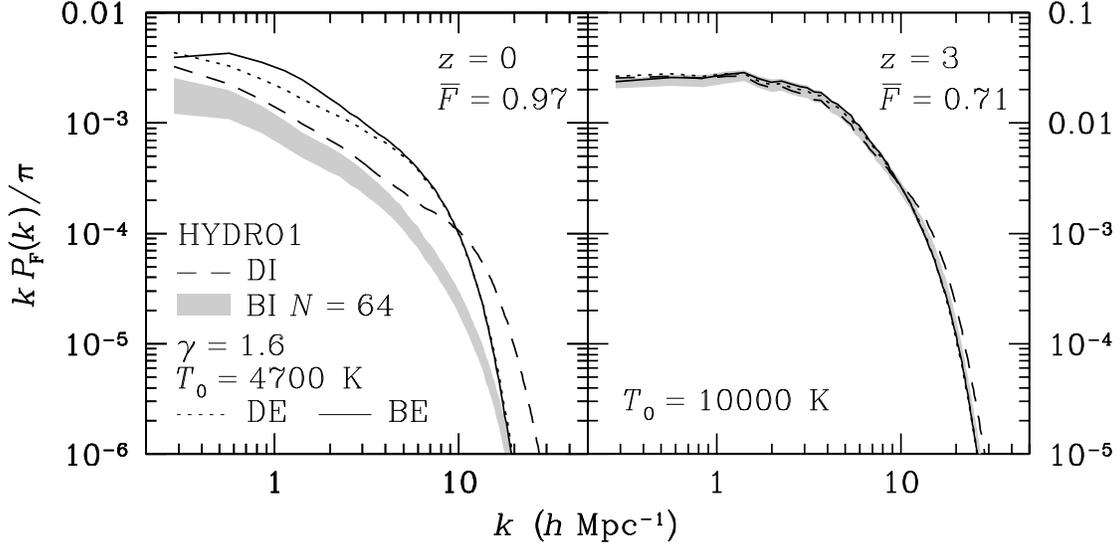}
\caption[Flux power spectra of \lya{} forests generated from baryon and 
dark matter distributions at $z = 0$ and $z = 3$]{
Flux PS's of \lya{} forests at $z = 0$ and $z = 3$. The \lya{} forests
are produced using the four methods listed in Table \ref{tab:psh}. Grey 
bands represent $1\sigma$ dispersions of flux PS's of \lya{} forests 
generated from baryons in ionization equilibrium. The dispersions are
calculated among 1000 groups, each of which consists of 64 LOS's.
\label{fig:fpshionT}}
\end{figure}

\clearpage
Fig.~\ref{fig:fpshionT} evaluates the statistical performance of 
pseudo-hydro techniques using flux PS's. Grey bands are $1\sigma$ 
dispersions of flux PS's of \lya{} forests produced using the method BI. 
The dispersions are calculated among 1000 groups, each of which contains 
64 LOS's randomly selected with no repetition. 
There is a good agreement among all methods at $z = 3$, but all the 
three pseudo-hydro methods, BE, DE, and DI fail to converge on BI at 
$z = 0$. The method DI seems to work better than the other two on scales 
$k < 5$ \mpci{} at $z = 0$. The excess of flux power for
pseudo-hydro techniques is expected because they all underestimate the 
IGM temperature and produce more flux fluctuations than the full-hydro 
method (see Figs.~\ref{fig:eos} and \ref{fig:fdmba0}). 
In fact, methods BI and BE have 
identical baryon distributions, so that the difference in their flux 
PS's can only be attributed to the IGM temperature. Hence, one  
concludes that the temperature structure of the IGM is critical to the 
\lya{} forest and flux PS at low redshift.

\begin{figure}
\centering
\includegraphics[width=5.8in]{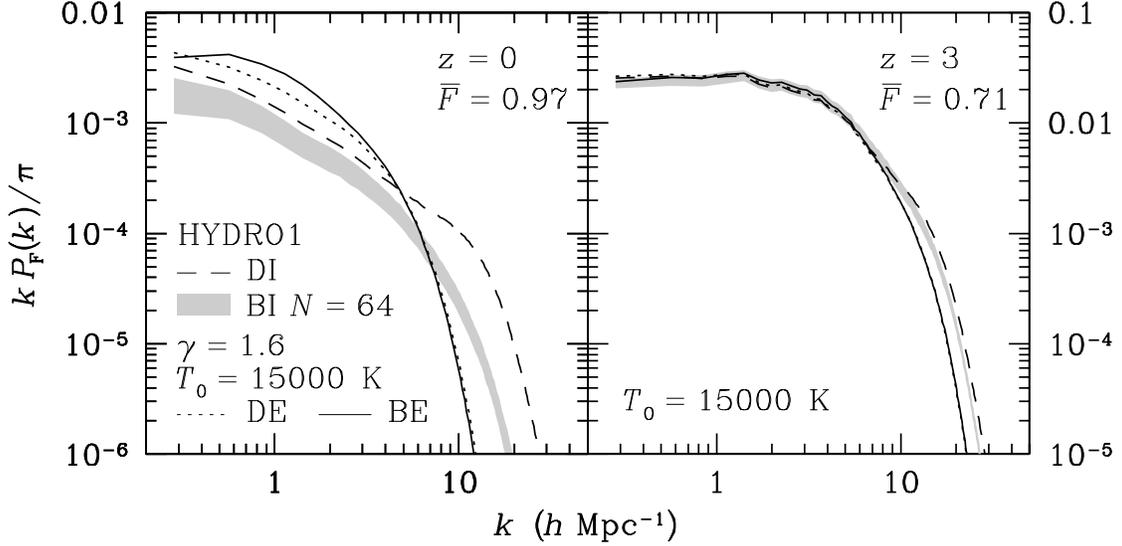}
\caption[Effect of thermal broadening on flux power spectra] {
The same as Fig.~\ref{fig:fpshionT}, except that for methods BE and DE
\lya{} forests are generated with $T_0 = 15000$ K. Methods BI and DI 
are not affected by $T_0$.
\label{fig:fpshion}}
\end{figure}

The mean-density temperature of the IGM $T_0$ does not affect the 
optical depth in methods BE and DE because it is absorbed into the 
parameter $A$ in the approximation 
$F= \exp\left[-A (\rho/\bar{\rho})^\gamma\right]$, which 
is adjusted to fit the observed mean flux. However, $T_0$ can affect 
simulated \lya{} forests through thermal broadening as indicated by
the fast drop of the flux PS's for methods BE and DE. To test this, 
I reproduce \lya{} forests using $T_0 = 15000$ K, which is roughly
three times (1.5 times) the mean-density temperature of the IGM in HYDRO1 
at $z = 0$ ($z = 3$). Flux PS's of these \lya{} forests are shown in 
Fig.~\ref{fig:fpshion}. One sees that the higher
mean-density temperature reduces more flux power on small scales while 
leaving flux PS's unchanged on large scales. As such, the flux PS is 
less robust on small scales as a constraint for cosmology. 
The small-scale excess of
flux power for the method DI with respect to BI is also due to
thermal broadening because the ionization-equilibrium temperature is 
almost always lower than the density-weighted temperature of SPH 
particles. 

\clearpage

\uatochack
\section[Tuning the \lya{} Forest]{Tuning the 
Ly$\boldsymbol{\alpha}$ Forest}
It is already seen in last section that pseudo-hydro techniques do not
work well at low redshift and their performances are not all equal. 
Simulated \lya{} forests are affected by many elements including, for 
instance, the EOS (for methods BE and DE) and the mean flux. If they are 
to be compared with the statistics of observed \lya{} forests and to be
used to determine cosmological parameters, one must understand the 
dependence of the statistics of simulated \lya{} forests on the 
above-mentioned elements.

\subsection{Equation of State}
The EOS maps density fluctuations to flux fluctuations by relating optical 
depths to densities. For a given density and mean flux, different EOS's will 
assign different optical depths, which alters amplitudes of flux 
fluctuations and, therefore, the flux PS. 

\begin{figure}
\centering
\includegraphics[width=5.5in]{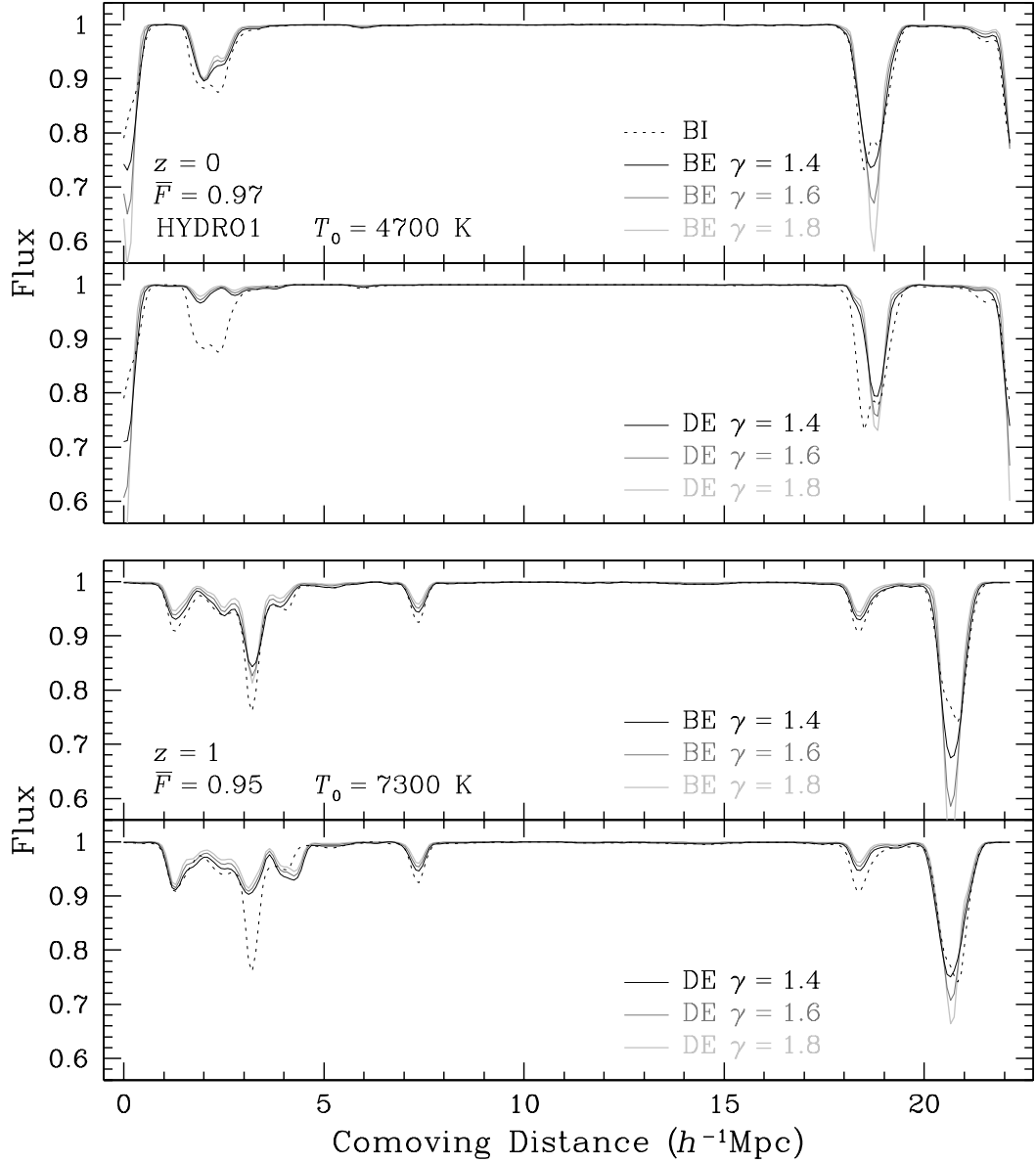}
\caption[\lya{} Forests generated from baryon and dark matter 
distributions with different equations of state at $z = 0$ and $1$]{
\lya{} Forests generated from baryon and dark matter distributions with 
different EOS's. The upper two panels are for $z = 0$, and 
the lower two $z = 1$. The \lya{} forests are produced from baryon 
densities and dark matter densities using 
$F = \exp[-A(\rho/\bar{\rho})^\gamma]$, where $A$ is adjusted to fit
the mean flux $\bar{F}$. Thermal broadening is included with temperature
given by $T = T_0(\rho/\bar{\rho})^\alpha$.
\label{fig:feos01}}
\end{figure}

\begin{figure}
\centering
\includegraphics[width=5.5in]{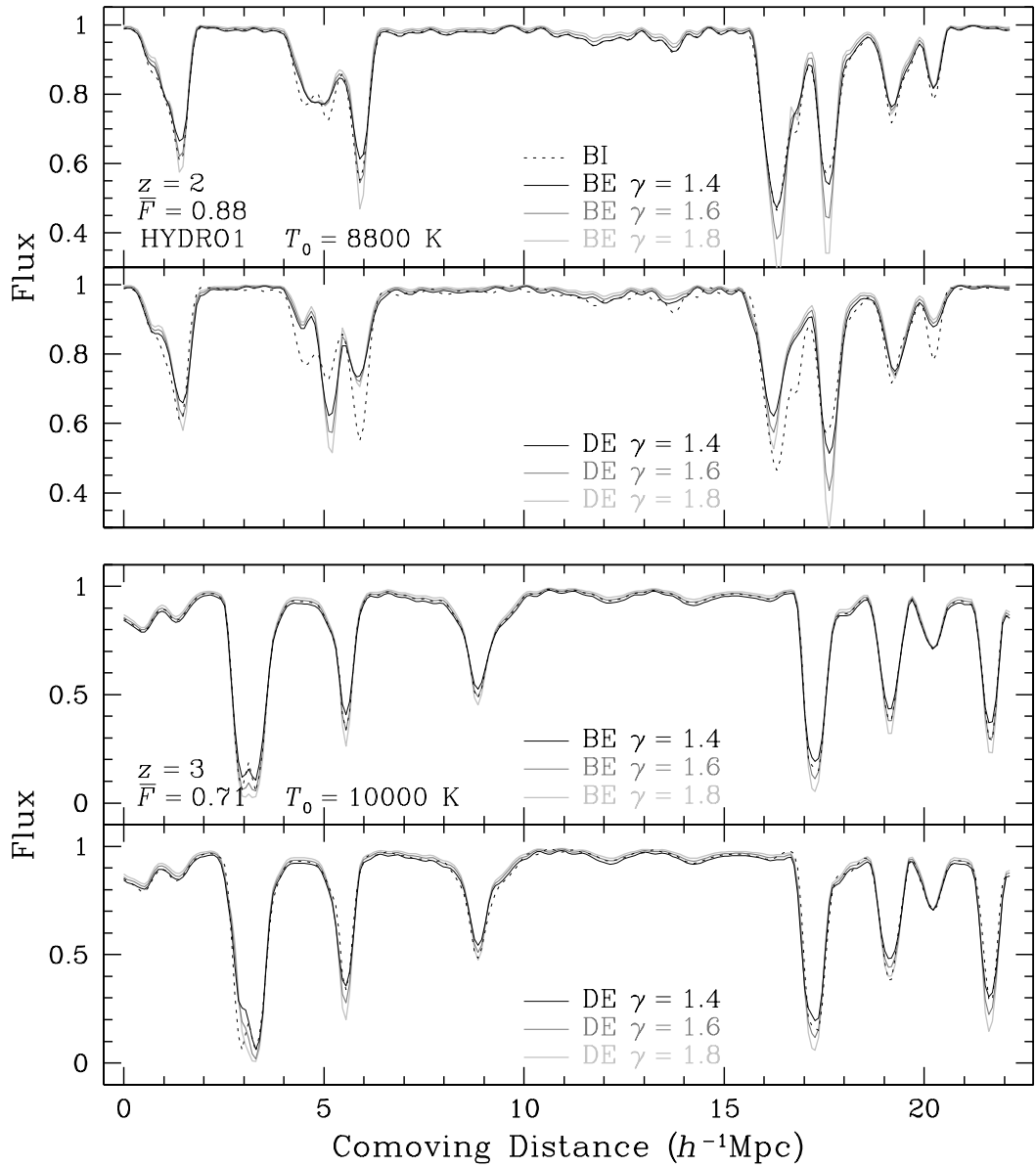}
\caption[\lya{} Forests generated from baryon and dark matter 
distributions with different equations of state at $z = 2$ and $3$]{
The same as Fig.~\ref{fig:feos01}, except that the upper two panels 
are for $z = 2$, and the lower two $z = 3$. 
\label{fig:feos23}}
\end{figure}

For a stiff EOS, i.e.~a small value of $\gamma$, high-density regions 
have to absorb less \lya{} flux, while, in compensation, low-density 
regions have to absorb more. In terms of flux, a stiff EOS leads to higher 
fluxes in deep (or large-equivalent-width) absorptions and lower fluxes in 
shallow absorptions than a soft EOS does. This expectation is confirmed in 
Figs.~\ref{fig:feos01} and \ref{fig:feos23}, where \lya{} forests generated
using methods BE and DE are compared with those using the method BI at 
$z = 0$, 1, 2, and 3. The mean flux is kept the same for all the methods 
used here at each epoch, while only the EOS is varied. The value of 
$\gamma = 1.4$ in the figures corresponds to an unrealistically 
stiff EOS, and it is provided only for the purpose of comparison.

Figs.~\ref{fig:feos01} and \ref{fig:feos23} show that low-amplitude and 
small-scale fluctuations in the flux are likely to be suppressed for 
methods BE and DE, while high-amplitude fluctuations are likely to be 
amplified. Overall, this boosts the flux PS on large scales with respect 
to that of the method BI as observed in Fig.~\ref{fig:fpshionT}. 
Although methods BE and DE are not good approximations at low redshift, 
they work remarkably well with $\gamma = 1.6$ at $z =3$. 

\ualofhack

\begin{figure}
\centering
\includegraphics[width=5.5in]{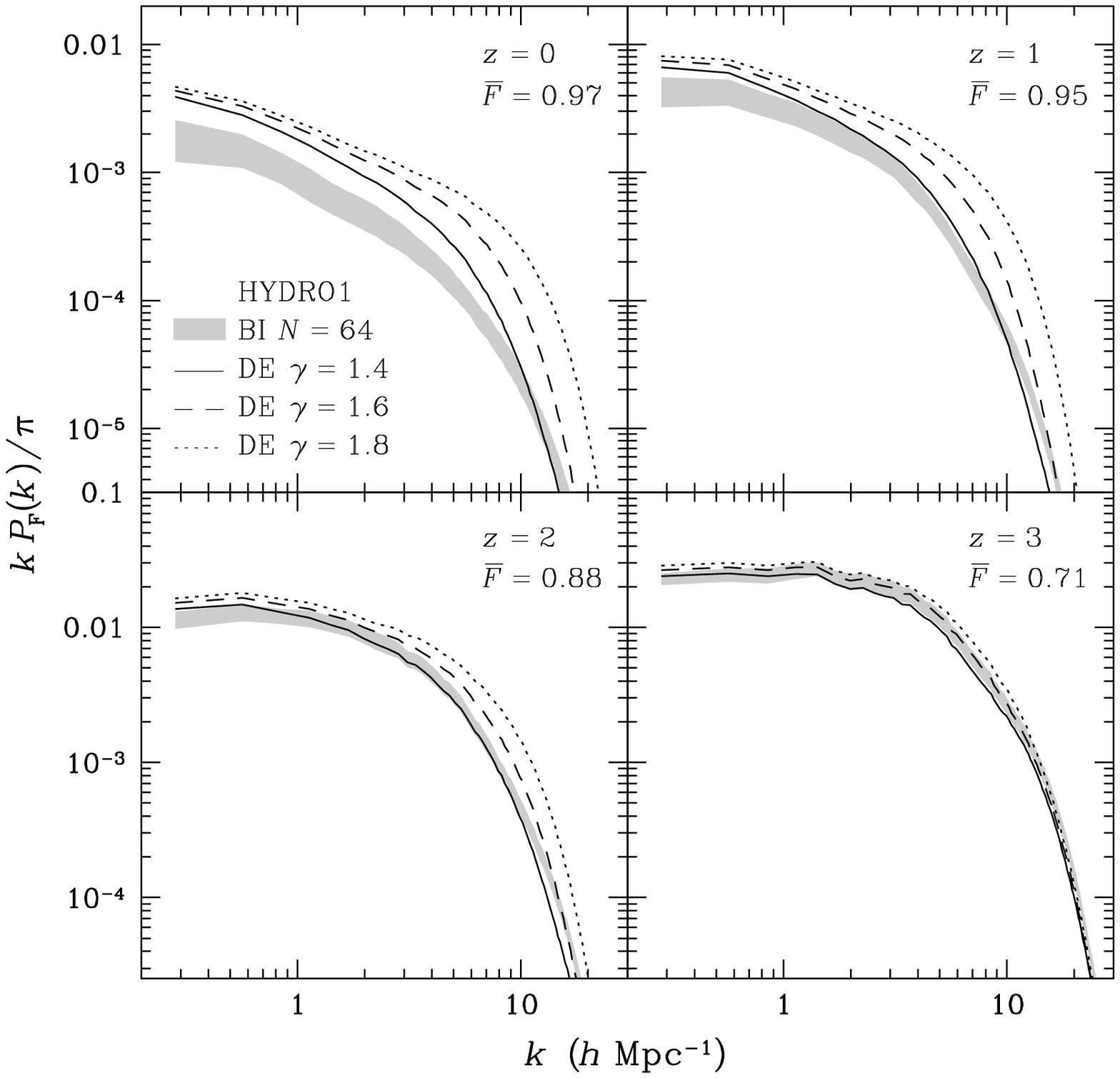}
\caption[Flux power spectra of \lya{} forests generated from 
dark matter densities using different equations of state]{
Flux PS's of \lya{} forests generated from dark-matter-converted 
baryon densities using different EOS's. Grey bands represent 
$1\sigma$ dispersions of the flux PS's of \lya{} forests generated from 
baryons in ionization equilibrium. The dispersions are calculated in the 
same way as in Fig.~\ref{fig:fpshionT}.
\label{fig:fpseosdm}}
\end{figure}

Since the amplitude of flux fluctuations increases with $\gamma$ in 
Figs.~\ref{fig:feos01} and \ref{fig:feos23}, a smaller value of $\gamma$
must lead to a lower flux PS. This is observed in Fig.~\ref{fig:fpseosdm}, 
where flux PS's of \lya{} forests obtained using the method DE with different
EOS's are compared with those using the method BI. Methods DE and BE produce 
very similar Flux PS's, so flux PS's of method BE are not shown separately.
Fig.~\ref{fig:fpseosdm} demonstrates that one cannot tune the EOS to make 
pseudo-hydro techniques work at low redshift. Meanwhile, pseudo-hydro 
techniques appear to be a good approximation for studies of the flux PS 
at $z = 3$. 
The difference among different EOS's is also less pronounced at $z = 3$, 
because the dynamical range of $\rho / \bar{\rho}$ is much smaller.

\subsection{Mean Flux}
The mean flux affects the \lya{} forest and the flux PS in a simple way. 
Low-density regions of the IGM cannot absorb much 
\lya{} flux no matter what mean flux is required. Thus, the mean flux
mostly affects regions where absorption is already significant. Specifically,
a higher mean flux weakens existing absorptions and decreases the flux 
PS over all scales. The \lya{} forest should be more sensitive to the mean
flux at lower redshift when \lya{} absorptions often arise from denser 
regions. 

\begin{figure}
\centering
\includegraphics[width=5.5in]{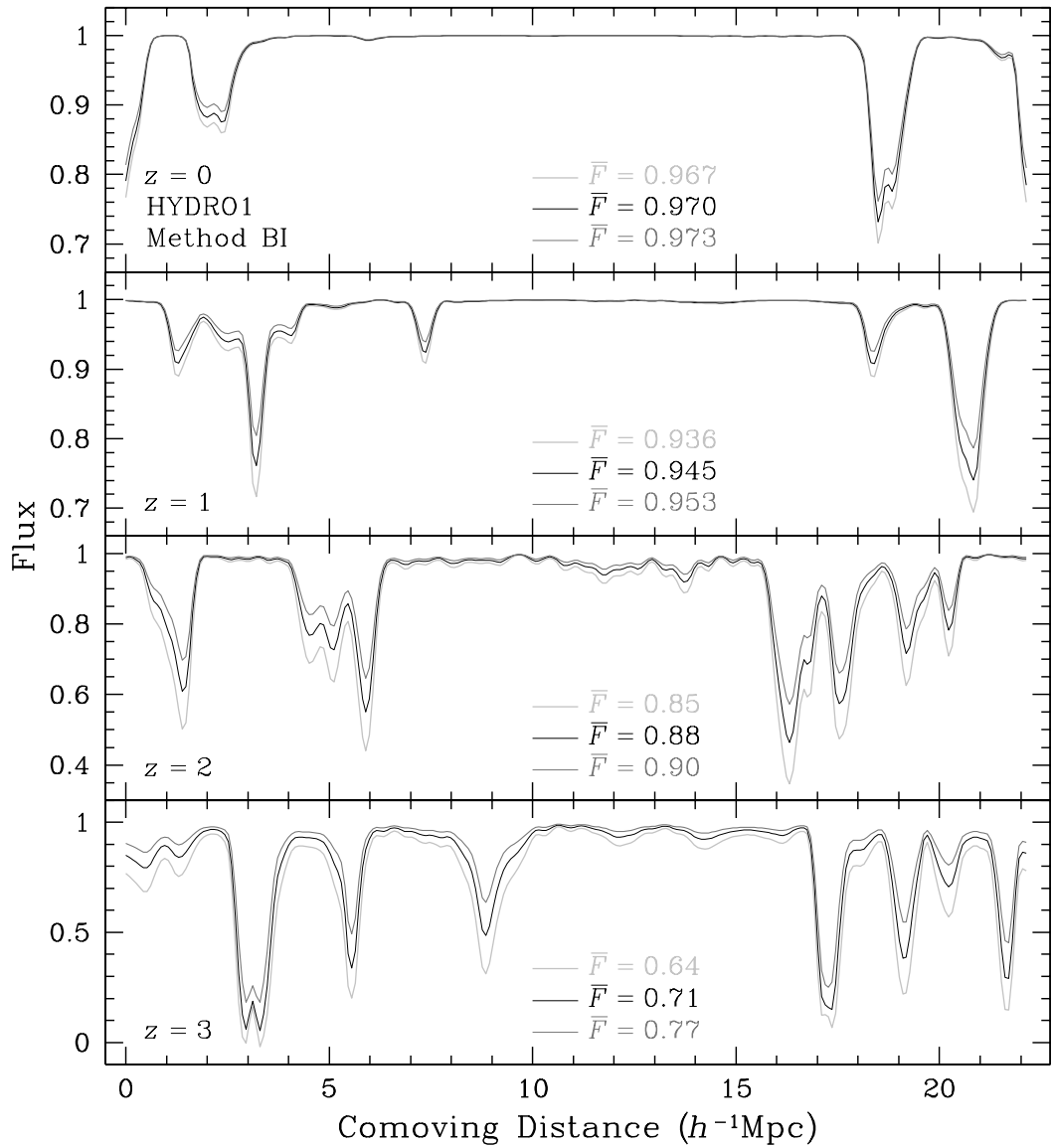}
\caption[Simulated \lya{} forests with different mean fluxes]{
Simulated \lya{} forests with mean fluxes from equation 
(\ref{eq:mflx}). The \lya{} forests are generated from baryons in 
ionization equilibrium. Note that the difference in mean fluxes increases 
with redshift.
\label{fig:fmflx}}
\end{figure}

Fig.~\ref{fig:fmflx} compares \lya{} forests obtained along the same LOS 
but with different mean fluxes given by observational and simulational 
uncertainties [see equation (\ref{eq:mflx})]. Since all the four methods
in Table \ref{tab:psh} have the same dependence on the mean flux, I only 
show flux PS's of \lya{} forests generated using the full-hydro method.
As expected, the mean flux
monotonically alters \lya{} forests in all regions with greater changes in 
deeper absorptions, and a lower mean flux gives rise to stronger 
fluctuations in the flux. If one takes into account that the difference in 
the mean fluxes at $z = 0$ is about 20 times smaller than that at $z = 3$, 
the low-redshift \lya{} forest does seem to be more sensitive to the 
mean flux.

\begin{figure}
\centering
\includegraphics[width=5.5in]{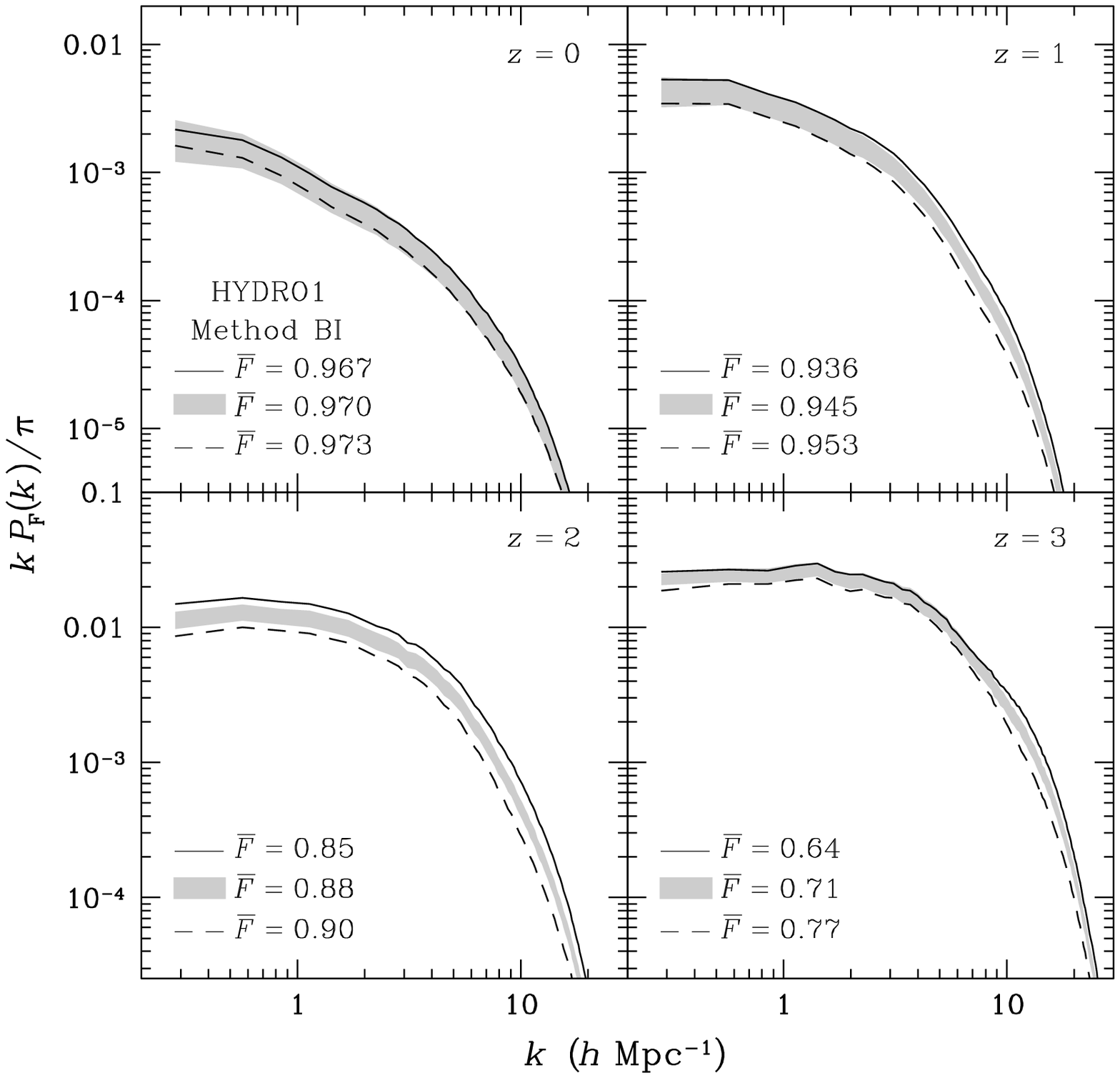}
\caption[Flux power spectra of simulated \lya{} forests with different
mean fluxes]{
Flux PS's of simulated \lya{} forests with mean fluxes from 
equation (\ref{eq:mflx}). The \lya{} forests are generated from baryons in 
ionization equilibrium. The grey band in each panel is the $1\sigma$ 
dispersion of the flux PS with the medium mean flux. The dispersions are 
calculated among 1000 groups, each of which consists of 64 LOS's.
\label{fig:fpsmflx}}
\end{figure}

Fig.~\ref{fig:fpsmflx} shows that flux PS's are also monotonically 
altered by the mean flux. A lower mean flux (more absorptions) leads to 
a higher flux PS on all scales. 
Unlike the EOS, the mean flux does not change the shape of the flux PS much. 
This implies that the mean flux can uniquely determine the normalization of 
the flux PS. In other words, the mean flux is a relatively robust 
constraint on simulated \lya{} forests.

\clearpage

\section{Mass Statistics vs. Flux Statistics}
\subsection{Power Spectrum}
The \lya{} forest has been used to infer the linear mass PS of the cosmic
density field. The nonlinear transform of the density to the flux has made
it difficult to derive the mass PS from the flux PS analytically. One way
to circumvent the difficulty is to use simulations to map the flux PS to 
the mass PS \citep{cwb02}. As such, it is important to compare the 
flux PS with the mass PS. 

\begin{figure}
\centering
\includegraphics[width=5.7in]{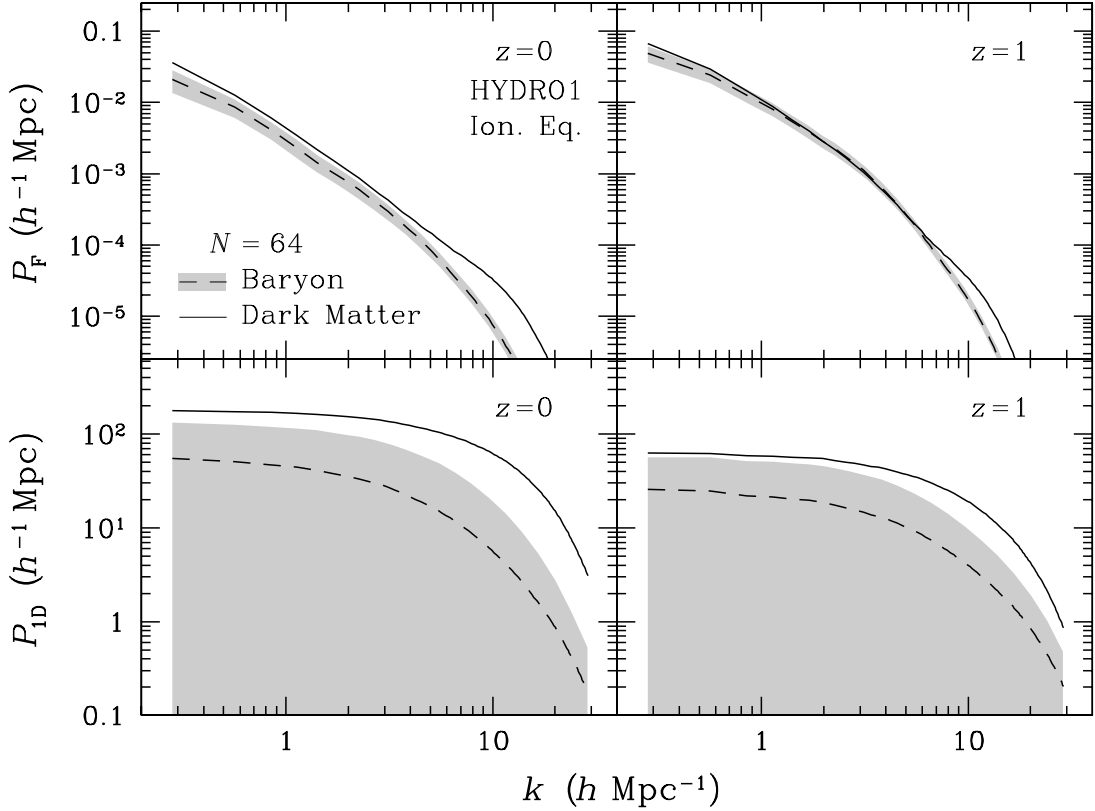}
\caption[Flux power spectra of simulated \lya{} forests and \oned{} mass 
power spectra of the underlying density fields at $z = 0$ and $1$]{
Flux PS's of simulated \lya{} forests and \oned{} mass PS's 
of the underlying density fields at $z = 0$ and $1$.
\label{fig:mfps01}}
\end{figure}

\begin{figure}
\centering
\includegraphics[width=5.7in]{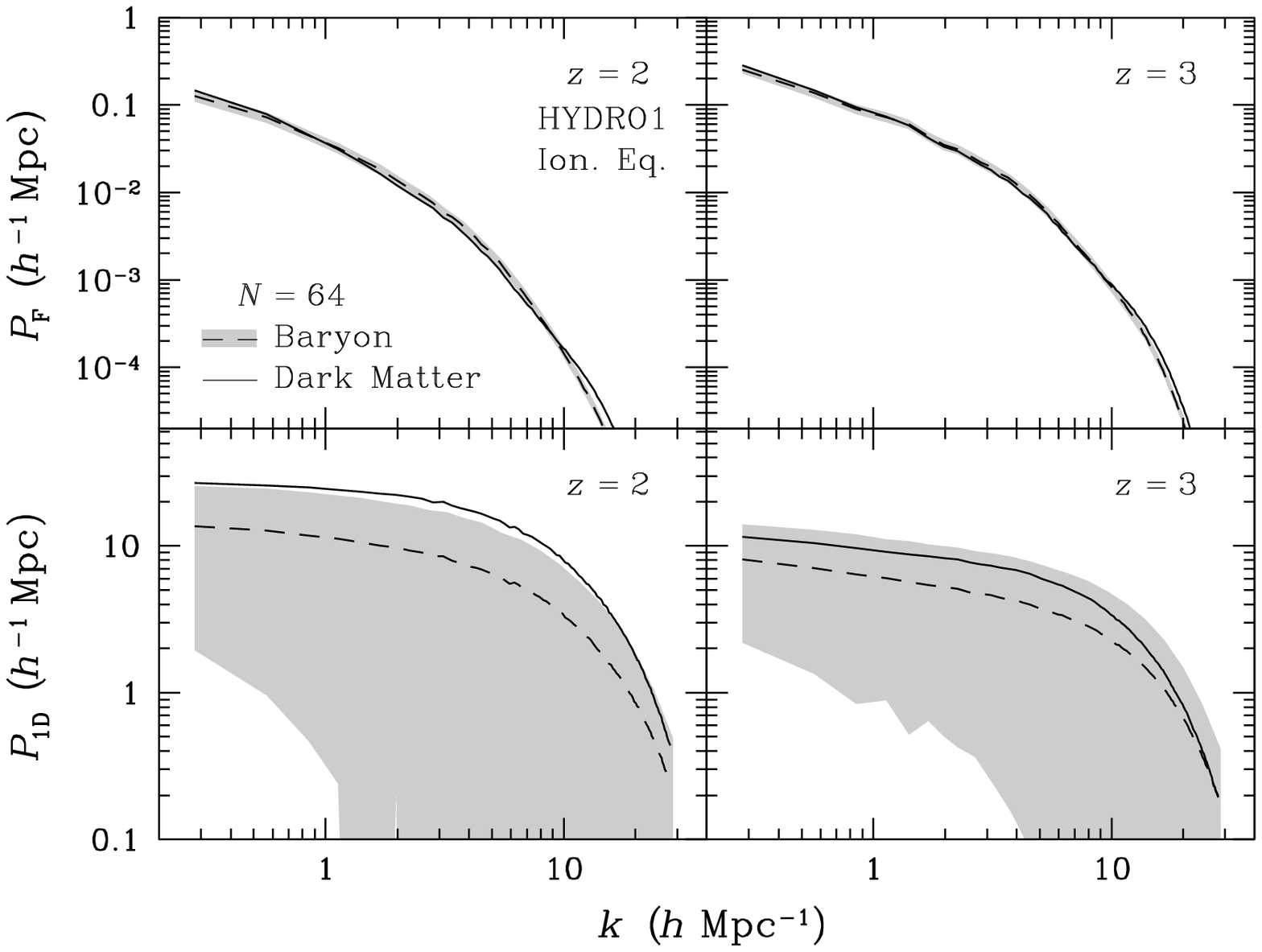}
\caption[Flux power spectra of simulated \lya{} forests and \oned{} mass 
power spectra of the underlying density fields at $z = 2$ and $3$]{
Flux PS's of simulated \lya{} forests and \oned{} mass PS's 
of the underlying density fields at $z = 2$ and $3$.
\label{fig:mfps23}}
\end{figure}

Plotted in Figs.~\ref{fig:mfps01} and \ref{fig:mfps23} are flux PS's 
of \lya{} forests produced from baryons and dark-matter-converted 
baryons in ionization 
equilibrium along with \oned{} mass PS's of baryons and dark matter. 
One $\sigma$ dispersions of baryon flux PS's and mass PS's are shown in 
grey. The most prominent feature
is that \oned{} mass PS's have much larger dispersions than flux PS's. As
discussed in Chapter \ref{ch:mps}, the variance in the \oned{} mass PS is 
severely inflated by the trispectrum of the cosmic density field because 
of the non-Gaussianity. 

An interesting observation is that unlike the mass PS the flux PS decreases
with time. This is due to the thinning of the \lya{} forest and the higher 
mean flux toward lower redshift that reduce fluctuations in the \lya{} flux.

The nonlinear transform between baryon density and 
flux greatly suppresses the fluctuations. For example, the overdensity 
$\delta$ may vary from -1 to hundreds (tens) 
at $z = 0$ ($z = 3$), but the flux can only be between 0 and 1. 
With a mean flux on the order of unity, fluctuations in the flux are 
$10$--$10^2$ times weaker than those in the cosmic density field. Hence, 
the flux PS is a factor of $10^2$ ($z = 3$) to $10^4$ ($z = 0$) times lower 
than the \oned{} mass PS. Moreover, the non-Gaussianity in the cosmic density 
field is even more suppressed in the flux because it is a higher-order effect. 
Thus, the flux trispectrum is much closer to zero as compared to the mass 
trispectrum of the cosmic density field, and the variance of the flux PS 
becomes much smaller than the variance of the \oned{} mass PS.

The near-Gaussian \lya{} flux is probably the reason that many 
simulations and techniques are able to reproduce lower-order statistics 
of the observed \lya{} forest, especially at high redshift. A potential 
problem arises because of Figs.~\ref{fig:mfps01} and \ref{fig:mfps23}. 
That is one could produce \lya{} forests from wildly different density 
fields but still have almost identical flux PS. For instance, even though 
baryons and dark matter differ considerably in terms of mass PS (see
also Fig.~\ref{fig:psdmba}), they are not so distinguishable from each 
other in flux PS's at $z \geq 2$. Conversely, we are able to measure 
the flux PS extremely well, but how much confidence do we have in 
recovering the underlying mass PS?

\subsection{Covariance}
Higher-order statistics 
%such as wavelet scale-scale correlation \citep{fangfeng00} 
may be able to break the degeneracy. Here I employ the
covariance of the \oned{} PS to explore the difference between \lya{} 
forests generated using the full-hydro method BI and those using the 
pseudo-hydro method DI.

\begin{figure}
\centering
\includegraphics[width=5.5in]{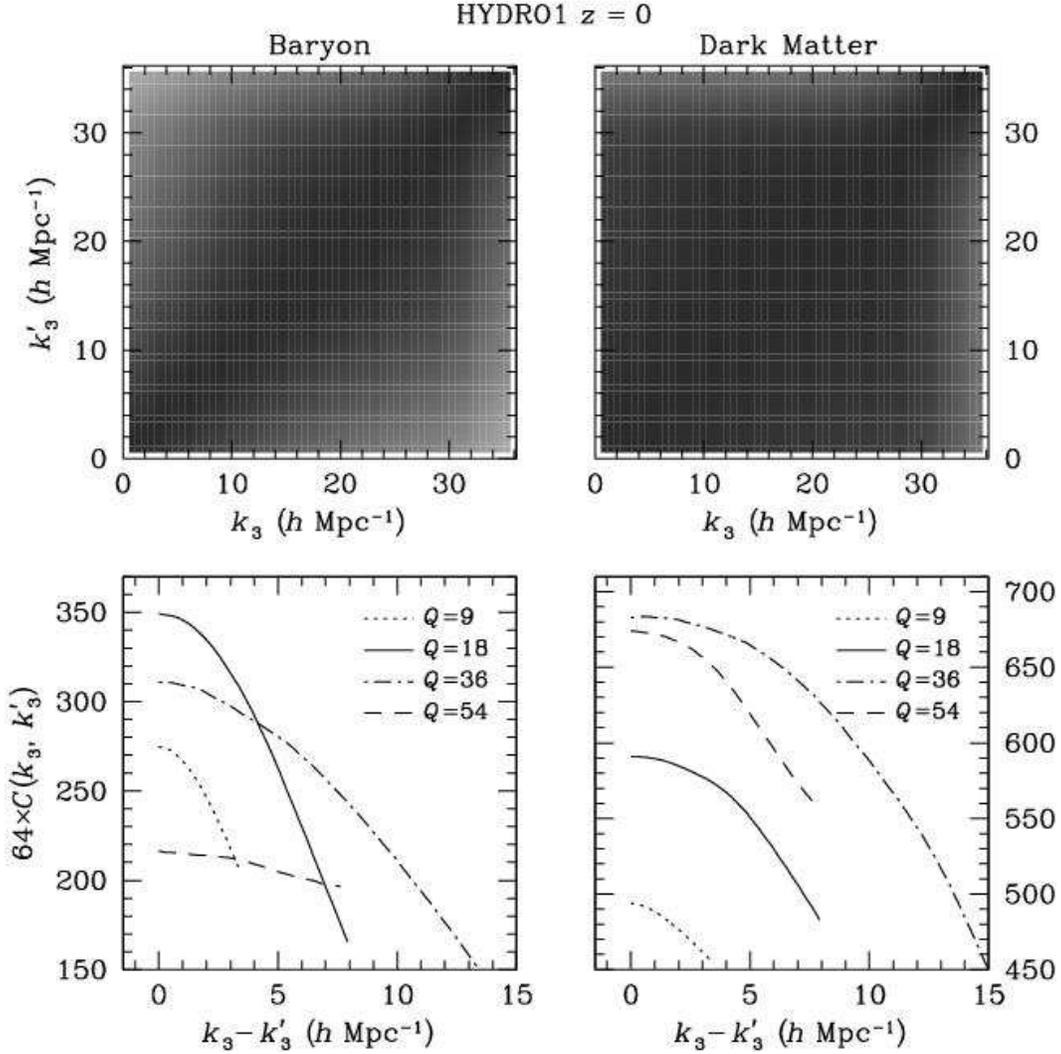}
\caption[Normalized covariances of \oned{} baryon and dark matter power 
spectra at $z = 0$]{
Normalized covariances of \oned{} baryon and dark matter PS's
at $z = 0$. The upper panels are covariances
$\hat{C}(k_3, k'_3)$ in the same grey scale as Fig.~\ref{fig:cov64}. 
The lower panels are cross sections of covariances
$C(k_3, k'_3)$ along $Q = (k_3 + k'_3) / \mpc$. The covariances 
$C(k_3, k'_3)$ are multiplied by 64 for better comparison with that of
GRFs, which follows $64\,C(k_3, k'_3) = \delta^{\rm K}_{n_3, n'_3}$.
All the covariances are calculated from 1000 groups, each 
of which consists of 64 LOS's ($N=64$) randomly selected from the density 
grid of HYDRO1. 
\label{fig:cv0db64}}
\end{figure}

\begin{figure}
\centering
\includegraphics[width=5.5in]{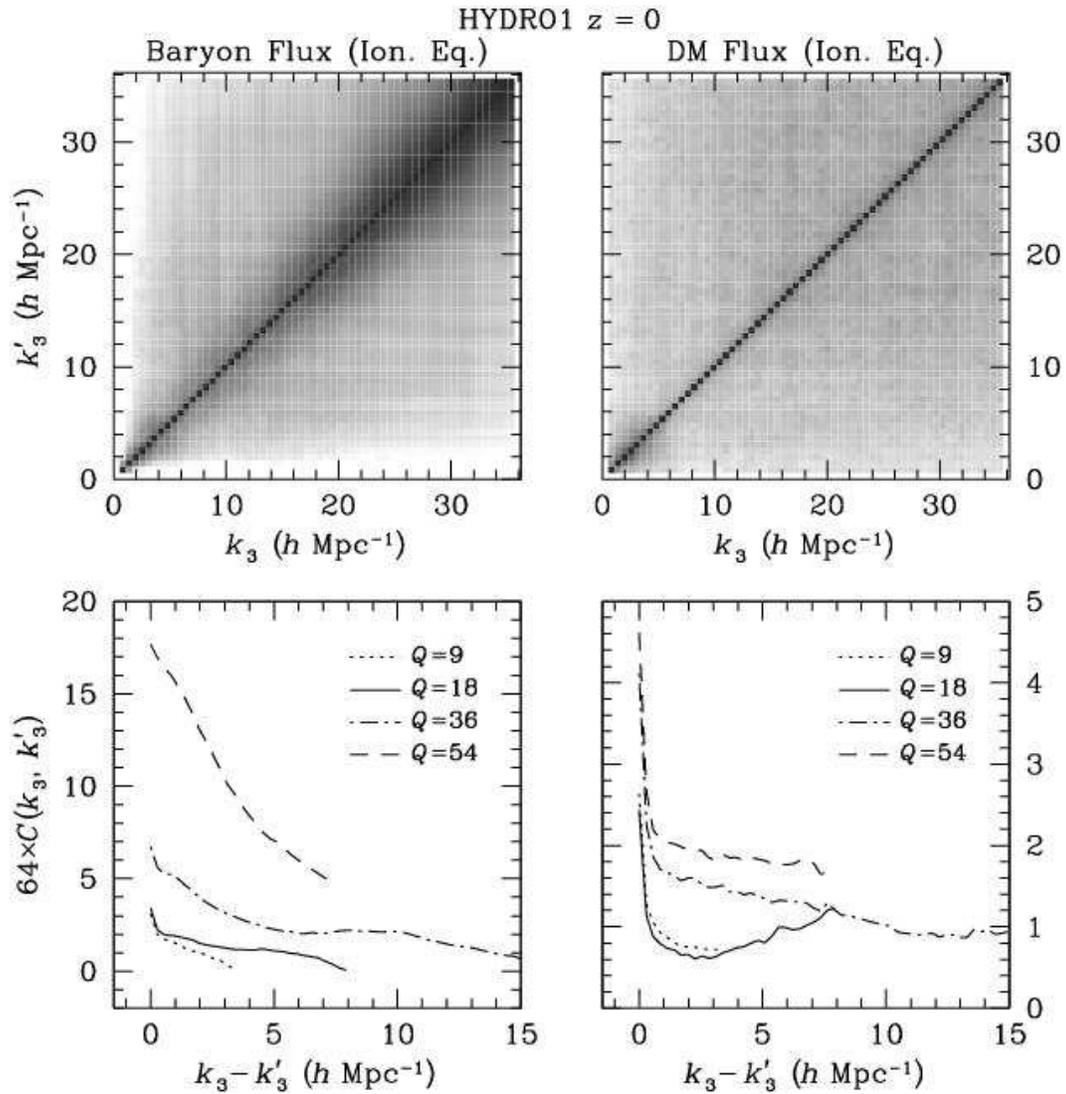}
\caption[Normalized covariances of flux power spectra based on baryon and 
dark matter distributions at $z = 0$]{
The same as Fig.~\ref{fig:cv0db64}, but for flux PS's of \lya{} forests
generated from baryons and dark matter using methods BI and DI.
\label{fig:cv0fdb64}}
\end{figure}

Figs.~\ref{fig:cv0db64} and \ref{fig:cv0fdb64} illustrate normalized 
covariances $\hat{C}(k_3, k'_3)$ and $C(k_3, k'_3)$ of \oned{} mass 
PS's and flux PS's, respectively, at $z = 0$. The covariances are 
calculated from 1000 groups, each of which consists of 64 LOS's ($N=64$) 
randomly selected from the density grid of HYDRO1. For GRFs, both 
$\hat{C}(k_3, k'_3)$ and $C(k_3, k'_3)$ are 
diagonal with $\hat{C}(k_3, k'_3) = \delta^{\rm K}_{n_3, n'_3}$ and 
$C(k_3, k'_3) = N^{-1}\delta^{\rm K}_{n_3, n'_3}$. For better comparison, 
the covariances $C(k_3, k'_3)$ are multiplied by $N$, so that the
Gaussian case has $N\,C(k_3, k'_3) = \delta^{\rm K}_{n_3, n'_3}$.
As already seen in Chapter \ref{ch:mps}, the covariances of 
\oned{} mass PS's are starkly non-Gaussian. The normalized variances 
$C(k_3, k_3)$ are two orders of magnitude higher than expected for GRFs.
The covariances of baryons are roughly a factor of 2 lower
than those of dark matter. This is likely due to the larger smoothing 
radii of SPH particles that filter out more small-scale non-Gaussian 
fluctuations.
The covariances of flux PS's have a dominant diagonal, though they are
still not Gaussian. The difference in the covariances between the 
full-hydro method BI and the pseudo-hydro method DI is comparable to the 
difference in their flux PS's. The method BI gives rise to stronger 
correlations between high-$k$ modes in the flux PS than the method DI
does because of the WHIM.

%\clearpage

\begin{figure}
\centering
\includegraphics[width=5.5in]{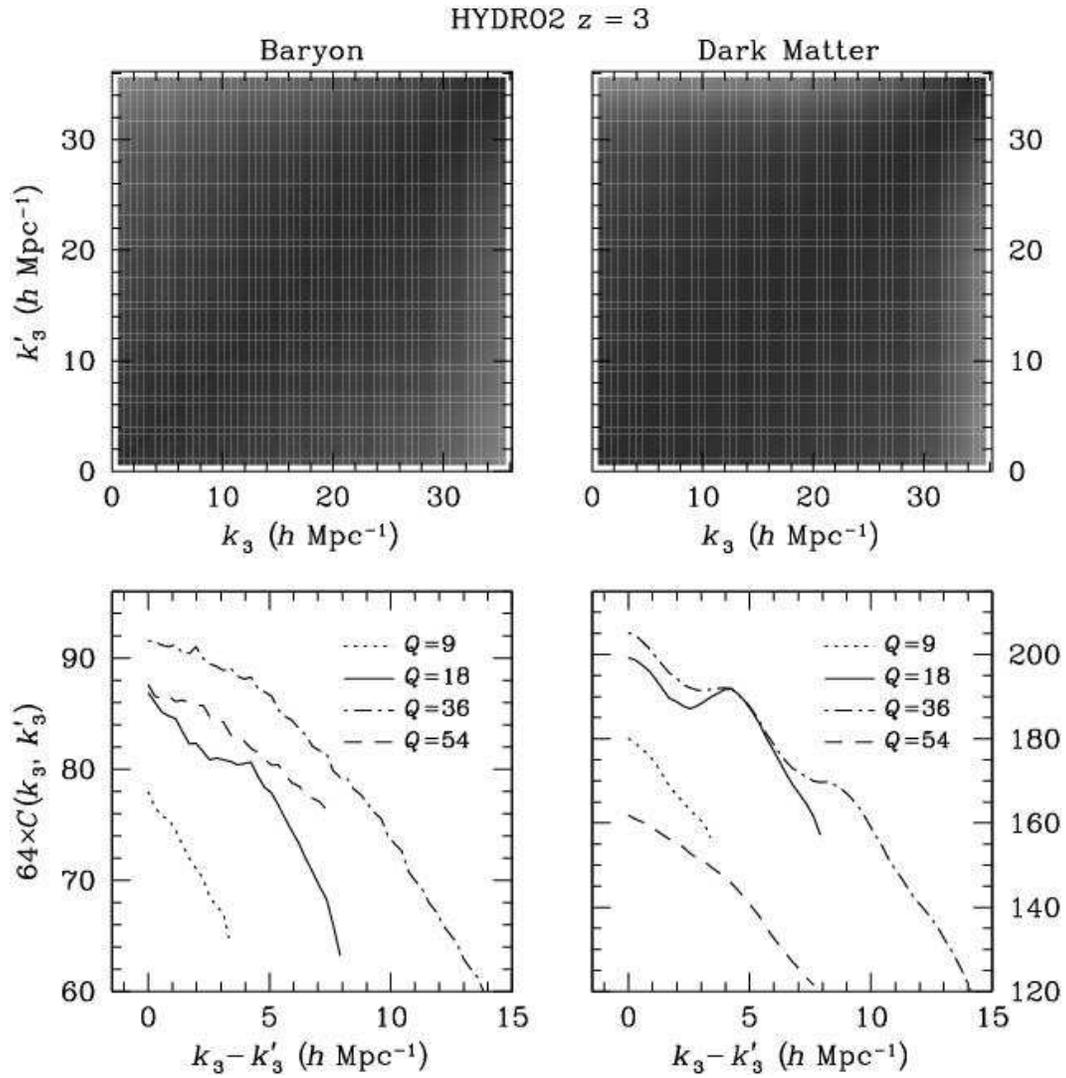}
\caption[Normalized covariances of \oned{} baryon and dark matter power 
spectra at $z = 3$]{
The same as Fig.~\ref{fig:cv0db64}, but for the simulation HYDRO2 
at $z = 3$.
\label{fig:cv3db64w}}
\end{figure}

\begin{figure}
\centering
\includegraphics[width=5.5in]{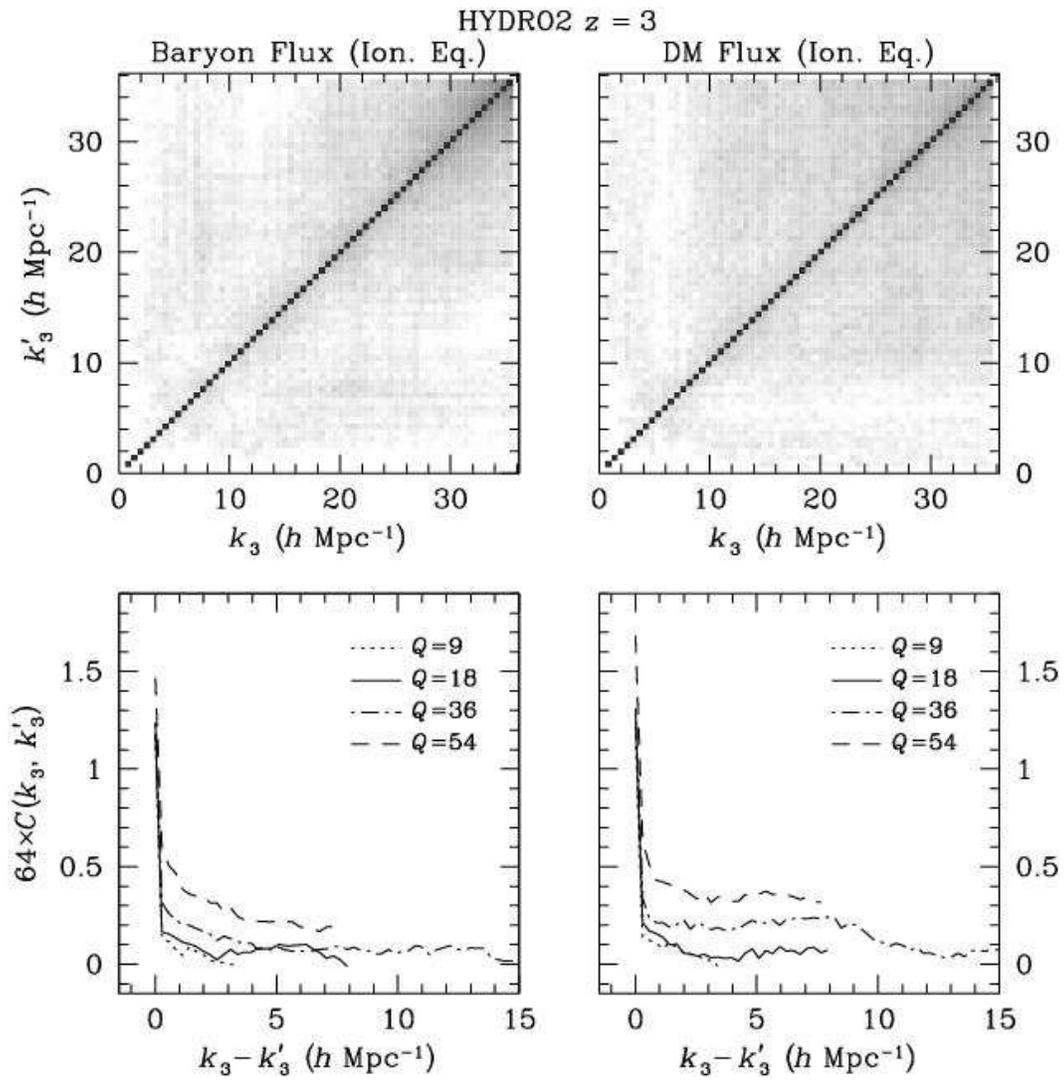}
\caption[Normalized covariances of flux power spectra based on baryon and 
dark matter distributions at $z = 3$]{
The same as Fig.~\ref{fig:cv3db64w}, but for flux PS's.
\label{fig:cv3fdb64w}}
\end{figure}

Figs.~\ref{fig:cv3db64w} and \ref{fig:cv3fdb64w} present normalized 
covariances 
of \oned{} mass PS's and flux PS's for the simulation HYDRO2 at $z = 3$, 
which are very similar to those for HYDRO1. At this redshift, the covariances
of \oned{} mass PS's are reduced by a factor of a few, but they are still 
highly non-Gaussian. Whereas, the covariances of flux PS's are very close to 
Gaussian, and the difference between the two methods BI and
DI is greatly reduced compared to that at $z = 0$. 

\begin{figure}
\centering
\includegraphics[width=5.5in]{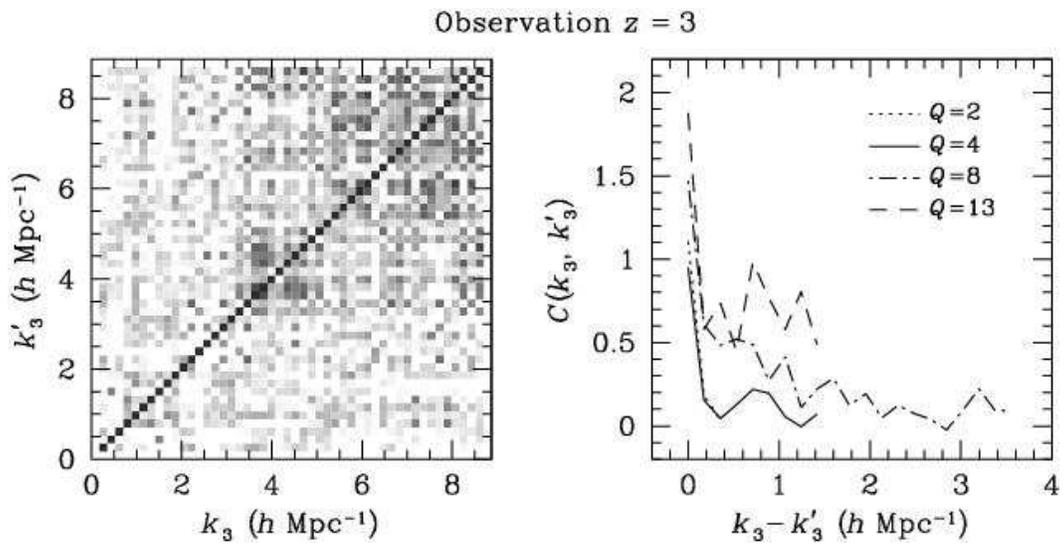}
\caption[Normalized covariances of observed flux power spectra at $z = 3$]{
The same as Fig.~\ref{fig:cv3fdb64w}, but for observed \lya{} forests. The
covariances are calculated from 27 segments of \lya{} forests
with $N=1$. Note that the resolution of the observed \lya{} forests
is about four times lower than that in the simulations.
\label{fig:cv3fobs}}
\end{figure}

In addition to simulations, I show in Fig.~\ref{fig:cv3fobs} normalized 
covariances of observed flux PS's at $z = 3$. The sample of \lya{} 
forests includes 20 QSO spectra from \citet{b94} and 
\citet{db96}. The QSO spectra 
are selected so that each contains at least one good chunk of spectrum 
that has no bad pixels or strong metal lines and spans 64 \AA{} anywhere
within $z = 2.9$--$3.1$. The spectral resolution is 1 \AA{}, which is about
four times lower than that in the simulations. In all, there are 27 
segments of \lya{} forests for the analysis. I do not re-group the 
segments, i.e.~$N = 1$, in calculating the covariances.

The main characteristics of the observed covariances are in good 
agreement with simulated ones. Namely, the normalized covariance matrices 
have a strongly dominant diagonal, and they are very close to Gaussian. The 
values of diagonal elements seem to match those in simulations, but 
the off-diagonal elements are a little noisier because of the small 
sample size. With a large number of high-resolution \lya{} forests, one
will be able to study the behavior of the covariance on smaller scales
(larger $k$) and reduce statistical uncertainties.

A general observation of covariances of flux PS's is that the correlation 
between two LOS modes decreases away from the diagonal, since two 
neighboring modes are more likely to be correlated than two distant modes. 
Beyond this point, however, 
the behavior of the covariances is not quantitatively understood. 
It also seems 
that the covariance of the flux PS does not provide more leverage for 
differentiating the underlying density field than the flux PS itself.
This echos the finding by \citet{mms03} that the flux trispectrum does not 
provide much extra information than the flux PS.

\clearpage

\section{Cosmology}
Because of the difficulty in deriving density statistics from flux 
statistics, one often resorts to numerical simulations and constrains 
cosmology by comparing observed flux statistics directly to simulated 
flux statistics. In addition, one utilizes fast $N$-body simulations 
and pseudo-hydro techniques in order to explore a large cosmological 
parameter space in manageable time. This necessitates an examination of
the accuracy of pseudo-hydro techniques and the sensitivity of 
flux statistics to cosmology.

\begin{figure}
\centering
\includegraphics[width=5.8in]{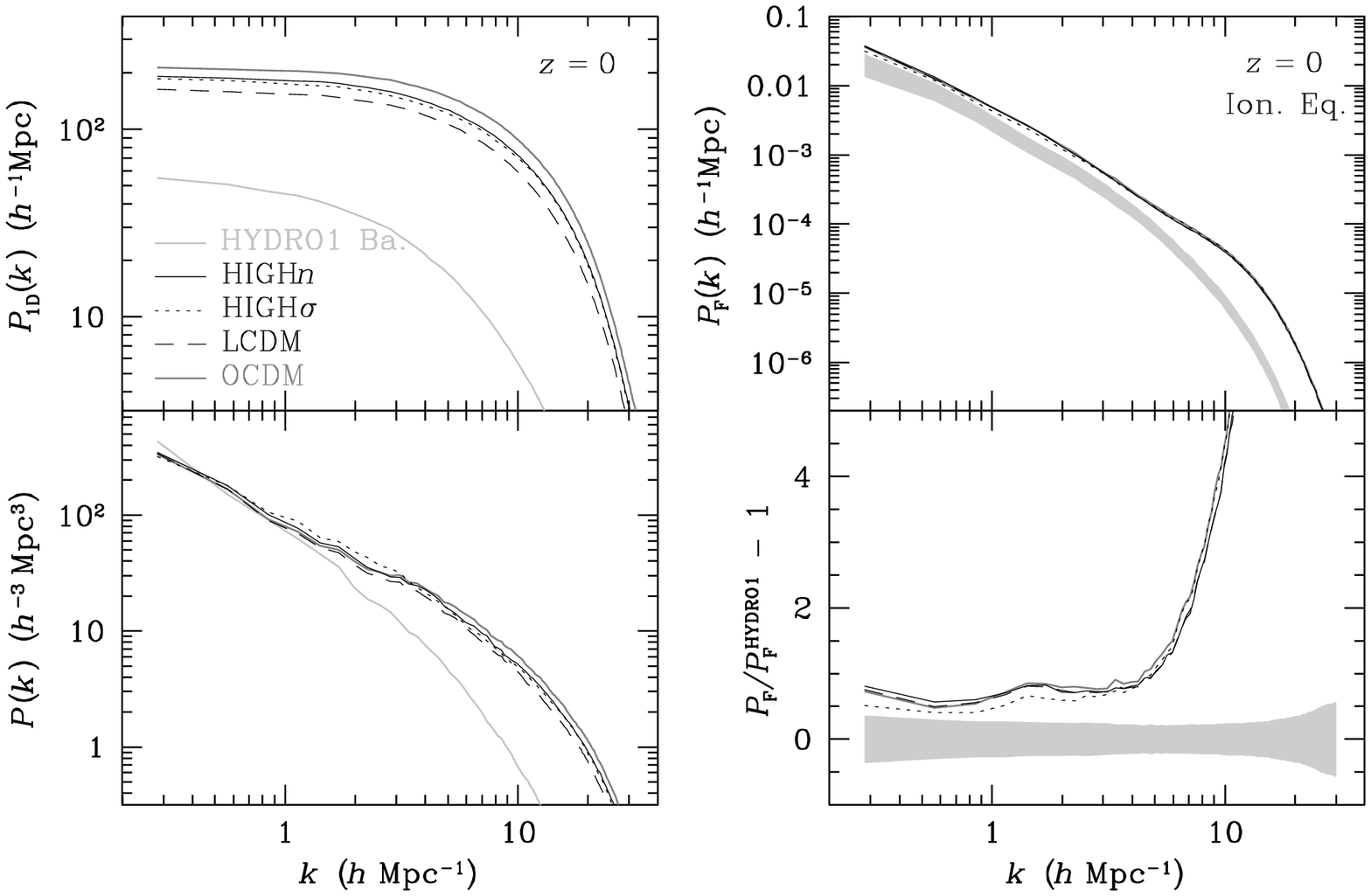}
\caption[Mass power spectra of baryons and dark matter and flux power 
spectra of simulated \lya{} forests for different 
cosmological models at $z = 0$]{
Mass PS's of baryons and dark matter and flux power spectra of 
simulated \lya{} forests for five cosmological models at $z = 0$. The 
upper left panel shows \oned{} mass PS's, and the lower panel 
\thrd{} mass PS's. The upper right panel shows flux PS's, and the lower panel
residuals of flux PS's with respect to the flux PS of \lya{} forests generated
with baryons in ionization equilibrium from HYDRO1 (light
grey lines and bands). All others flux PS's are from $N$-body simulations 
using the method DI.
\label{fig:cosmfps0}}
\end{figure}

\begin{figure}
\centering
\includegraphics[width=5.8in]{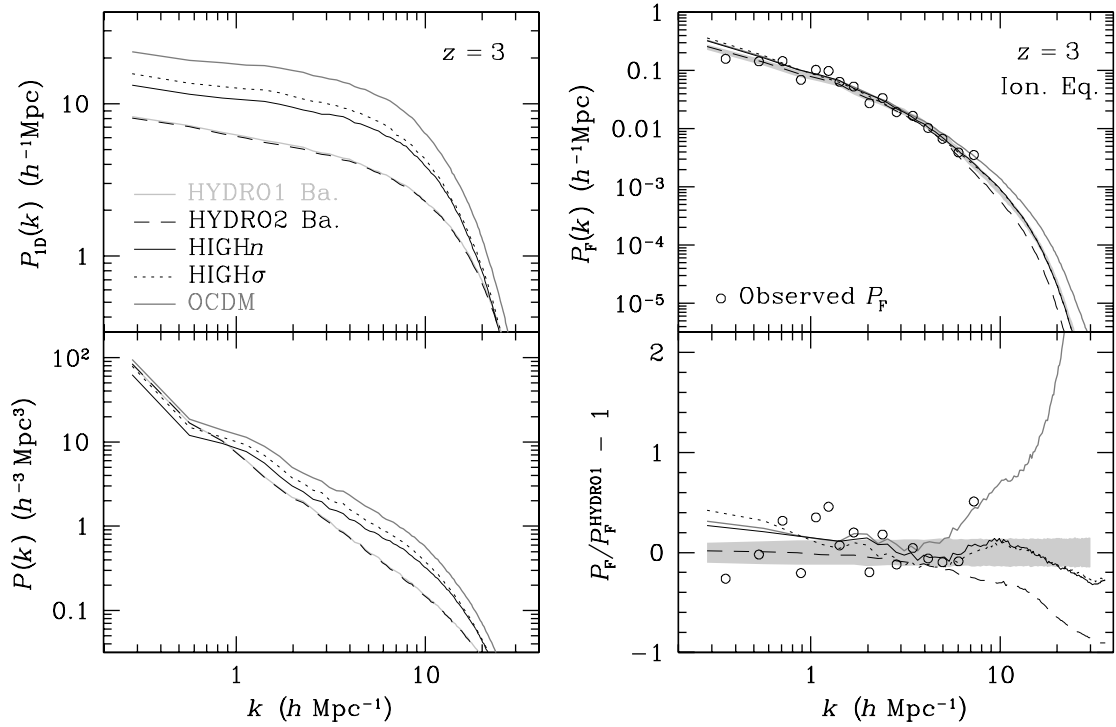}
\caption[Mass power spectra of baryons and dark matter and flux power 
spectra of simulated \lya{} forests for different 
cosmological models at $z = 3$]{
The same as Fig.~\ref{fig:cosmfps0} except that dark-matter-based PS's 
from the simulation LCDM is replaced by baryon-based PS's from HYDRO2, and
that all PS's are at $z = 3$. The observed flux PS is measured from 27
segments of \lya{} forests between $z = 2.9$ and $3.1$.
\label{fig:cosmfps3}}
\end{figure}

Figs.~\ref{fig:cosmfps0} and \ref{fig:cosmfps3} compare mass PS's and 
flux PS's for six simulations: HYDRO1, HYDRO2, LCDM, high-$n$ LCDM (HIGH$n$),
high-$\sigma_8$ LCDM (HIGH$\sigma$), and open CDM (OCDM). Table 
\ref{tab:para1} lists parameters for all the models. The $N$-body 
simulations all have the same box size of $22.222$ \mpc{} on each side and
evolve $128^3$ CDM particles from $z = 49$ to 0 using {\sc gadget}. Note that 
the HIGH$n$ model has an opposite tilt than HYDRO1 and HYDRO2. Not all the
simulations are consistent with most recent observations, and they are 
provided only for testing the cosmological dependence of the flux PS.

Pseudo-hydro techniques have already been proven inaccurate at low redshift 
by several tests above. I include the results at $z = 0$ here only to show 
that all the flux PS's based on the method DI are nearly 
indistinguishable from each other except the HIGH$\sigma$ model. 

At $z = 3$, the LCDM model is replaced 
by HYDRO2. Interestingly, HYDRO1 and HYDRO2 simulations have identical 
\oned{} mass PS's and \thrd{} mass PS's. Their flux PS's differ on small 
scales. This is due to the factor that HYDRO2 has, on average, a higher 
IGM temperature, and it might not be directly resulted from the 
difference in cosmological parameters. There is also a considerable 
difference of flux PS's between $N$-body simulations and hydrodynamical 
simulations at $k < 1$ \mpci, but such difference is already present in 
Fig.~\ref{fig:fpshionT} where pseudo-hydro techniques are applied to 
dark matter in the simulation HYDRO1 itself. Therefore, it cannot be 
attributed to the cosmological model. The only $N$-body simulation that 
significantly deviates from HYDRO1 is OCDM---assuming that the method DI 
works equally well for OCDM. 

On scales below a few \mpc{} ($k > $ a few \mpci), Fig.~\ref{fig:fpshion}
suggests that detailed knowledge of the state of the IGM is needed in
order to correctly reproduce the flux PS. On large scales,
Fig.~\ref{fig:cosmfps3} reveals significant differences between 
cosmological models and between the full-hydro method and pseudo-hydro 
techniques. Therefore, to constrain cosmology using the flux PS and 
$N$-body simulations, one 
should have precise calibrations of pseudo-hydro techniques and 
focus on scales above a few \mpc{}.

It is seen that the flux PS of the 27 \lya{} forest segments is roughly 
matched by all the simulations at $z = 3$. The scatter of the observed 
flux PS is too large to be useful for determining cosmological parameters
because of the small size of the sample. But with a much larger number of 
\lya{} forests, it is possible to reduce the sample variance error 
of the flux PS and place meaningful constraints on cosmology.

\chapter[Inverting the Ly$\alpha$ Forest]{Inverting the 
Ly$\boldsymbol{\alpha}$ Forest} \label{ch:inv}
\noindent
There are two ways of obtaining density statistics from the \lya{} forest. 
One is to measure flux statistics and then map them into density 
statistics. The other is to measure statistics of densities inverted 
from the \lya{} forest. As discussed in previous
chapters, pseudo-hydro techniques are good approximations at $z=3$, but 
they still require precise calibrations using hydrodynamical simulations. 
On the other hand, it is not practical to search 
the parameter space using hydrodynamical simulations that incorporate
all important astrophysical processes. Therefore, it is worthwhile  
exploring methods of inverting the \lya{} forest.

Baryon densities $\rho_{\mathrm{b}}$ and mass densities $\rho$ may be 
extracted from transmitted fluxes $F$ of Ly$\alpha$ forests 
using the same equation that is used in the pseudo-hydro technique DE, i.e.
\begin{equation} \label{eq:flux-dens}
F = e^{-\tau} \simeq e^{-A(\rho_\mathrm{b}/\bar{\rho}_{\rm b})^\gamma} 
\simeq e^{-A(\rho/\bar{\rho})^\gamma}.
\end{equation}
More accurately, one should also include ionization equilibrium, thermal 
broadening, etc. A question may be raised here: given the uncertainty of 
pseudo-hydro techniques, what can be gained by inverting the \lya{} 
forest using equation (\ref{eq:flux-dens})? 
It is seen in Figs.~\ref{fig:fdmba0} and \ref{fig:fpshionT} that the 
lack of temperature information is the major source of error for 
pseudo-hydro techniques. Whereas, one could determine temperatures of
observed \lya{} forests from line profiles and obtain relatively 
accurate baryon densities. Then, the mass PS of baryons on large scales 
can be used to constrain cosmology.

When the density 
is high enough, the spectrum is saturated, i.e.~$F \simeq 0$. With noises
and uncertainties in the spectrum, a direct inversion using equation 
(\ref{eq:flux-dens}) is very unreliable in saturated regions. Despite 
the difficulty, methods of direct inversion are systematically developed,
for example, with Lucy's method by \citet{nh99}, and with Bayesian method 
for a three-dimensional inversion by \citet{pvr01}. One may also use 
higher order lines to recover the optical depth and the underlying density
\citep{cs98,ast02}, even though the contamination by lower order lines needs
to be carefully removed. Once LOS densities are obtained, many 
statistics, such as the \oned{} mass PS, can be measured.

The saturation problem is avoided if one maps the mass PS 
directly from the flux PS of the Ly$\alpha$ forest 
without an inversion \citep{cwk98,cwp99,cwb02}.
However, a close examination of the Fourier transform of equation 
(\ref{eq:flux-dens}) shows that Fourier modes on different scales are mixed 
by the nonlinear density--flux relation (see Section \ref{sec:map}). 
The mixing depends on the underlying density field, 
and it is hard to predict analytically.

If the inversion is necessary, a proper treatment in saturated regions
has to be developed. In many physical systems, sizes are often correlated
with other quantities such as masses and densities. For example, more 
massive stars or dark matter halos have larger sizes, but 
lower mean densities \citep{bm98,nfw96}. One may expect a similar 
trend for the saturated Ly$\alpha$ absorption.
On the contrary, the mean density is found to increase with the 
width of saturation. This is due to the fact that the IGM is very 
diffuse and far from virialization, while the other objects mentioned above
are the opposite. 

In this chapter, densities are expressed in units of 
their corresponding cosmic mean densities, and the subscript for LOS 
wavenumber is dropped for convenience.

\section[Correlation between the Mean Density and Width of Saturated 
\lya{}\\ Absorptions]{Correlation between the Mean 
Density and Width of Saturated Ly$\boldsymbol{\alpha}$ Absorptions} 
\label{sec:scaling}
Neglecting the probability that two physically separate absorption systems 
fall in the same redshift coordinate, one can associate a saturated 
absorption in the Ly$\alpha$ forest with a single high density region. 
If I assume further that there is no substructure present, and the IGM 
evolves more or less the same way everywhere, then the size of the saturated 
region has to be tightly correlated with its mean density 
\citep[see also][]{s01}. In 
reality, the neglected elements above and uncertainties elsewhere will 
introduce a spread to the correlation. 

\subsection{Simulations} \label{sec:sim}
Figs.~\ref{fig:fdmba3}, \ref{fig:mfps23}, and 
additional tests at $z = 4$ demonstrate that pseudo-hydro techniques 
work very well at $z \gtrsim 2$. Therefore,
I choose $N$-body simulations to investigate the relation between the
mean density and width of saturated \lya{} absorptions. 

A standard particle-particle-particle-mesh \citep[P$^3$M,][]{he81} 
code developed by \citet{jf94} is used to evolve
128$^3$ dark matter particles in a cubic box of 12.8 \mbox{$h^{-1}$Mpc} 
(comoving) each side. The initial PS is given by the 
fitting formula from 
\citet{b86}. The model parameters are listed in Table \ref{tab:para}. All 
the models start from $z = 15$ and stop at $z = 3$ in 950 steps.
\begin{table}
\centering
\caption[Parameters of the $N$-body simulations]
{Parameters of the $N$-body simulations.}
\label{tab:para}
\begin{tabular}{cccccc}
\hline
Model & $\Omega$ & $\Omega_\Lambda$ & $h$ & $\Gamma^{\ a}$ & 
	$\sigma_\mathrm{8}$ \\
\hline
LCDM & 0.3 & 0.7 & 0.7 & 0.21 & 0.85 \\
OCDM & 0.3 & 0   & 0.7 & 0.21 & 0.85 \\
SCDM & 1.0 & 0   & 0.5 & 0.5  & 0.67 \\
TCDM & 1.0 & 0   & 0.5 & 0.25 & 0.60 \\
\hline
\end{tabular}

%\medskip
\parbox{0.43 \textwidth}{
\begin{itemize}
\setlength{\leftskip}{-0.2in}
\setlength{\itemsep}{0.0ex}
\item[$^a$] $\Gamma$ is the shape parameter of the PS.
\end{itemize}}
\end{table}

\begin{figure} 
\centering
\includegraphics[width=5.5in]{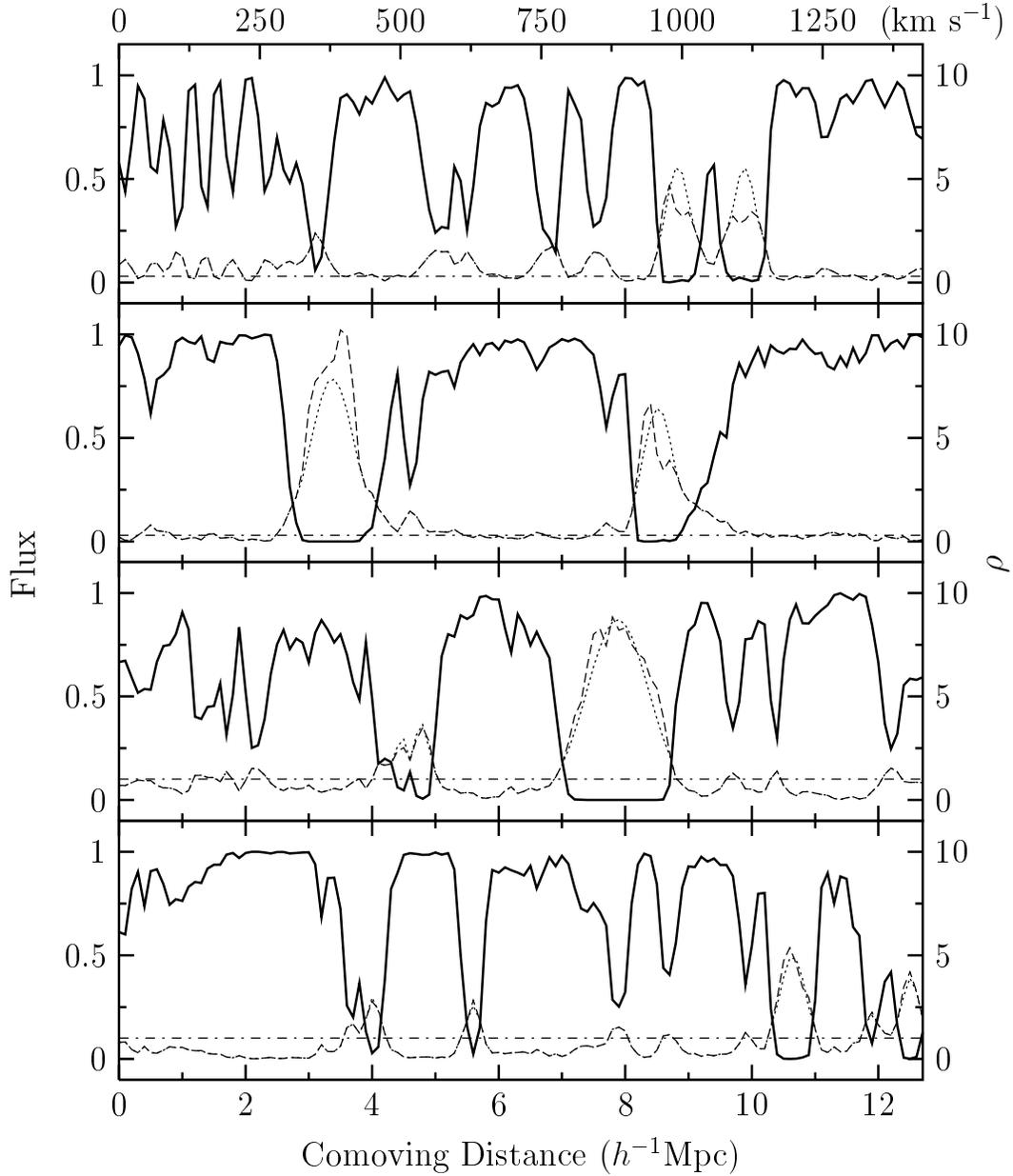}
\caption[Simulated Ly$\alpha$ forests and line-of-sight densities from 
the LCDM model at $z=3$]{Simulated Ly$\alpha$ forests and LOS 
densities from the LCDM model at $z=3$. The solid
lines are the flux $F$, the dashed lines are the matter density $\rho$, 
the dotted lines are the recovered density, and the horizontal dash-dotted
lines indicate the threshold flux level $\eta=0.03$ 
(see Section \ref{sec:test}) in the upper 2 panels, and $\eta =0.10$ in the 
lower 2 panels. }
\label{fig:sample}
\end{figure}

The parameter $A$ in equation (\ref{eq:flux-dens}) is chosen to fit
the mean flux equation (\ref{eq:mflx}). I extrapolate the mean flux up 
to $z = 4.5$, which is not critical to the analysis in Section 
\ref{sec:rho-w} but is, nevertheless, supported by simulations 
(Riediger et al.~1998). %\citep{rpm98}. 
The constant $\gamma$ in equation (\ref{eq:flux-dens}) is set to
$1.6$ since it is the best fit in Fig.~\ref{fig:feos23}. 
Four samples of the Ly$\alpha$ forest are shown in 
Fig.~\ref{fig:sample}. They are drawn from the LCDM 
simulation at $z=3$. The corresponding LOS densities are plotted along
with the fluxes. It is evident that most of the Ly$\alpha$ lines arise 
where $\rho \lesssim 10$.

\subsection{The Correlation} \label{sec:rho-w}
I define the width $w$ of the saturation as the distance between the two 
points that bracket the absorption at a given threshold flux level $\eta$.
In other words, it is the width of a region in which $F<\eta$. 
Since the flux $F$ in the saturated region is dominated by noise, one 
should set the threshold above the noise level. 
That is why I have not added any noise in the simulations: noise 
outside the saturation can be routinely dealt with, while noise
in the saturation, being below the threshold, does not matter.
Within a reasonable range of spectral quality, I choose the 
threshold $\eta=0.03$ and $0.10$ for tests below.

\begin{figure} 
\centering
\includegraphics[width=4.5in]{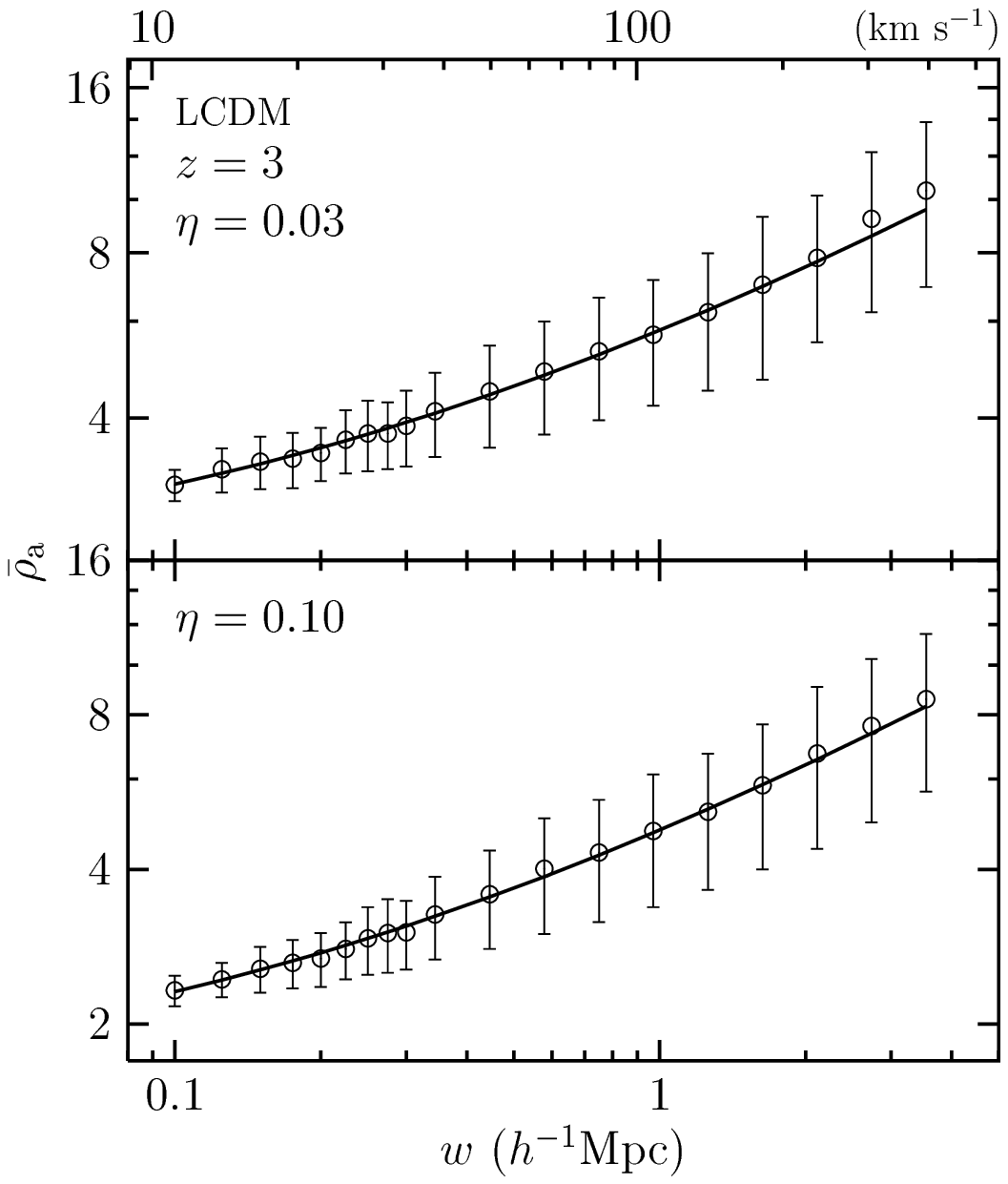}
\caption[Correlation between the mean density 
$\bar{\rho}_{\rm a}$ and width $w$ of saturated regions in the 
Ly$\alpha$ forest]{The correlation between the mean density 
$\bar{\rho}_{\rm a}$ and width $w$ of saturated regions in the 
Ly$\alpha$ forest. The trend lines fit the data well beyond 
$2\ h^{-1}$Mpc, even though they are obtained by fitting only the data 
with $w \leq 2\ h^{-1}$Mpc. The data above $0.3\ h^{-1}$Mpc are binned in 
logarithmic intervals for readability. Similar treatments apply
to Figs.~\ref{fig:powerspect}, \ref{fig:powerspect-ost}, 
\ref{fig:powerspect-moc}, and \ref{fig:bdk} as well. 
\label{fig:scaling}}
\end{figure}

Once the width is determined, the mean density $\bar{\rho}_{\rm a}$ can be
readily calculated from the corresponding LOS density field. 
Since the 
density and the flux are assigned only on grid points, interpolation is 
needed to find $w$ and $\bar{\rho}_{\rm a}$. Fig.~\ref{fig:scaling} shows the
scaling relation for the LCDM model at $z=3$. It clearly
demonstrates that $\bar{\rho}_{\rm a}$ increases with $w$. Furthermore, the 
correlation is reasonably tight for narrow saturations.
Note that the trend lines are obtained by fitting only the 
data with $w \leq 2\ h^{-1}$Mpc, because wider saturations are very rare
as compared to the rest.

The $\bar{\rho}_{\rm a}$--$w$ relation is well fitted by
\begin{equation}
\bar{\rho}_{\rm a}=\rho_\mathrm{0} + a\ (w/h^{-1}\mbox{Mpc})^b.
\end{equation}
The term $\rho_\mathrm{0}$ sets a baseline for the scaling, because the 
matter density is non-vanishing even if there is no saturation.
The parameter $a$ is a scaling coefficient, which reflects the overall
amplitude of the density fluctuation, and the exponent $b$ is more or
less determined by the nature of hierarchical clustering. The width $w$
is in units of $h^{-1}$Mpc. Table 
\ref{tab:scaling} lists the values for the LCDM model at $z = 3$ and $4.5$.
\begin{table} \centering
\caption[Parameters in the fitting $\bar{\rho}_{\rm a}=\rho_\mathrm{0} + 
a \ (w/h^{-1}\mbox{Mpc})^b$ for the LCDM model]{Parameters in the fitting 
$\bar{\rho}_{\rm a}=\rho_\mathrm{0} + a \ (w/h^{-1}\mbox{Mpc})^b$
for the LCDM model.}
\label{tab:scaling}
\begin{tabular}{ccccc} 
\hline
$z$ & $\eta$ & $\rho_\mathrm{0}$ & a & b \\
\hline
   3    & 0.03 & 1.93  & 3.85 & 0.545 \\
        & 0.10 & 1.39  & 3.39 & 0.564 \\
  4.5   & 0.03 & 0.11  & 2.78 & 0.455 \\
        & 0.10 & 0.24  & 2.04 & 0.556 \\
\hline
\end{tabular}
\end{table}
The reason why $\rho_\mathrm{0}$ decreases with redshift is that the universe 
is more uniform early on, so that even low density regions have to absorb 
a substantial amount of Ly$\alpha$ flux to produce the low mean flux. 
This is possible because the neutral fraction of the IGM at $z=4.5$ is 
higher than that at $z=3$. As the universe evolves, density fluctuations 
grows stronger and stronger, and the scaling coefficient $a$ becomes larger 
and larger. The exponent
$b$ has changed little over $z = 3$--$4.5$. The parameters also show a 
dependence on the threshold flux $\eta$, because $\eta$ sets the
threshold density $\rho_\eta$, above which the $\bar{\rho}_{\rm a}$--$w$ 
relation is explored.

\begin{figure} 
\centering
\includegraphics[width=5.5in]{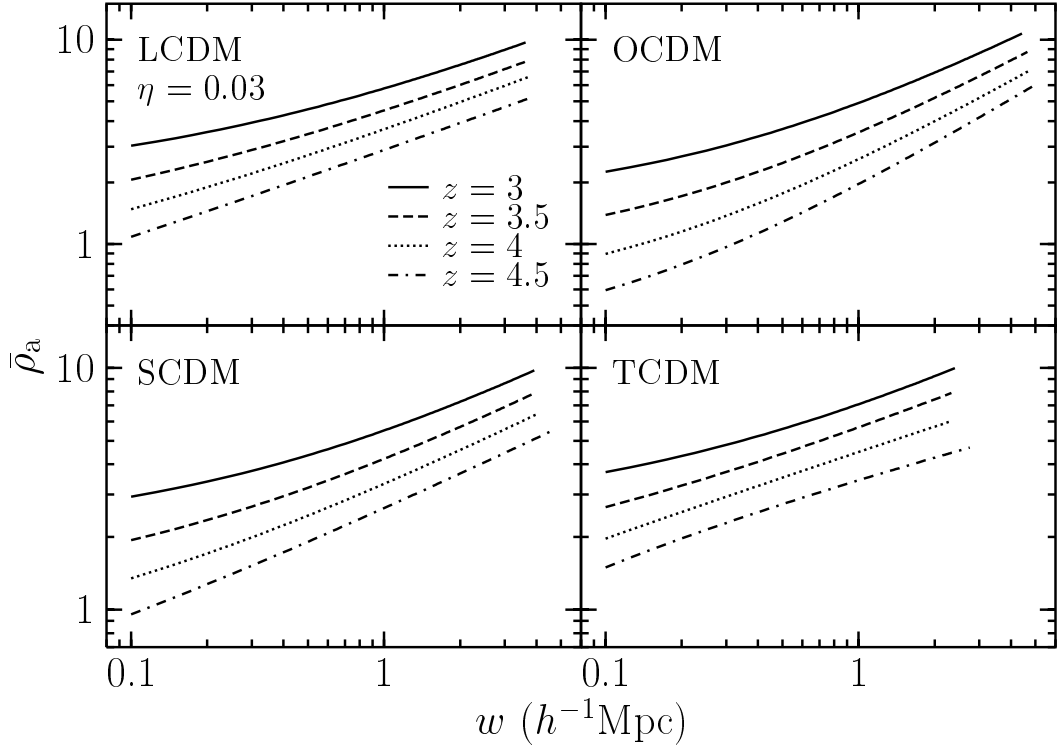}
\caption[Best-fittings of the $\bar{\rho}_{\rm a}$--$w$ relation
for four cosmological models from $z=3$ to $4.5$]
{Best-fittings of the $\bar{\rho}_{\rm a}$--$w$ relation
for the four models at $z=3$--$4.5$. The threshold $\eta = 0.03$. 
The parameters are obtained by fitting the saturations with 
$w \leq 2\ h^{-1}$Mpc.}
\label{fig:scaling-lost}
\end{figure}

Fig.~\ref{fig:scaling-lost} shows the evolution of the 
$\bar{\rho}_{\rm a}$--$w$ relation for the four models. The difference between 
models is mostly due to the amplitude of fluctuations---in other words, the PS.
\ualofhack

\subsection{A Simple Model}
The $\bar{\rho}_{\rm a}$--$w$ relation is analogous---but not completely 
equivalent---to the 
curve of growth \citep[e.g.][]{pr93}, which studies the correlation between 
the \mbox{H\,{\sc i}} column density $N_{\rm HI}$ and the 
equivalent width $W$ of Ly$\alpha$ absorptions. The similarity is as 
follows. The width $w$ is approximately the same as $W$, 
and the mean matter density $\bar{\rho}_{\rm a}$ is proportional to 
$w^{-1}N_{\rm HI}$ if $\rho \propto \rho_\mathrm{b}$ holds true. Since
$N_{\rm HI} \propto W$ at small values of $W$, it is not 
surprising to see $\bar{\rho}_{\rm a}$ grow slowly, 
i.e.~$\mathrm{d}\ln \bar{\rho}_{\rm a} /\mathrm{d} \ln w \sim 0$, 
for small $w$.

On the other hand, the $\bar{\rho}_{\rm a}$--$w$ relation addresses saturated
absorptions where, according to the curve-of-growth analysis, $W$ is almost 
a constant independent of the \mbox{H\,{\sc i}} column density. 
Therefore, $\bar{\rho}_{\rm a}$ would have risen steeply against $w$, 
i.e.~$b \gg 1$. This 
apparent inconsistency arises from the cosmological context, because the 
width $w$ is determined not only by $N_{\rm HI}$ but also by the physical
extent of the (relatively) dense region through Hubble expansion and 
peculiar velocities. 

Although it is not obvious why the exponent $b$ is less than unity, I can make 
an order-of-magnitude estimate using the density profile 
$\rho(r) \propto r^{-1}$ 
from a spherical self-similar infall. I modify the profile to avoid the
singularity at $r=0$ by adding a smoothing length $\epsilon$, so that 
$\rho(r) \propto (r^2+\epsilon^2)^{-1/2}$. The mean density within the 
radius $R$, at which $\rho=\rho_\eta$, is 
\begin{eqnarray} \label{eq:rho-R}
\bar{\rho}_{\rm a}(R) &=& \frac{3}{R^3}\int_0^R \rho(r) r^2dr = 
\frac{3\rho_\eta\sqrt{R^2+\epsilon^2}}{2R^3} \times \\ \nonumber
& & \left[R\sqrt{R^2+\epsilon^2}
+\epsilon^2\log \left(\frac{\epsilon}{R+\sqrt{R^2+\epsilon^2}}\right)\right].
\end{eqnarray}
From equation (\ref{eq:rho-R}), one gets an estimate 
$b\sim \mathrm{d}\ln\bar{\rho}_{\rm a}/\mathrm{d}\ln R = 0.39,\ 0.42$, and $0.43$ for 
$\epsilon=0.2,\ 0.3$, and $0.4\ h^{-1}$Mpc respectively at $R=2\ h^{-1}$Mpc. 
The same quantity for the $\bar{\rho}_{\rm a}$--$w$ relation is 
$\mathrm{d}\ln \bar{\rho}_{\rm a} /\mathrm{d} \ln w=0.41$, where I have used the 
parameters of the
LCDM model at $z=3$ with $\eta = 0.03$ (see Table \ref{tab:scaling}). It 
should be emphasized that the profile $\rho(r) \propto r^{-1}$ is not quite 
justified for the IGM at $z = 3$, and without any modification, it gives the 
same mean density within the boundary $\rho=\rho_\eta$ regardless the size 
of the system, i.e.~$b=0$. Therefore, equation (\ref{eq:rho-R}) is not 
expected to give a good approximation of the $\bar{\rho}_{\rm a}$--$w$ 
relation.

\section{Inversion with A Gaussian Density Profile} \label{sec:test}

If the flux level is above the threshold, equation (\ref{eq:flux-dens}) 
can be used to find the density. When it is below the 
threshold, one must provide a density profile that matches $\bar{\rho}_{\rm a}(w)$
and $\rho_\eta$ to fill in the missing information in the saturation. 

While the density profile is worth studying in its own right, I simply 
choose the Gaussian profile 
\begin{equation}
\rho(s) = \frac{B}{\sigma \sqrt{2\pi}}
	\exp\left[-\frac{1}{2}\bigl(\frac{s-s_\mathrm{0}}{\sigma}\bigr)^2\right],
\end{equation}
where $s$ is the coordinate in redshift space, $s_\mathrm{0}$ is the center of 
the saturation, and $B$ and $\sigma$ are solved simultaneously from
\begin{eqnarray}
\bar{\rho}_{\rm a}(w) &=& \frac{1}{w} \int^{s_\mathrm{0}+w/2}_{s_0-w/2} \rho(s) ds = 
\frac{B}{w}\mbox{ erf}\Bigl(\frac{1}{2\sqrt{2}}\frac{w}{\sigma}\Bigr), \\
\rho_\eta &=& \rho(s_\mathrm{0} \pm w/2) = \frac{B}{\sigma \sqrt{2\pi}}
	\exp\left[-\frac{1}{8}\bigl(\frac{w}{\sigma}\bigr)^2\right], \nonumber
\end{eqnarray}
where erf($x$) is the error function.
The Gaussian profile has the advantage that it does
not introduce any artificial power on small scales. However, it is
arguable that the right amount of small-scale power should be added
through the profile, and so a more realistic profile may be needed. 

\begin{figure} 
\centering
\includegraphics[width=5.in]{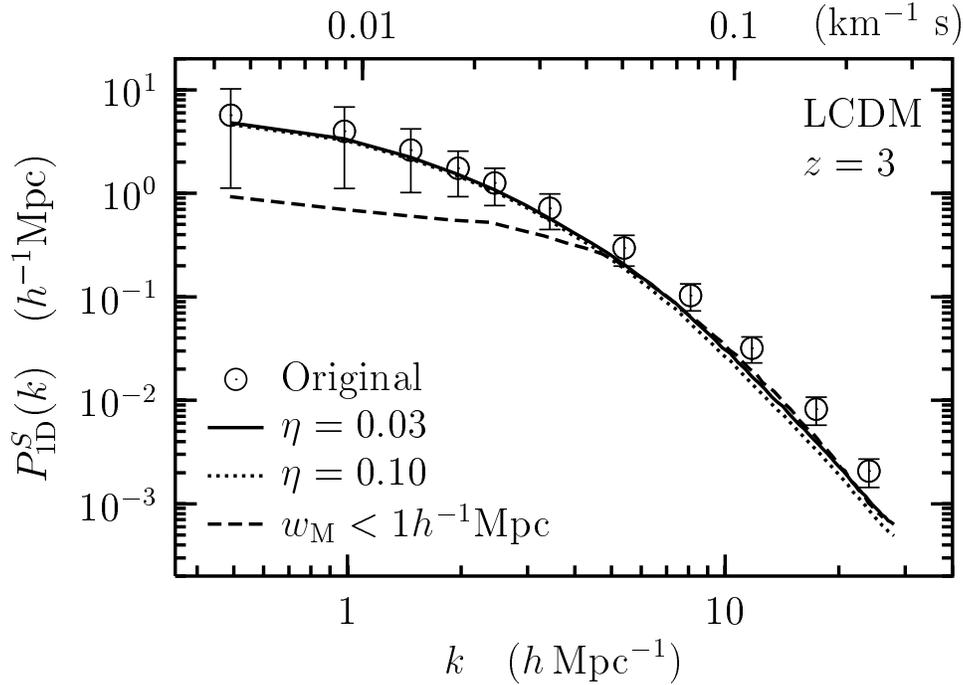}
\caption[One-dimensional mass power spectra of original densities and 
inverted densities in redshift space for the LCDM model]
{One-dimensional mass PS's of original densities and 
inverted densities in redshift space. 
Circles include all LOS's from the N-body simulation, and the 
dashed line contains only the ones that have a maximum width of 
saturation $w_\mathrm{M} < 1 \ h^{-1}$Mpc. 
The recovered densities are inverted from fluxes with thresholds 
$\eta = 0.03$ (solid line) and $\eta = 0.1$ (dotted line).
The error bars are 1$\sigma$ dispersions 
among 610 groups, each of which consists of 20 LOS's. The error 
bars of the recovered densities, which are not plotted, are comparable to 
that of the original densities. 
\label{fig:powerspect}}
\end{figure}

The recovered LOS densities with $\eta = 0.03$ and 
$\eta = 0.10$ are shown 
along with the four original LOS densities and fluxes in 
Fig.~\ref{fig:sample}. Since a universal density profile is 
employed in the inversion, the recovered densities do not necessarily
match the original densities.

\begin{figure} 
\centering
\includegraphics[width=5in]{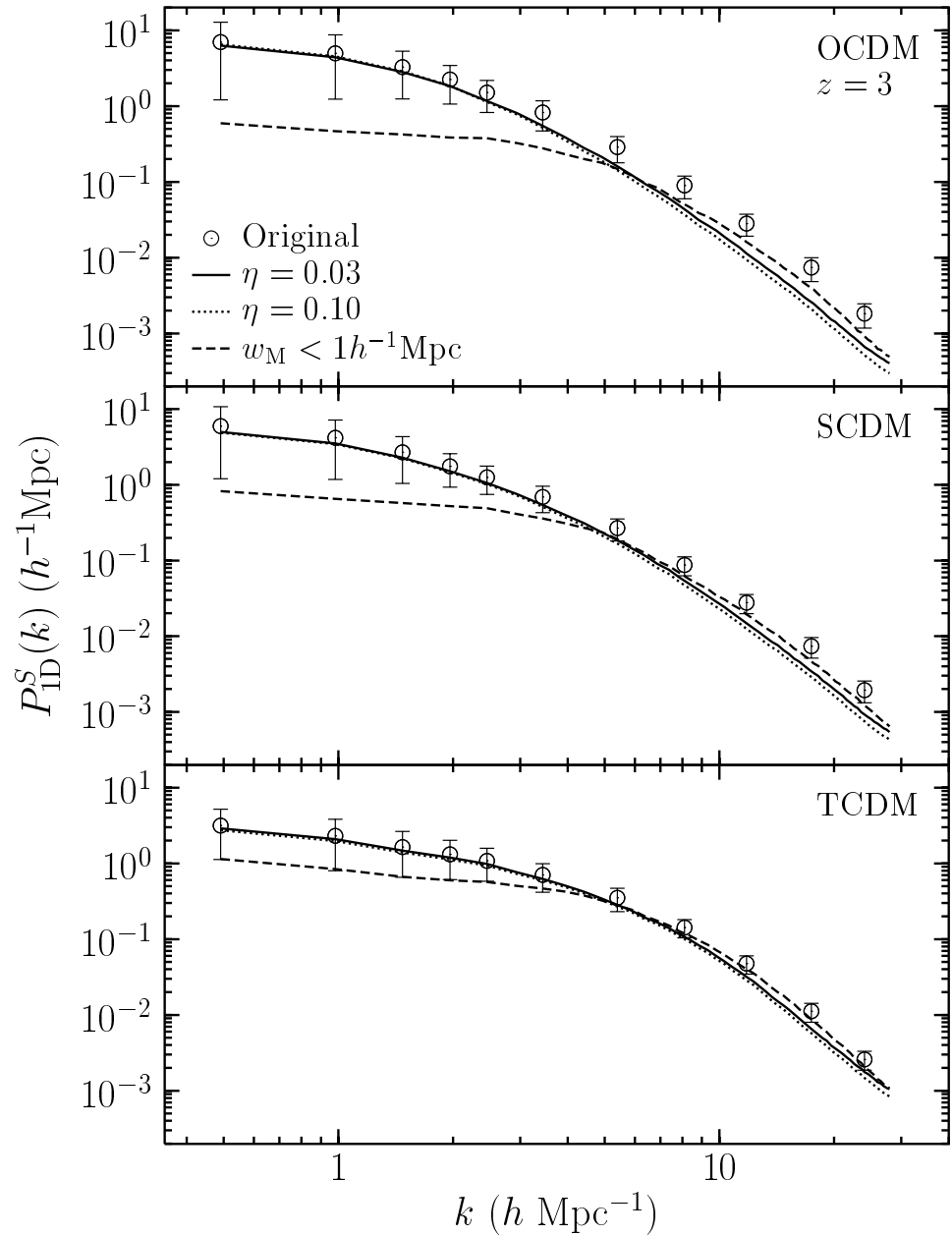}
\caption[One-dimensional mass power spectra of original densities and 
inverted densities in redshift space for OCDM, SCDM, and TCDM models]
{The same as Fig.~\ref{fig:powerspect}, but for OCDM, 
SCDM, and TCDM models.
\label{fig:powerspect-ost}}
\end{figure}

To assess the statistical quality of the inversion, I plot in 
Fig.~\ref{fig:powerspect} the original and recovered mass PS's of 
the LCDM model with different flux thresholds. Other models are shown 
in Fig.~\ref{fig:powerspect-ost}. The recovered PS has good agreement 
with the original PS on large scales 
($k \lesssim 3\ h$\,Mpc$^{-1}$), but it is underestimated on smaller scales, 
where the Gaussian profile essentially has 
no power. The signal to noise ratio, or the threshold flux, has little 
influence on large scales, but a low noise level does slightly improve the 
recovery on small scales. 

\begin{figure} 
\centering
\includegraphics[width=5.in]{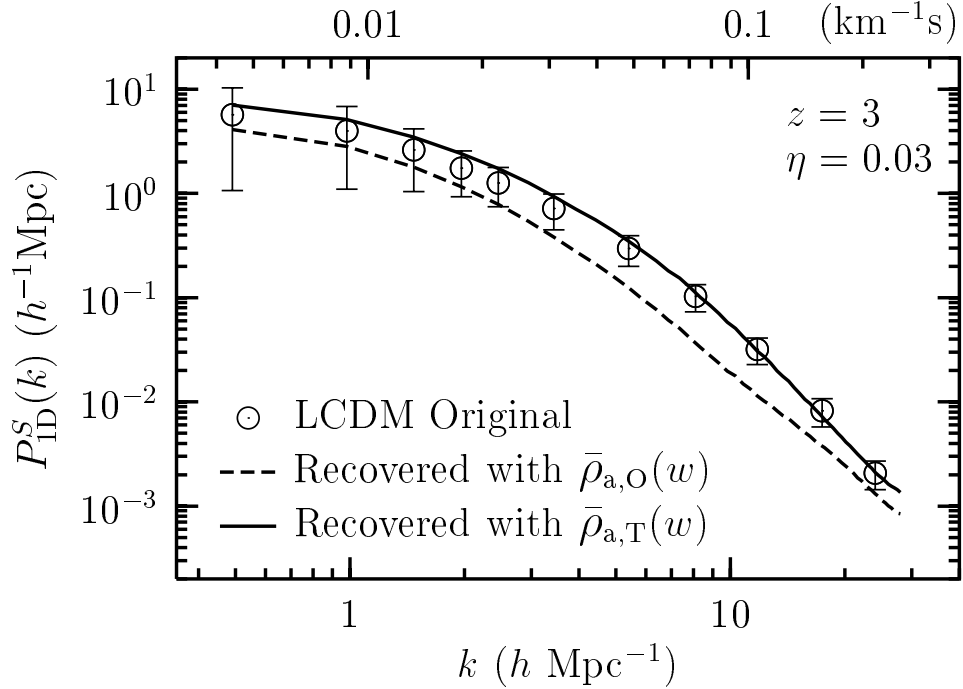}
\caption[Recovery of the \oned{} mass power spectrum with uncertainties 
in the $\bar{\rho}_{\rm a}$--$w$ 
relation]{The same as Fig.~\ref{fig:powerspect}, except that the 
fluxes are drawn from the LCDM simulation, while the densities are 
recovered with $\bar{\rho}_{\rm a}$--$w$ relations from OCDM and TCDM 
simulations.
\label{fig:powerspect-moc}}
\end{figure}

For comparison,
I show in Figs.~\ref{fig:powerspect} and \ref{fig:powerspect-ost} the 
original PS of the LOS densities that have a maximum width of 
saturation $w_\mathrm{M} < 1 \ h^{-1}$Mpc. This is equivalent to 
removing---or giving 
less weight to---the saturated regions when one measures the PS.
It is seen that the saturated regions are very important 
to scales $\gtrsim 2\ h^{-1}$Mpc, while the unsaturated regions give a good
estimate of the small-scale power. Thus one may improve the recovery of 
PS as follows. First, invert the Ly$\alpha$ forest with 
equation (\ref{eq:flux-dens}) and the $\bar{\rho}_{\rm a}$--$w$ relation. Second, 
do the inversion after removing the saturated regions. Finally, the 
best-estimate of the PS is just 
the common envelope of the PS's of the two recovered densities. 

Fig.~\ref{fig:scaling-lost} indicates that the $\bar{\rho}_{\rm a}$--$w$ 
relation varies from model to model. Thus it is necessary to check if
the recovery is sensitive to $\bar{\rho}_{\rm a}(w)$---in other words, if it is
model dependent. In Fig.~\ref{fig:powerspect-moc} 
I plot the PS's of LCDM densities 
recovered with $\bar{\rho}_{\rm a,O}(w)$ and $\bar{\rho}_{\rm a,T}(w)$ from 
OCDM and tilted cold dark matter (TCDM) 
models respectively. It seems that by boosting $\bar{\rho}_{\rm a}(w)$ a small
amount [$\bar{\rho}_{\rm a,T}(w)$ lies a little higher than $\bar{\rho}_{\rm a,L}(w)$], one 
gets even better estimate of the PS on small scales. However,
lowering $\bar{\rho}_{\rm a}(w)$ could underestimate the power by a factor of 2
on large scales, and even more on small scales. The effect of this model 
dependence could be reduced with the constraint on small scales, since it
is possible to 
recover the small-scale power well by removing the saturations.

\begin{figure} 
\centering
\includegraphics[width=5.in]{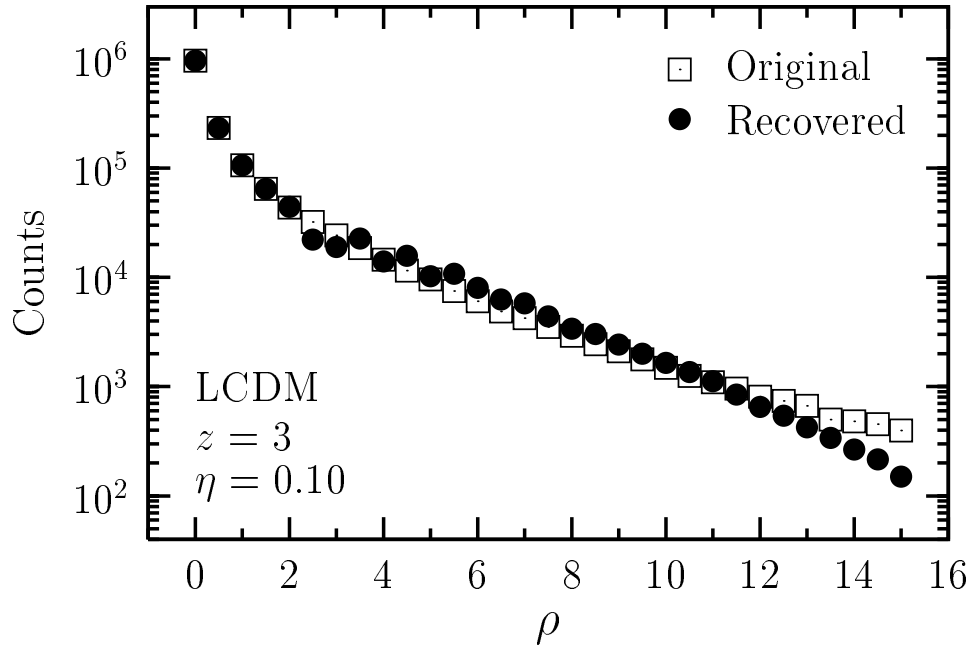}
\caption[One-point distribution function of inverted densities 
and that of the original densities]
{The one-point distribution 
function of inverted densities (filled circles) and that of the 
original densities (open squares).
\label{fig:one-point}}
\end{figure}

Fig.~\ref{fig:one-point} tests a different statistics---the one-point 
distribution function for the LCDM model. It is evident that the Gaussian 
profile 
leads to a drop of the probability at high densities. In other words, 
it statistically reduces the heights of density peaks. 

\section{Mapping the Power Spectra}
\label{sec:map}
It seems that the inversion is no longer needed at least for determining
the mass PS if one establishes a direct mapping between 
the flux and the mass PS's \citep{cwb02}. It is shown that given
a set of cosmological parameters, there is a statistical mapping, which 
reliably recovers the mass PS from the flux PS
\citep{gh02}, even though the mapping is model dependent.

\begin{figure} 
\centering
\includegraphics[width=5.in]{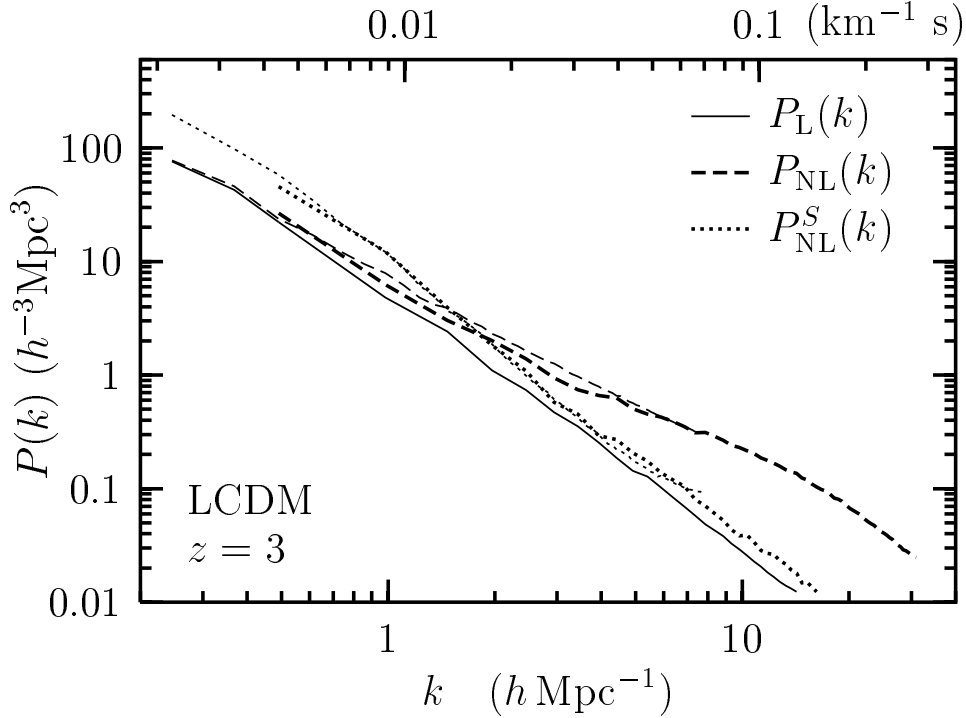}
\caption[Three-dimensional mass power spectra of the LCDM model at $z=3$]
{Three-dimensional mass PS's of the LCDM model at $z=3$. The
dashed lines are the real-space PS's, the dotted lines are
the redshift-space PS's averaged over solid angles, 
and the thin solid line is the 
initial PS (at $z=15$) evolved to $z=3$ by linear theory. 
The thin dashed line and the thin dotted line are from an additional 
simulation with the same parameters as the LCDM model except that the 
box is 51.2 \mbox{$h^{-1}$Mpc} each side.
\label{fig:nonl}}
\end{figure}

The physical links between the three-dimensional linear mass PS 
$P_\mathrm{L}(k)$ and the flux PS can be 
summarized by the flowchart:
\[
P_\mathrm{L}(k) \longrightarrow P_\mathrm{NL}(k) 
\longrightarrow P_\mathrm{NL}^S(\mathbf{k}) 
\longrightarrow P^S_{\rm 1D}(k) \buildrel d^2(k)\over
\longrightarrow P_\mathrm{F}(k),
\]
where $P_\mathrm{NL}(k)$ is the \thrd{} nonlinear mass PS 
\citep{pd96, th96}, $P_\mathrm{NL}^S(\mathbf{k})$ is 
$P_\mathrm{NL}(k)$ 
in redshift-space \citep{k87, p99, zf03}, and 
$d^2(k) = P_\mathrm{F}(k) / P^S_{\rm 1D}(k)$. 
Note that $P_\mathrm{NL}^S(\mathbf{k})$ is anisotropic.
I show in Fig.~\ref{fig:nonl} 
that the nonlinear evolution and redshift distortion are already
significant at $z=3$. The fact that the departure from linear evolution 
is very similar in both large and small 
box simulations indicates that such departure is real, and the cosmic 
density field has already gone nonlinear below 10 \mbox{$h^{-1}$Mpc} 
($k > 0.6$ \mpci) at $z=3$ %\citep[see also][]{zjf01,pvr01}. 
(see also Pichon et al.~2001; Zhan et al.~2001).
The angularly averaged \thrd{} PS $P_\mathrm{NL}^S(k)$, i.e.~the monopole, 
does not give a complete view of the
difference between the real-space PS and redshift-space PS, but it does 
show that peculiar velocities boost the power on 
large scales and reduce it substantially on small scales as seen in 
Fig.~\ref{fig:psdmba}. A two-dimensional projection of 
$P_\mathrm{NL}^S(\mathbf{k})$ can be found in \citet{p01}.

In contrast, the mapping in \citet{cwb02} follows a simplified path:
\[
P_\mathrm{L}(k) \longrightarrow P_\mathrm{1D,L}(k) \buildrel b^2(k)\over 
\longrightarrow P_\mathrm{F}(k),
\]
where $P_\mathrm{1D,L}(k)=(2\pi)^{-1}\int_k^\infty P_\mathrm{L}(y) y dy$,
and $b^2(k)= P_\mathrm{F}(k)/P_\mathrm{1D,L}(k)$. 
Given a one-dimensional density field in redshift space 
at $z=3$, one can only
measure $P^S_{\rm 1D}(k)$, and so $P_\mathrm{1D,L}(k)$ is not observable unless the density
field is linear on all scales and no peculiar velocity is present. In
other words, $P^S_{\rm 1D}(k)$ is directly connected to $P_\mathrm{F}(k)$, while $P_\mathrm{1D,L}(k)$ 
is not.

Around the cosmic mean matter density, i.e.~$|\delta| \ll 1$, equation 
(\ref{eq:flux-dens}) is approximately
\begin{equation} \label{eq:delta=0}
F(s) \simeq e^{-A}-A\gamma e^{-A}\delta(s),
\end{equation}
where I have included the dependence on the redshift coordinate $s$ 
explicitly. It is seen in the linearized equation (\ref{eq:delta=0}) that 
the Fourier modes of the flux are proportional to those of the overdensity 
when $k \neq 0$. Hence, the flux PS is proportional to the 
mass PS
\begin{equation}
P_\mathrm{F}(k) \simeq A^2\gamma^2e^{-2A}P^S_{\rm 1D}(k), \quad k \neq 0.
\end{equation}
Caution should be taken when using equation (\ref{eq:delta=0}), because 
it is valid only if one smoothes the Ly$\alpha$ forest over large scales
%\citep[see, for example,][]{zjf01, zf02}.
(see, for example, Zhan et al.~2001; Zhan \& Fang 2002).

It is obvious from equation (\ref{eq:flux-dens}) and 
Fig.~\ref{fig:sample} that the condition $|\delta(s)| \ll 1$ is not 
satisfied when $F\simeq 1$ or $F\simeq 0$. Therefore, one has to include 
higher order terms of $\delta(s)$. Taking the simplest case, in which 
$\delta(s) = \sin(k_\mathrm{0}s)$ and only $\delta^2(s)$ term is added to 
equation (\ref{eq:delta=0}), one immediately finds that $P_\mathrm{F}(k)$ 
contains spurious power on the mode $2k_\mathrm{0}$ that $P^S_{\rm 1D}(k)$ 
does not contain. In general, the 
nonlinear density--flux relation distorts, spreads, and mixes power in 
the cosmic density field over different scales in 
the flux PS. I have not focused other sources of distortion such as 
linear filtering \citep{gh98}, line profile \citep[see][]{g92}, and 
instrumentation, since they are well studied in the literature.

\begin{figure} 
\centering
\includegraphics[width=5.in]{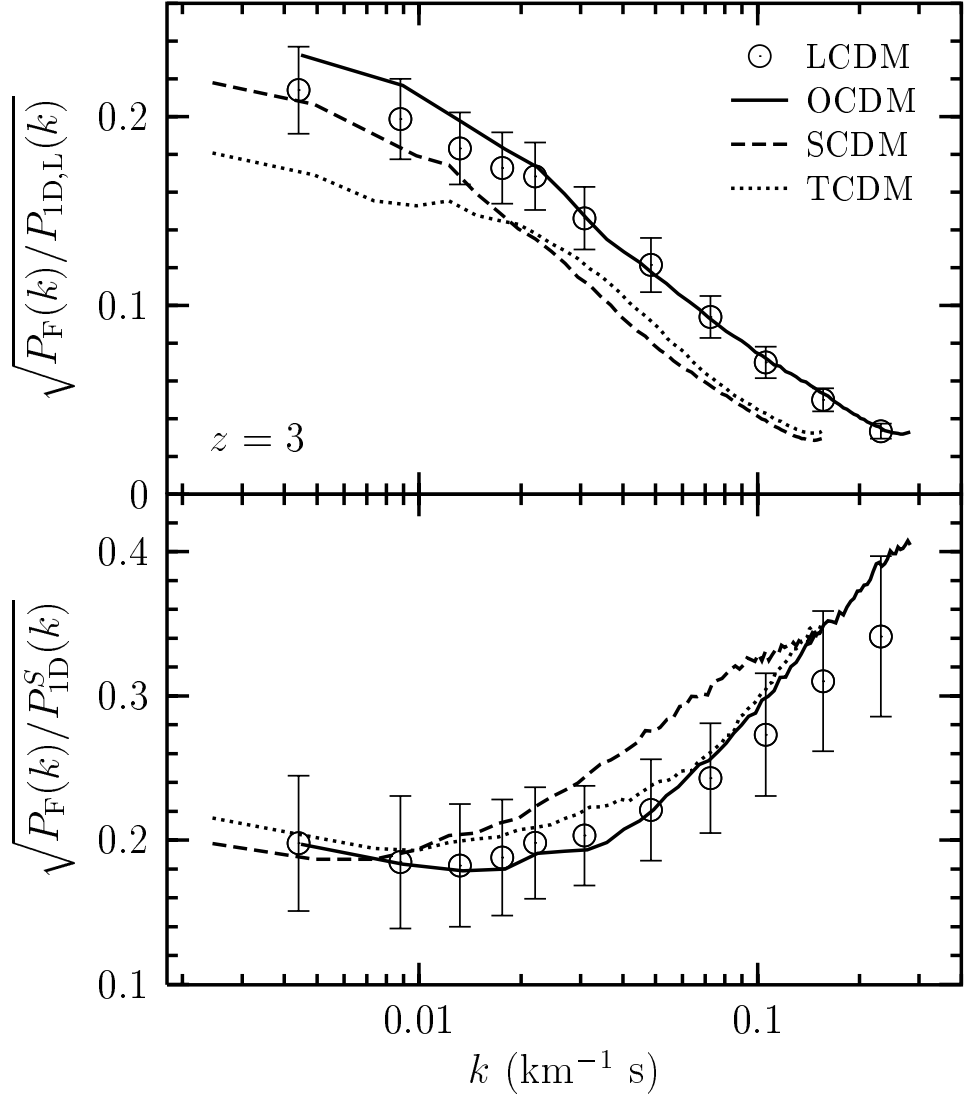}
\caption[Ratios $b(k)=\sqrt{P_\mathrm{F}(k)/P_\mathrm{1D,L}(k)}$
and $d(k)=\sqrt{P_\mathrm{F}(k)/P^S_{\rm 1D}(k)}$]
{The ratio $b(k)=\sqrt{P_\mathrm{F}(k)/P_\mathrm{1D,L}(k)}$ (upper panel)
and $d(k)=\sqrt{P_\mathrm{F}(k)/P^S_{\rm 1D}(k)}$ (lower panel). The Ly$\alpha$ forests 
are divided into groups, each of which consists of 20 LOS's.
The PS's are averaged within each group except $P_\mathrm{1D,L}(k)$, which 
is theoretical. The error bars of the LCDM model are 1$\sigma$ 
dispersions among 610 groups. Error bars of 
other models are comparable to that of the LCDM model. 
\label{fig:bdk}}
\end{figure}

Fig.~\ref{fig:bdk} shows the ratios $b(k)$ and $d(k)$.
The dispersion in $d(k)$ is larger than that of $b(k)$, because both
$P_\mathrm{F}(k)$ and $P^S_{\rm 1D}(k)$ contribute to the dispersion 
of $d(k)$, while $b(k)$ receives only a single contribution from 
$P_\mathrm{F}(k)$. Even so, the error propagation would cause a large 
scatter in determining $P_\mathrm{1D,L}(k)$ and $P_\mathrm{L}(k)$. In fact, 
the scatter can be large enough that one would not be able to determine the 
cosmological model based only on the recovered $P_\mathrm{L}(k)$. 
On the other hand, the model dependence of $b(k)$ is strong enough, so that 
one may have to assume a cosmological model to recover
the mass PS from the flux PS. The difference among
the models is not solely due to $\sigma_\mathrm{8}$. 
For example, the standard cold dark matter (SCDM) model has a slightly 
higher $\sigma_\mathrm{8}$ than the TCDM 
model, so one would expect $b(k)$ of the SCDM model to have the same shape 
as that of the TCDM model. This is not observed in 
Fig.~\ref{fig:bdk}, however. It is also evident from the behavior of $b(k)$ 
and 
$d(k)$ that the PS of the underlying density field $P^S_{\rm 1D}(k)$ is 
significantly lower than the linear mass PS $P_\mathrm{1D,L}(k)$ on scales
$k>0.02$ \mbox{(km s$^{-1}$)$^{-1}$}. This is due mostly to peculiar 
velocities, or redshift distortion.

\chapter{Conclusions} \label{ch:con} 
\noindent
I have investigated the covariance of the one-dimensional mass PS for
both GRFs and simulated density fields. The nonlinear evolution and 
non-Gaussianity of the cosmic density field on small scales 
%\citep[see also][]{zjf01, spj03, zsh03, z03} 
(see also Zhan et al.~2001; Smith et al.~2003; Zaldarriaga et al.~2003; 
Zhan 2003) have caused correlations between the
fluctuations on different scales and increased the cosmic variance of the
one-dimensional mass PS. Because of this, a large number of LOS's are 
needed to accurately measure the one-dimensional mass PS and recover the 
three-dimensional mass PS.

The length of LOS's introduces a window function in the LOS 
direction, 
which mixes neighboring Fourier modes in the cosmic density field. 
Fig.~\ref{fig:covlen} has demonstrated for simulations that the length 
of LOS's does affect the covariance of the one-dimensional mass PS. 
The covariance of the observed one-dimensional mass PS will receive a 
significant contribution from this effect, because in practice the 
length of LOS's suitable for study is always much less than the size 
of the observable universe [see equation (\ref{eq:windlen})].

One may reduce the cosmic variance by binning \emph{independent}
Fourier modes. However, since the modes of the cosmic density field are
strongly correlated, binning will be less effective in reducing the sample
variance error. On the other hand, the non-Gaussian behavior of the 
covariance provides important information of the field such as the 
trispectrum.

Using hydrodynamical simulations and $N$-body simulations, I find 
that pseudo-hydro techniques are able to reproduce 
the flux and flux PS that are obtained using the full-hydro method
at $z \gtrsim 2$. There is also a good match between observed and 
simulated flux statistics such as the flux PS and its 
covariance at $z =3$.

The accuracy of pseudo-hydro techniques does not seem to be high 
enough for determining cosmology at percent level. 
One needs to precisely calibrate pseudo-hydro techniques with
hydrodynamical simulations. Moreover, it is better to constrain 
cosmology using the flux PS on scales above a few \mpc{} in order to 
reduce the uncertainties caused by the IGM temperature and thermal 
broadening. At low redshift, the shock-heated WHIM has greatly 
altered the temperature--density relation of the IGM, so that fluctuations
in the \lya{} flux are strongly suppressed by thermal broadening. 
As a result, pseudo-hydro techniques overpredict the flux PS at low 
redshift.

The transform from density to flux quenches fluctuations by orders of 
magnitude and leads to near-Gaussian \lya{} fluxes. Hence, the variance
of the flux PS is much less than that of the \oned{} mass PS. In other 
words, one can measure the flux 
PS to a high precision with a relatively small number of LOS's, but the
underlying mass PS cannot be determined as precisely as the flux PS.
Attempts to recover the \thrd{} mass PS should take into account the 
sample variance error of the \oned{} mass PS. 

It is possible to invert the \lya{} forest and measure statistics of
\oned{} density fields directly. 
I provide a scaling relation between the mean matter density 
$\bar{\rho}_{\rm a}$ and the width of saturated Ly$\alpha$ absorptions 
$w$, which helps invert saturated \lya{} absorptions. 
The inversion with the $\bar{\rho}_{\rm a}$--$w$ relation is able to recover 
the redshift-space \oned{} mass PS fairly well on scales above 
$2\ h^{-1}$Mpc, but it
underestimates the power on small scales due to the use of the Gaussian 
profile. An improvement is suggested based on
the observation that the small-scale \oned{} mass PS can be better 
recovered if saturated absorptions are removed. Thus, by combining both the 
$\bar{\rho}_{\rm a}$--$w$ inversion and the result after removing 
saturated absorptions, one could get a good estimate of the \oned{} mass PS 
over a wide range of scales.

The $\bar{\rho}_{\rm a}$--$w$ relation provides an important constraint to 
the inversion of the Ly$\alpha$ forest. To incorporate it in any 
inversion scheme, one needs to determine the
$\bar{\rho}_{\rm a}$--$w$ relation according to spectral 
resolution and noise. The threshold flux can be conveniently set 
to a small value above the noise level. The recovered redshift-space 
\oned{} mass PS is not very sensitive to the $\bar{\rho}_{\rm a}$--$w$ 
relation on large scales. However, one would greatly underestimate the
large-scale \oned{} mass PS if saturated regions were excluded from the sample.

As seen in Chapters \ref{ch:mps} and \ref{ch:fps}, \oned{} density fields 
have a large cosmic variance in their PS due to non-Gaussianity. 
Nevertheless, this variance can be reduced by a large number of 
LOS's. With thousands of \lya{} forests from  
QSO surveys, it is possible to subdue the variance to 
a few percent level, and the \lya{} forest may eventually become 
comparable to other 
fields of precision cosmology such as the CMB and weak lensing. 

\sloppy

\end{document}